\documentclass[reqno,11pt]{amsart}
\usepackage[utf8]{inputenc}
\usepackage{enumitem}
\usepackage{graphicx}
\usepackage{amscd}
\usepackage{slashed}
\usepackage{amssymb}
\usepackage{mathtools} 
\usepackage{pstricks}
\usepackage[mathscr]{eucal}
\numberwithin{equation}{section}
\allowdisplaybreaks[1]

\title[Complex Structures on Jet Spaces and Bosonic Fock Space Dynamics]{Complex Structures on Jet Spaces
and Bosonic Fock Space Dynamics for Causal Variational Principles}

\author[F.\ Finster]{Felix Finster}
\address{Fakult\"at f\"ur Mathematik \\ Universit\"at Regensburg \\ D-93040 Regensburg \\ Germany}
\email{finster@ur.de}

\author[N.\ Kamran]{Niky Kamran \\ \\ August 2018 / November 2019}
\address{Department of Mathematics and Statistics \\ McGill University \\ Montr{\'e}al \\ Canada}
\email{nkamran@math.mcgill.ca}

\newtheorem{Def}{Definition}[section]
\newtheorem{Thm}[Def]{Theorem}
\newtheorem{Prp}[Def]{Proposition}
\newtheorem{Lemma}[Def]{Lemma}

\newtheorem{Corollary}[Def]{Corollary}

\newcommand{\Thanks}{\vspace*{.5em} \noindent \thanks}
\newcommand{\beq}{\begin{equation}}
\newcommand{\eeq}{\end{equation}}
\newcommand{\Proof}{\begin{proof}}
\newcommand{\QED}{\end{proof} \noindent}

\newcommand{\la}{\langle}
\newcommand{\ra}{\rangle}
\newcommand{\bra}{\mathopen{<}}
\newcommand{\ket}{\mathclose{>}}

\newcommand{\C}{\mathbb{C}}
\newcommand{\R}{\mathbb{R}}
\newcommand{\1}{\mbox{\rm 1 \hspace{-1.05 em} 1}}

\newcommand{\N}{\mathbb{N}}

\newcommand{\h}{\mathfrak{h}}
\newcommand{\J}{\mathfrak{J}}
\newcommand{\Jin}{\mathfrak{J}^\text{\rm{\tiny{in}}}}
\newcommand{\Ctest}{C^\text{\rm{\tiny{test}}}}
\newcommand{\Jtest}{\mathfrak{J}^\text{\rm{\tiny{test}}}}
\newcommand{\Jvary}{\mathfrak{J}^\text{\rm{\tiny{vary}}}}

\newcommand{\vac}{\text{\rm{vac}}}
\newcommand{\Jlin}{\mathfrak{J}^\text{\rm{\tiny{lin}}}}
\newcommand{\Gtest}{\Gamma^\text{\rm{\tiny{test}}}}
\newcommand{\Gvary}{\Gamma^\text{\rm{\tiny{vary}}}}
\newcommand{\Glin}{\Gamma^\text{\rm{\tiny{lin}}}}
\newcommand{\Jdiff}{\mathfrak{J}^\text{\rm{\tiny{diff}}}}

\newcommand{\bep}{\begin{pmatrix}}
\newcommand{\enp}{\end{pmatrix}}

\renewcommand{\O}{\mathscr{O}}

\newcommand{\F}{{\mathscr{F}}}

\newcommand{\K}{{\mathcal{K}}}
\newcommand{\Pert}{{\mathscr{P}}}
\newcommand{\Phol}{\mathscr{P}_\text{\rm{\hol}}}

\renewcommand{\O}{{\mathscr{O}}}
\renewcommand{\L}{{\mathcal{L}}}
\newcommand{\Sact}{{\mathcal{S}}}
\newcommand{\s}{{\mathfrak{s}}}
\newcommand{\symm}{\text{s}}
\newcommand{\Lin}{\text{\rm{L}}}

\newcommand{\Cisc}{C^\infty_{\text{\rm{sc}}}}

\newcommand{\Fock}{{\mathcal{F}}}
\newcommand{\scrM}{\myscr M}
\newcommand{\scrN}{\myscr N}
\newcommand{\scrH}{\myscr H}

\renewcommand{\u}{\mathfrak{u}}
\renewcommand{\v}{\mathfrak{v}}
\newcommand{\w}{\mathfrak{w}}
\newcommand{\n}{\mathfrak{n}}
\newcommand{\hol}{\text{\rm{hol}}}
\newcommand{\ah}{\text{\rm{ah}}}
\newcommand{\calB}{{\mathcal{B}}}
\newcommand{\wick}{\,\text{\bf{:}}\,}
\newcommand{\Gdiff}{\Gamma^\text{\rm{\tiny{diff}}}}
\newcommand{\tin}{t_\text{\rm{in}}}
\newcommand{\tout}{t_\text{\rm{out}}}
\newcommand{\itemD}{\item[{\raisebox{0.125em}{\tiny $\blacktriangleright$}}]}
\newcommand{\bitem}{\begin{itemize}[leftmargin=2.5em]}
\newcommand{\eitem}{\end{itemize}}
\renewcommand{\div}{{\rm{div}}\,}
\newcommand{\scrt}{T}
\newcommand{\loc}{\text{\rm{loc}}}
\renewcommand{\sc}{\text{\rm{sc}}}
\newcommand{\x}{\mathbf{x}}

\DeclareFontFamily{OT1}{rsfso}{}
\DeclareFontShape{OT1}{rsfso}{m}{n}{ <-7> rsfso5 <7-10> rsfso7 <10-> rsfso10}{}
\DeclareMathAlphabet{\myscr}{OT1}{rsfso}{m}{n}

\DeclareMathOperator{\re}{Re}
\DeclareMathOperator{\im}{Im}

\DeclareMathOperator{\supp}{supp}

\begin{document}
\maketitle

\begin{abstract}
Based on conservation laws for surface layer integrals for critical points of causal variational principles,
it is shown how jet spaces can be endowed with an almost-complex structure.
We analyze under which conditions the almost-complex structure can be integrated to
a canonical complex structure.
Combined with the scalar product expressed by a surface layer integral, we obtain a complex
Hilbert space~$(\h, \la .|. \ra)$. The Euler-Lagrange equations of the causal variational principle
describe a nonlinear time evolution on~$\h$.
Rewriting multilinear operators on~$\h$ as linear operators on corresponding tensor products
and using a conservation law for a nonlinear surface layer integral, we obtain a linear
norm-preserving time evolution on bosonic Fock spaces.
The so-called holomorphic approximation is introduced, in which the dynamics is described by a unitary time
evolution on the bosonic Fock space. The error of this approximation is quantified.
Our constructions explain why and under which assumptions critical points of causal variational principles
give rise to a second-quantized, unitary dynamics on Fock spaces.
\end{abstract}
\tableofcontents

\section{Introduction} \label{secintro}
The purpose of this paper is to work out the connection
between two mathematical concepts which at first sight might seem unrelated:
causal variational principles and bosonic Fock spaces.
{\em{Bosonic Fock spaces}} are complex Hilbert spaces which arise
in the mathematical formulation of many-particle quantum
systems. The dynamics of such systems is described by a unitary time evolution
on the Fock space. More precisely,
\beq \label{fockdyn}
\Psi(t) = e^{-i t H}\: \Psi_0 \:,
\eeq
where the Hamiltonian~$H$ is a symmetric operator on the Fock space~$(\Fock, \la .|. \ra_\Fock)$.
{\em{Causal variational principles}}, on the other hand, were introduced in~\cite{continuum}
as a mathematical generalization of the causal action principle,
being the analytical core of the physical theory of causal fermion systems
(see the textbook~\cite{cfs} or the introductions~\cite{dice2014, review}).
In general terms, 
given a manifold~$\F$ together with a non-negative function~$\L : \F \times \F \rightarrow \R^+_0$,
in a causal variational principle one minimizes the action~$\Sact$ given by
\[ \Sact (\rho) = \int_\F d\rho(x) \int_\F d\rho(y)\: \L(x,y) \]
under variations of the measure~$\rho$ on~$\F$, keeping the total volume~$\rho(\F)$ fixed
(for the precise mathematical setup see Section~\ref{seccvp} below).
Working with measures on a manifold, there is a-priori no Hilbert space structure,
making the connection to bosonic Fock spaces far from obvious.
Here we make use of two key observations: First, variations of the measure~$\rho$
can be described by so-called {\em{jets}} consisting of scalar functions and vector fields
in spacetime $M:= \supp \rho$ (see~\cite{jet} or Section~\ref{secwEL} below,
where~$\supp$ denotes the support of the measure~$\rho$).
The resulting jet spaces are real vector spaces.
The second observation is that on the jet spaces one can introduce bilinear forms
which have the structure of so-called surface layer integrals
\[ \int_\Omega \bigg( \int_{M \setminus \Omega} (\cdots)\: \L(x,y)\: d\rho(y) \bigg)\, d\rho(x) \:, \]
where $(\cdots)$ stands for a differential operator involving jets.
A surface layer integral generalizes the concept of a surface integral over~$\partial \Omega$
to the setting of causal fermion systems (for the general idea see~\cite[Section~2.3]{noether}).
Moreover, as a consequence of the Euler-Lagrange (EL)  equations corresponding to the
causal variational principle, there are jet spaces for which the surface layer integrals
do not depend on the choice of the set~$\Omega$
(see~\cite{noether, jet, osi} or the summary in Section~\ref{secosi} below).

Starting from these structures, we here analyze how to endow the jet spaces with
a complex structure. Moreover, we study the question of whether and how the interaction
as described by the causal variational principle can be formulated in terms of
a time evolution on the resulting complex vector spaces.
For non-interacting, so-called {\em{linear systems}}, the conserved surface layer integrals
induce a canonical complex structure, giving rise to a
 {\em{complex Hilbert space}} of jets~$(\h, \la .|. \ra)$ (see~\eqref{Hprod} in Section~\ref{seccomplexlinear}).
The nonlinear dynamics as described by the EL equations
corresponding to the causal variational principle yields a complicated mixing of these jets
and their complex conjugates. We rewrite this nonlinear dynamics as a {\em{linear dynamics}}
on a suitable tensor product. More precisely, due to the mixing of jets and their complex conjugates,
the time evolution is not an operator
on the bosonic Fock space~$\Fock := \oplus_{n=0}^\infty \h^n$, but instead it is real-linear
operator on the tensor product of~$\Fock$ with its dual space~$\Fock^*$.
An essential step of our construction is to show that this time evolution
preserves the norm on~$\Fock^* \otimes \Fock$ (see Theorem~\ref{thmNunit2}).
This result is based on a new conservation law
for surface layer integrals which, because of its nonlinear dependence on the jets, we refer to
as the {\em{nonlinear surface layer integral}} (see Section~\ref{secosinonlin} and Appendix~\ref{appconserve}).
The conservation law for this nonlinear surface layer integral is intimately related to
so-called {\em{inner solutions}} of the linearized field equations 
(see Sections~\ref{secinner}).
Although these inner solutions are very small (see Section~\ref{secinnersmall} and Appendix~\ref{secapproxinner}),
they do contribute to the surface layer integrals
via flux integrals (see~\eqref{innerfluxdef} in Section~\ref{seclinear}),
which in turn give rise to a rescaling of the Fock space norm as needed for the Fock norm to be preserved.
We also derive an approximate dynamics,
the so-called {\em{holomorphic approximation}}, described by a unitary time evolution
on~$\Fock$ of the form~\eqref{fockdyn} (see Theorem~\ref{thmHsymm} and Definition~\ref{defhol}).
The error of the holomorphic approximation is quantified by working out the corrections
(see Theorems~\ref{thmerror} and~\ref{thmcommute} as well as Appendix~\ref{seccompare}).

The paper is organized as follows.
In Section~\ref{secprelim} we give the necessary background
on causal variational principles. 
In Section~\ref{secinner} we introduce inner solutions and collect a few important properties.
Section~\ref{secosinonlin} is devoted to the non-linear surface layer integral and the
corresponding conservation law. In Section~\ref{secinnersmall} the approximation
of small inner solutions is introduced.
In Section~\ref{secscatter} we specify how to describe a scattering process
in Minkowski space. Moreover, the conservation laws for surface layer integrals are adapted to this setting,
and the freedom in choosing complex structures are analyzed.
In Section~\ref{secfock}, the Fock space description is introduced.
After recalling the basics on Fock spaces (Section~\ref{secfockprelim}),
we introduce field operators and work out their commutation relations (Section~\ref{secccr}).
Then we rewrite the time evolution as a linear operator on Fock spaces.
We begin with the case in which the time evolution is compatible with the complex structure
(as is made precise by the notion of holomorphic connections; see Definition~\ref{defcc}).
In this case, expanding the nonlinear dynamics as described by the EL equations of the causal variational principle
in a perturbation series and rewriting the resulting $p$-multilinear operators as linear operators
on the $p$-fold tensor product, we obtain a unitary time evolution on the Fock space~$\Fock
:= \oplus_{n=0}^\infty \h^n$ (Section~\ref{secfockholo}).
In the general case that the time evolution is {\em{not}} compatible with the complex structure,
we obtain instead a norm-preserving complex-linear time evolution on~$\Fock^* \otimes \Fock$
(Section~\ref{secfockFsF}).
Section~\ref{secholo} is devoted to the holomorphic approximation, where the
time evolution on~$\Fock^* \otimes \Fock$ is approximated by a unitary time evolution on~$\Fock$.
In preparation, we need to analyze the conservation laws and the complex structure
at intermediate times (Section~\ref{secvolterms}).
Then the holomorphic approximation is introduced (Section~\ref{secholomorphic}) and its corrections
are worked out (Section~\ref{secerror}).
In Section~\ref{secphi4} we illustrate our constructions by explaining the analogies and differences
to classical field theory in the example of $\phi^4$-theory in Minkowski space.
The appendices provide some background material. In Appendix~\ref{appconserve}
the nonlinear conservation law is considered from a more abstract perspective, and it is shown
how the corresponding conservation law can be arranged.
In Appendix~\ref{secapprox} it is explained how to justify the approximations made in the
article. The approximation of small inner solutions
is justified in Appendix~\ref{secapproxinner} by considering the scalings
for Dirac systems based on the computations in~\cite{action} and~\cite[Appendix~A]{jacobson}.
in Appendix~\ref{seccompare} it is explained in words how the
holomorphic approximation is related to the concepts of microscopic mixing
as introduced in~\cite{qft}. A detailed justification of the holomorphic approximation
is not given here, but will be worked out in a separate paper.

We close with two remarks. First, we point out that 
we here restrict attention to {\em{bosonic}} Fock spaces; the additional constructions
giving rise to fermionic Fock spaces will be developed separately in~\cite{fockfermionic}.
Second, we note that the connection between causal variational principles and Fock spaces
was first established in~\cite{qft}, however only for causal fermion systems
and based on the classical equations obtained in the continuum limit
(a limiting case giving an interaction via classical bosonic fields 
in Minkowski space worked out in detail in~\cite{cfs}).
In contrast to this work, we here analyze directly the EL equations corresponding to the
causal variational principle. Moreover, we work closely with the conservation laws for
surface layer integrals.
In this way, the constructions in the present paper give a more general and more fundamental
connection to bosonic Fock spaces.

\section{Preliminaries} \label{secprelim}
\subsection{Causal Variational Principles in the Non-Compact Setting} \label{seccvp}
We consider causal variational principles in the non-compact setting as
introduced in~\cite[Section~2]{jet}. Thus we let~$\F$ be a (possibly non-compact)
smooth manifold of dimension~$m \geq 1$
and~$\rho$ a (positive) Borel measure on~$\F$ (the {\em{universal measure}}).
Moreover, we are given a non-negative function~$\L : \F \times \F \rightarrow \R^+_0$
(the {\em{Lagrangian}}) with the following properties:
\begin{itemize}[leftmargin=2em]
\item[\rm{(i)}] $\L$ is symmetric: $\L(x,y) = \L(y,x)$ for all~$x,y \in \F$.\label{Cond1}
\item[\rm{(ii)}] $\L$ is lower semi-continuous, i.e.\ for all sequences~$x_n \rightarrow x$ and~$y_{n'} \rightarrow y$,
\[ \L(x,y) \leq \liminf_{n,n' \rightarrow \infty} \L(x_n, y_{n'})\:. \]\label{Cond2}
\end{itemize}
The {\em{causal variational principle}} is to minimize the action
\beq \label{Sact} 
\Sact (\rho) = \int_\F d\rho(x) \int_\F d\rho(y)\: \L(x,y) 
\eeq
under variations of the measure~$\rho$, keeping the total volume~$\rho(\F)$ fixed
({\em{volume constraint}}).
Here the  notion {\em{causal}} in ``causal variational principles'' refers to the fact that
the Lagrangian induces on~$M$ a causal structure. Namely, two spacetime points~$x,y \in M$
are said to be timelike and space-like separated if~$\L(x,y)>0$ and~$\L(x,y)=0$, respectively.
For more details on this notion of causality, its connection to the causal structure
in Minkowski space and to general relativity we refer to~\cite[Chapter~1]{cfs}, \cite{nrstg}
and~\cite[Sections~4.9 and~5.4]{cfs}.

If the total volume~$\rho(\F)$ is finite, one minimizes~\eqref{Sact}
over all regular Borel measures with the same total volume.
If the total volume~$\rho(\F)$ is infinite, however, it is not obvious how to implement the volume constraint,
making it necessary to proceed as follows.
We need the following additional assumptions:
\begin{itemize}[leftmargin=2em]
\item[\rm{(iii)}] The measure~$\rho$ is {\em{locally finite}}
(meaning that any~$x \in \F$ has an open neighborhood~$U$ with~$\rho(U)< \infty$).\label{Cond3}
\item[\rm{(iv)}] The function~$\L(x,.)$ is $\rho$-integrable for all~$x \in \F$, giving
a lower semi-continuous and bounded function on~$\F$. \label{Cond4}
\end{itemize}
Given a regular Borel measure~$\rho$ on~$\F$, we then vary over all
regular Borel measures~$\tilde{\rho}$ with
\[ 
\big| \tilde{\rho} - \rho \big|(\F) < \infty \qquad \text{and} \qquad
\big( \tilde{\rho} - \rho \big) (\F) = 0 \]
(where~$|.|$ denotes the total variation of a measure).
These variations of the causal action are well-defined.
It is shown in~\cite[Lemma~2.3]{jet} that a minimizer
satisfies {\em{Euler-Lagrange (EL) equations}}
stating that for a suitable value of the parameter~$\s>0$,
the lower semi-continuous function~$\ell : \F \rightarrow \R_0^+$ defined by
\beq \label{ldef}
\ell(x) := \int_\F \L(x,y)\: d\rho(y) - \s
\eeq
is minimal and vanishes on spacetime~$M:= \supp \rho$,
\beq \label{EL}
\ell|_M \equiv \inf_\F \ell = 0 \:.
\eeq
For further details we refer to~\cite[Section~2]{jet}.

\subsection{The Weak Euler-Lagrange Equations and Jet Spaces} \label{secwEL}
\hspace*{0.05cm}
The EL equations~\eqref{EL} are nonlocal in the sense that
they make a statement on~$\ell$ even for points~$x \in \F$ which
are far away from spacetime~$M$.
It turns out that for the applications we have in mind, it is preferable to
evaluate the EL equations locally in a neighborhood of~$M$.
This leads to the {\em{weak EL equations}} introduced in~\cite[Section~4]{jet}.
We here give a slightly less general version of these equations which
is sufficient for our purposes. In order to explain how the weak EL equations come about,
we begin with the simplified situation that the function~$\ell$ is smooth.
In this case, the minimality of~$\ell$ implies that the derivative of~$\ell$
vanishes on~$M$, i.e.\
\beq \label{ELweak}
\ell|_M \equiv 0 \qquad \text{and} \qquad D \ell|_M \equiv 0
\eeq
(where~$D \ell(p) : T_p \F \rightarrow \R$ is the derivative).
In order to combine these two equations in a compact form,
it is convenient to consider a pair~$\u := (a, u)$
consisting of a real-valued function~$a$ on~$M$ and a vector field~$u$
on~$T\F$ along~$M$, and to denote the combination of 
multiplication and directional derivative by
\beq \label{Djet}
\nabla_{\u} \ell(x) := a(x)\, \ell(x) + \big(D_u \ell \big)(x) \:.
\eeq
Then the equations~\eqref{ELweak} imply that~$\nabla_{\u} \ell(x)$
vanishes for all~$x \in M$.
The pair~$\u=(a,u)$ is referred to as a {\em{jet}}.

In the general lower-continuous setting, one must be careful because
the directional derivative~$D_u \ell$ in~\eqref{Djet} need not exist.
Our method for dealing with this difficulty is to restrict attention to vector fields
for which the directional derivative is well-defined.
Moreover, we must specify the regularity assumptions on~$a$ and~$u$.
To begin with, we always assume that~$a$ and~$u$ are {\em{smooth}} in the sense that they
have a smooth extension to the manifold~$\F$. Thus the jet~$\u$ should be
an element of the jet space
\beq \label{Jdef}
\J_\rho := \big\{ \u = (a,u) \text{ with } a \in C^\infty(M, \R) \text{ and } u \in \Gamma(M, T\F) \big\} \:,
\eeq
where~$C^\infty(M, \R)$ and~$\Gamma(M,T\F)$ denote the space of real-valued functions and vector fields
on~$M$, respectively, which admit a smooth extension to~$\F$.
We remark that the question on whether a function or vector field on~$M$ can be
extended smoothly to~$\F$ is rather subtle. The needed conditions are made precise
by Whitney's extension theorem (see for example the more recent account in~\cite{fefferman}).
Here we do not enter the details of these conditions, but use them as implicit
assumptions entering our definition~\eqref{Jdef}.
We remark that these conditions will be fulfilled in the setting of Definition~\ref{defsms} in which $M:=\supp \rho$ carries itself a manifold structure.
Clearly, the fact that a jet~$\u$ is smooth does not imply that the functions~$\ell$
or~$\L$ are differentiable in the direction of~$\u$. This must be ensured by additional
conditions which are satisfied by suitable subspaces of~$\J_\rho$
which we now introduce.
First, we let~$\Gdiff_\rho$ be those vector fields for which the
directional derivative of the function~$\ell$ exists,
\beq \label{Gdiffdef}
\Gdiff_\rho = \big\{ u \in C^\infty(M, T\F) \;\big|\; \text{$D_{u} \ell(x)$ exists for all~$x \in M$} \big\} \:.
\eeq
This gives rise to the jet space
\beq \label{Jdiffdef}
\Jdiff_\rho := C^\infty(M, \R) \oplus \Gdiff_\rho \;\subset\; \J_\rho \:.
\eeq
For the jets in~$\Jdiff_\rho$, the combination of multiplication and directional derivative
in~\eqref{Djet} is well-defined. 
We choose a linear subspace~$\Jtest_\rho \subset \Jdiff_\rho$ with the property
that its scalar and vector components are both vector spaces,
\beq\label{Gammatest}
\Jtest_\rho = \Ctest(M, \R) \oplus \Gtest_\rho \;\subseteq\; \Jdiff_\rho \:,
\eeq
and the scalar component is nowhere trivial in the sense that
\beq \label{Cnontriv}
\text{for all~$x \in M$ there is~$a \in \Ctest(M, \R)$ with~$a(x) \neq 0$}\:.
\eeq
Then the {\em{weak EL equations}} read (for details cf.~\cite[eq.~(4.10)]{jet})
\beq \label{ELtest}
\nabla_{\u} \ell|_M = 0 \qquad \text{for all~$\u \in \Jtest_\rho$}\:.
\eeq
The purpose of introducing~$\Jtest_\rho$ is that it gives the freedom to restrict attention to the portion of
information in the EL equations which is relevant for the application in mind.
For example, if one is interested only in the macroscopic dynamics, one can choose~$\Jtest_\rho$
to be composed of jets pointing in directions where the 
microscopic fluctuations of~$\ell$ are disregarded.

We finally point out that the weak EL equations~\eqref{ELtest}
do not hold only for minimizers, but also for critical points of
the causal variational principle. Therefore, all methods and results of this paper do not apply only to
minimizers, but more generally to critical points.
For brevity, we also refer to a measure with satisfies the weak EL equations~\eqref{ELtest}
as a {\em{critical measure}}.

We conclude this section by introducing a few jet spaces 
and specifying differentiability conditions
which will be needed later on.
We begin with the spaces~$\J^\ell_\rho$, where~$\ell \in \N_0 \cup \{\infty\}$ can be
thought of as the order of differentiability if the derivatives act  simultaneously on
both arguments of the Lagrangian:
\begin{Def} \label{defJvary}
For any~$\ell \in \N_0 \cup \{\infty\}$, the jet space~$\J_\rho^\ell \subset \J_\rho$
is defined as the vector space of test jets with the following properties:
\begin{itemize}[leftmargin=2em]
\item[\rm{(i)}] For all~$y \in M$ and all~$x$ in an open neighborhood of~$M$,
directional derivatives
\beq \label{derex}
\big( \nabla_{1, \v_1} + \nabla_{2, \v_1} \big) \cdots \big( \nabla_{1, \v_p} + \nabla_{2, \v_p} \big) \L(x,y)
\eeq
(computed componentwise in charts around~$x$ and~$y$)
exist for all~$p \in \{1, \ldots, \ell\}$ and all~$\v_1, \ldots, \v_p \in \J_\rho^\ell$.
\item[\rm{(ii)}] The functions in~\eqref{derex} are $\rho$-integrable
in the variable~$y$, giving rise to locally bounded functions in~$x$. More precisely,
these functions are in the space
\[ L^\infty_\text{\rm{loc}}\Big( M, L^1\big(M, d\rho(y) \big); d\rho(x) \Big) \:. \]
\item[\rm{(iii)}] Integrating the expression~\eqref{derex} in~$y$ over~$M$
with respect to the measure~$\rho$,
the resulting function (defined for all~$x$ in an open neighborhood of~$M$)
is continuously differentiable in the direction of every jet~$\u \in \Jtest_\rho$.
\end{itemize}
\end{Def} \noindent
Here and throughout this paper, we use the following conventions for partial derivatives and jet derivatives:
\begin{itemize}[leftmargin=2em]
\itemD Partial and jet derivatives with an index $i \in \{ 1,2 \}$, as for example in~\eqref{derex}, only act on the respective variable of the function $\L$.
This implies, for example, that the derivatives commute,
\beq\label{ConventionPartial}
\nabla_{1,\v} \nabla_{1,\u} \L(x,y) = \nabla_{1,\u} \nabla_{1,\v} \L(x,y) \:.
\eeq
\itemD The partial or jet derivatives which do not carry an index act as partial derivatives
on the corresponding argument of the Lagrangian. This implies, for example, that
\[ \nabla_\u \int_\F \nabla_{1,\v} \, \L(x,y) \: d\rho(y) =  \int_\F \nabla_{1,\u} \nabla_{1,\v}\, \L(x,y) \: d\rho(y) \:. \]
\end{itemize}
We point out that, in contrast to the method and conventions used in~\cite{jet},
{\em{jets are never differentiated}}.

In order for all integral expressions to be well-defined, we impose throughout the paper
that the space~$\Jtest_\rho$ has the following properties (for details see~\cite[Section~3.5]{osi}).
\begin{Def} \label{defslr}
The jet space~$\Jtest_\rho$ is {\bf{surface layer regular}}
if~$\Jtest_\rho \subset \J^2_\rho$ (see Definition~\ref{defJvary}) and
if for all~$\u, \v \in \Jtest_\rho$ and all~$p \in \{1, 2\}$ the following conditions hold:
\begin{itemize}[leftmargin=2em]
\item[\rm{(i)}] The directional derivatives
\beq \nabla_{1,\u} \,\big( \nabla_{1,\v} + \nabla_{2,\v} \big)^{p-1} \L(x,y) \label{Lderiv1}
\eeq
exist.
\item[\rm{(ii)}] The functions in~\eqref{Lderiv1} are $\rho$-integrable
in the variable~$y$, giving rise to locally bounded functions in~$x$. More precisely,
these functions are in the space
\[ L^\infty_\text{\rm{loc}}\Big( M, L^1\big(M, d\rho(y) \big); d\rho(x) \Big) \:. \]
\item[\rm{(iii)}] The $\u$-derivative in~\eqref{Lderiv1} may be interchanged with the $y$-integration, i.e.
\[ \int_M \nabla_{1,\u} \,\big( \nabla_{1,\v} + \nabla_{2,\v} \big)^{p-1} \L(x,y)\: d\rho(y)
= \nabla_\u \int_M \big( \nabla_{1,\v} + \nabla_{2,\v} \big)^{p-1} \L(x,y)\: d\rho(y) \:. \]
\end{itemize}
\end{Def}

\subsection{The Nonlinear Solution Space and Linearized Solutions}
In what follows, we shall be concerned with families of critical measures,
always for a fixed value of the Lagrange parameter~$\s$ in~\eqref{ldef}.
In order to obtain these families of solutions, we want to vary a given
measure~$\rho$ (typically a critical measure) without changing its general structure.
To this end, we multiply~$\rho$ by a weight function
and apply a diffeomorphism, i.e.
\beq \label{rhoFf}
\tilde{\rho} = F_* \big( f \,\rho \big) \:,
\eeq
where~$F \in C^\infty(M, \F)$ and~$f \in C^\infty(M, \R^+)$ are smooth mappings
(as defined before~\eqref{Jdef}).
We now consider a set of such measures which all satisfy the weak EL equations,
\beq
\calB \subset \left\{ \text{$\tilde{\rho}$ critical measure of the form~\eqref{rhoFf}} \right\} . \label{Bdef}
\eeq
In the smooth setting, $\calB$ can be given the structure of a Fr{\'e}chet manifold
(see~\cite[Section~3 and Appendix~A]{jet}). Here we do not assume smoothness, but we
work instead in the lower semi-continuous setting introduced in~\cite[Section~4]{jet}.
Nevertheless, it might be helpful for the reader to visualize~$\calB$ as
a manifold, to identify jets with scalar functions and vector fields,
conserved quantities with differential forms on~$\calB$, and so on.
For this reason, we always mention how our objects
can be understood geometrically in the smooth setting.

A variation of the measure~\eqref{rhoFf} is described by
a family~$(f_\tau, F_\tau)$ with~$\tau \in (-\delta, \delta)$ and~$\delta>0$. Infinitesimally, the variation
is again described by a jet
\beq \label{vinfdef}
\v = (b,v) := \frac{d}{d\tau} (f_\tau, F_\tau) \big|_{\tau=0}\:.
\eeq
If the corresponding curve~$\rho_\tau$ lies in~$\calB$ and~$\calB$ is a smooth manifold,
then the tangent vector~$\v$ is a vector in the tangent space~$T_\rho \calB$.
In the non-smooth setting, the fact that~$\rho_\tau$ lies in~$\calB$
means that the weak EL equations~\eqref{ELtest} are preserved by the variation.
Evaluating this condition infinitesimally in~$\tau$ gives rise to the
{\em{linearized field equations}}, which we now introduce.
Before beginning, we point out that for the analysis of the above functions~$f_\tau$ and~$F_\tau$,
we always work with Taylor expansions of the component functions
in given charts. Therefore, for any~$x \in M$ we choose a chart
of~$\F$ around~$x$  and work in components~$x^\alpha$.
For ease in notation, we usually omit the index~$\alpha$ as well as all vector and tensor indices.
But one should keep in mind that, from now on, we always work in given charts.

The property of the family of measures~$\tilde{\rho}_\tau$ of the form~\eqref{rhoFf}
for a given family~$(f_\tau, F_\tau)$ to be critical for all~$\tau$
means infinitesimally in~$\tau$ that the jet~$\v$ defined by~\eqref{vinfdef}
satisfies the {\em{linearized field equations}} (for the derivation see~\cite[Section~3.3]{perturb}
and~\cite[Section~4.2]{jet})
\beq \label{eqlinlip}
\la \u, \Delta \v \ra|_M = 0 \qquad \text{for all~$\u \in \Jtest_\rho$} \:,
\eeq
where
\beq \label{eqlinlip2}
\la \u, \Delta \v \ra(x)
:= \nabla_{\u} \bigg( \int_M \big( \nabla_{1, \v} + \nabla_{2, \v} \big) \L(x,y)\: d\rho(y) - \nabla_\v \:\s \bigg) \:.
\eeq
In order for the last expression to be well-defined, we always assume that~$\v \in \J^1_\rho$.
We denote the vector space of all solutions of the linearized field equations by~$\Jlin_\rho \subset \J^1_\rho$.
In the smooth setting, $\Jlin_\rho$ can be identified with the tangent space~$T_\rho \calB$.

\subsection{Green's Operators and the Causal Fundamental Solution} \label{secglobhyp}
In~\cite{linhyp} the existence theory for solutions of the linearized field equations was developed.
In particular, it was shown under suitable assumptions that advanced and retarded Green's operators
exist. We now recall a few notions and results from~\cite{linhyp}.
The {\em{inhomogeneous}} linearized field equations are obtained by adding an inhomogeneity
on the right side of~\eqref{eqlinlip}, i.e.\
\[ \la \u, \Delta \v \ra|_M = \la \u, \w \ra \qquad \text{for all~$\u \in \Jtest$} \:. \]
One way to give the right side of this equation a precise meaning is to
regard~$\w$ as a dual jet, so that~$\la \u, \w \ra$ is a dual pairing
(for details see~\cite[Sections~2.2. and~2.3]{linhyp}). In what follows, it is more suitable to identify jets and dual
jets by a scalar product (for details see~\cite[Section~3.2]{linhyp}). To this end,
we let~$\Gamma_x$ be the subspace of the tangent space spanned by the test jets,
\beq \label{Gxdef}
\Gamma_x := \big\{ u(x) \:|\: u \in \Gtest \big\} \;\subset\; T_x\F\:.
\eeq
We introduce a Riemannian metric~$g_x$ on~$\J_x$.
This Riemannian metric also induces a pointwise scalar product on the jets. Namely, setting
\beq \label{Jxdef}
\J_x := \R \oplus \Gamma_x \:,
\eeq
we obtain the scalar product on~$\J_x$
\beq
\la \v, \tilde{\v} \ra_x \,:\, \J_x \times \J_x \rightarrow \R \:,\qquad
\la \v, \tilde{\v} \ra_x := b(x)\, \tilde{b}(x) + g_x \big(v(x),\tilde{v}(x) \big) \:. \label{vsprod}
\eeq
We denote the corresponding norm by~$\|.\|_x$.
By integrating the scalar product~\eqref{vsprod} over~$M$ we obtain a scalar product
on the jets. We point out that the choice of the Riemannian metric is not canonical.
The freedom in choosing the Riemannian metric can be used in order to satisfy the
hyperbolicity conditions needed for proving existence of solutions (as explained
after~\cite[Definition~3.3]{linhyp}). Since we will use these existence results
later on, we assume that the Riemannian metric in~\eqref{vsprod} has been chosen
in agreement with these hyperbolicity conditions.

By integrating the pointwise scalar product~\eqref{vsprod} over~$M$ we obtain
Hilbert spaces of jets. To this end, we first consider jets~$\v=(b,v)$,
where~$v$ now is a measurable section of~$T\F$
along~$M$ with~$v(x) \in \J_x$ for all~$x \in M$. The jets of this form which are
square integrable form a Hilbert space denoted by~$L^2(M, d\rho)$.
Likewise, $L^2_\loc(M, d\rho)$ denotes the jets which are locally square integrable
(i.e.\ which are square integrable over every compact subset of~$M$).
The vector space~$L^2_0(M, d\rho)$ is the subspace of jets with compact support.

We always assume that spacetime is {\em{globally hyperbolic}} (see~\cite[Definition~4.19]{linhyp})
and that the Lagrangian has {\em{finite range}} (see~\cite[Definition~4.6]{linhyp}).
Under these assumptions there exist {\em{causal Green's operators}}
\beq \label{Scausaldef}
S^\wedge, S^\vee \::\: L^2_0(M, d\rho) \rightarrow L^2_\loc(M, d\rho)
\eeq
(see~\cite[Section~5.2]{linhyp}).
The operator~$S^\wedge$ is {\em{retarded}}
and~$S^\vee$ is {\em{advanced}} in the sense that the support of the~$S^\wedge f$
and~$S^\vee f$ lies in the causal future (respectively past) of the support of~$f$,
up to vectors in the orthogonal complement of the test jets (for details see~\cite[Section~4.4]{linhyp}).
For our purposes, it is most convenient to restrict the Green's operators to a smaller domain, denoted by
\beq \label{J0sdef}
\J^*_{\rho,0} := \big\{ \u \in L^2_0(M, d\rho) \:\big|\: S^\vee \u, S^\wedge \u \in \Jvary_\rho \big\} \:,
\eeq
where~$\Jvary_\rho$ is a suitably chosen subspace of~$\Jtest_\rho$
(for details see~\cite[Section~3.2]{linhyp}). Then the Green's operators satisfy the relations
\beq \label{Scausal}
S^\vee, S^\wedge \::\: \J^*_{\rho,0} \rightarrow \J_{\rho, \sc} \qquad \text{and} \qquad
\Delta S^\vee = \Delta S^\wedge = -\1 \:,
\eeq
where~$\J_{\rho, \sc}$ denotes the jets in~$\Jvary_\rho$ with spatially compact support
(for details see~\cite[Section~5.3]{linhyp}).
Moreover, the {\em{causal fundamental solution}} is defined by
\beq \label{Kdef}
G := S^\wedge - S^\vee \::\: \J^*_{\rho,0} \rightarrow \J_{\rho,\sc} \:.
\eeq
Using the right equation in~\eqref{Scausal},
one sees that~$G$ maps to linearized solutions. We denote those solutions by
\[ \Jlin_{\rho,\sc} := G\, \J^*_{\rho,0} \;\subset\; \J_{\rho,\sc} \;\subset\; \Jlin \cap \Jtest \:. \]
Moreover, the operator~$G$ gives rise to the short exact sequence
\beq \label{exact}
0 \rightarrow \Jvary_{\rho,0} \overset{\Delta}{\longrightarrow} \J^*_{\rho,0}
\overset{G}{\longrightarrow} \J_{\rho,\sc}
\overset{\Delta}{\longrightarrow} \J^*_{\rho,\sc} \rightarrow 0 \:,
\eeq
where~$\Jvary_{\rho,0}$ is a subspace of the compactly supported test jets, whereas~$\J^*_{\rho,\sc}$
is a space of spatially compact jets (for details see again~\cite[Section~5.3]{linhyp}).

\subsection{The Perturbation Map and its Linearization} \label{secpert}
The perturbation expansion developed in~\cite{perturb} provides a method
for constructing critical measures from a linearized solution. 
We now recall a few results from this paper in a form which is most convenient to us.
Having a scattering process in mind where the interaction takes place in a compact
spacetime region, we can assume that all inhomogeneities appearing in
the perturbation expansion have compact support. Assuming suitable regularity,
the inhomogeneities are even in the jet space~$\J^*_{\rho,0}$.
Then we can work with the Green's operators in~\eqref{Scausaldef}, obtaining jets
with spatially compact support in~$\J_{\rho, \sc}$.
Correspondingly, we consider linearized solutions in~$\Jlin_{\rho, \sc}$,
again with spatially compact support.
Then the constructions in~\cite{perturb} formally give rise to the so-called {\em{perturbation map}}
\beq \label{Pertdef}
\Pert_\rho : U \subset \Jlin_{\rho, \sc} \rightarrow \calB \:,
\eeq
where~$U$ is an open neighborhood of the origin
(the reason why this equation is only formal is that the perturbation expansion is not known to converge).
Clearly, the operator~$\Pert_\rho$ depends on the choice of the Green's operators.

In differential geometric language, the mapping~$\Pert_\rho$ can be regarded as a local chart
of~$\calB$ in a neighborhood of~$\rho$. We use the notation
\[ \tilde{\rho} = \Pert_\rho(\w) \qquad \text{with} \qquad \w \in U \:. \]
Always working with measures of the form~\eqref{rhoFf}, we can identify
the measure~$\tilde{\rho}$ with the pair~$(f,F)$. This allows us to introduce the short notation
where
\beq \label{shortnotation}
(f,F) \qquad \text{stands for the measure} \qquad F_*(f\, \rho) \:.
\eeq
Then the linearization of~$\Pert_\rho$ maps linearized solutions to linearized solutions, i.e.
\beq \label{DP}
D\Pert_\rho|_\w \::\: \Jlin_{\rho, \sc}
\rightarrow \Jlin_{\tilde{\rho}, \sc} \:,\qquad \u, \v \mapsto \tilde{\u}, \tilde{\v}
\eeq
(where~$D\Pert_\rho|_\w$ is the derivative at~$\w$ defined as a linear mapping).
In~\cite{perturb} explicit formulas for the perturbation map are derived to every order
in perturbation theory. For this perturbation expansion one uses~\eqref{shortnotation}
to identify~$\Pert_\rho(\w)$ with the pair~$(f,F)$, and then expands this pair in suitable charts
to obtain a formal power series of jets. In order to keep the notation as simple as possible,
here we write the perturbation expansion symbolically as
\beq \label{Pertexpand}
\Pert_\rho(\lambda \w) = \sum_{p=1}^\infty \lambda^p\: \Pert_\rho^{(p)}
\big( \!\underbrace{\w, \ldots, \w}_{\text{$p$ arguments}} \!\big) \:.
\eeq
The coefficients of the expansions have the properties that~$\Pert_\rho^{(1)}$ is the
identity and that 
\beq \label{Pp}
\Pert_\rho^{(p)} \::\: \big( \Jlin_{\rho, \sc} \big)^p \rightarrow
\J_{\rho, \sc} \quad \text{is $p$-multilinear and symmetric}\:.
\eeq
By differentiating, we obtain
\beq \label{dPertexpand}
D \Pert_\rho|_{\lambda \w} \,(\u) = \sum_{p=1}^\infty p\, \lambda^{p-1}\: \Pert_\rho^{(p)}
\big( \underbrace{\w, \ldots, \w}_{\text{$p-1$ arg.}} , \u \big) \:.
\eeq

\subsection{Conserved Surface Layer Integrals for Linearized Solutions} \label{secosi}
As explained in the introduction,
the main goal of this paper is to get a connection to quantum field theory in Minkowski space.
With this in mind, from now on we make the following simplifying assumptions:
\begin{Def} \label{defsms}
Spacetime~$M:= \supp \rho$ has a {\bf{smooth manifold structure}} if
the following conditions hold:
\bitem
\item[\rm{(i)}] $M$ is a $k$-dimensional  smooth, oriented and connected submanifold of~$\F$.
Equipped with a smooth atlas, we also denote it by~$\scrM$.
\item[\rm{(ii)}] In a chart~$(x,U)$ of~$\scrM$, the universal measure is absolutely continuous with respect
to the Lebesgue measure with a smooth, strictly positive weight function,
\beq \label{hdef}
d\rho = h(x)\: d^kx \qquad \text{with} \quad h \in C^\infty(\scrM, \R^+) \:.
\eeq
\eitem
If in addition, the manifold~$\scrM$ is topologically of the form~$\scrM = \R \times \scrN$
with a manifold~$\scrN$ which admits a complete Riemannian metric~$g_{\scrN}$,
then spacetime is said to admit a {\bf{global time function}}.
\end{Def} \noindent
We remark that the assumption of~$M$ being a smooth submanifold of~$\F$ also implies that the
conditions of the Whitney extension theorem mentioned after~\eqref{Jdef} are satisfied.
As a consequence, smoothness as defined after~\eqref{Jdef} is consistent with
the usual notion of smoothness in the coordinate charts.
Moreover, similar to our assumption that the scalar components of the test jets
is nowhere trivial~\eqref{Cnontriv}, it is sensible and useful to assume that the
subspace of the tangent space spanned by the test jets contains all
the tangent vectors to~$M$,
\beq \label{Tinclude}
T_x M \subset \Gamma_x \qquad \text{for all~$x \in M$}\:.
\eeq

We write the points of~$\scrM$ as~$(t,\x)$ with~$t \in \R$ and~$\x \in \scrN$
and interpret the first component as time. For clarity, we also denote the time coordinate by~$\scrt$, i.e.\
\beq \label{scrtdef}
\scrt \::\: M \rightarrow \R\:,\qquad (t, \x) \mapsto t
\eeq
and denote the surfaces of constant time and the past of these surfaces by
\begin{align}
N^t &:= \scrt^{-1}(t) \label{Ntdef} \\
\Omega^t &:= \{ x \in M \:|\: \scrt(x) \leq t \} \:. \label{Omegatdef}
\end{align}
It is often convenient to decompose the measure~$\rho$ into a time and spatial integration measure,
\beq \label{mutdef}
d\rho = dt\, d\mu_t \:,
\eeq
where the measure~$\mu_t$ is given in local coordinates~$(t, x^1,\ldots, x^{k-1})$ by
\[ d\mu_t = h(t,\x)\: dx^1 \cdots dx^{k-1} \]
with~$h$ as in~\eqref{hdef}. 
Clearly, the interpretation of~$\scrt$ as ``time'' also entails additional assumptions
(in particular, the distinction from a ``spatial coordinate''). These additional assumptions
will be supplemented in the course of the paper when we need them.
We also point out that, thinking of~$\scrM$ as Minkowski space,
our global chart does not distinguish a reference frame of Minkowski space.
Indeed, if the conditions in Definition~\ref{defsms} hold in one reference frame, they hold
just as well in any other reference frame. But we use the above diffeomorphism~$M \rightarrow \scrM$
and the global time function in order to distinguish a spatial orientation and a direction of time.

Let~$\rho \in \calB$ be a critical measure.
Then, as shown in~\cite{jet, osi}, there are various conservation laws for surface layer integrals.
We now collect those surface layer integrals and conservation laws which are of relevance for our constructions.
\begin{Def} The spatially compact jets~$\J_{\rho, \sc}$ are {\bf{surface layer finite}}
if for any~$t \in \R$ the following integrals are finite,
\begin{align*}
\sum_{a,b=1}^2
\int_{\Omega^t} d\rho(x) \int_{M \setminus \Omega^t} d\rho(y)\:
\Big( \big| \nabla_{a,\v} \L(x,y) \big| + \big| \nabla_{a,\v} \nabla_{b,\v} \L(x,y) \big|  \Big) &< \infty\:. \\
\sum_{a,b=1}^2
\int_{N^t} d\mu_t(x) \int_M d\rho(y)\:
\Big( \big| \nabla_{a,\v} \L(x,y) \big| + \big| \nabla_{a,\v} \nabla_{b,\v} \L(x,y) \big|  \Big) &< \infty\:.
\end{align*}
\end{Def}

\begin{Def} \label{defosi} Assuming that~$\J_{\rho, \sc}$ is surface layer finite, we define the
following surface layer integrals,
\begin{align}
\gamma^t_\rho \::\: \J_{\rho, \sc} &\rightarrow \R \qquad \text{(conserved one-form)} \notag \\
\gamma^t_\rho(\v) &= \int_{\Omega^t} d\rho(x) \int_{M \setminus \Omega^t} d\rho(y)\:
\big( \nabla_{1,\v} - \nabla_{2,\v} \big) \L(x,y) \label{gamma} \\
\sigma^t_\rho \::\: \J_{\rho, \sc} \times \J_{\rho, \sc} &\rightarrow \R  \qquad \text{(symplectic form)} \notag \\
\sigma^t_\rho(\u, \v) &= \int_{\Omega^t} d\rho(x) \int_{M \setminus \Omega^t} d\rho(y)\:
\big( \nabla_{1,\u} \nabla_{2,\v} - \nabla_{2,\u} \nabla_{1,\v} \big) \L(x,y) \label{sigma} \\
(.,.)^t_\rho \::\: \J_{\rho, \sc} \times \J_{\rho, \sc} &\rightarrow \R \qquad \text{(surface layer inner product)} \notag \\
(\u, \v)^t_\rho &= \int_{\Omega^t} d\rho(x) \int_{M \setminus \Omega^t} d\rho(y)\:
\big( \nabla_{1,\u} \nabla_{1,\v} - \nabla_{2,\u} \nabla_{2,\v} \big) \L(x,y) \:. \label{sprod}
\end{align}
\end{Def}
Here~\eqref{gamma} corresponds to the conservation law for the functional~$I^\Omega_1$
as established in~\cite[Theorem~3.1 and Section~3.3]{osi}; see also~\cite[Corollary~3.9]{osi}. 
The surface layer integral in~\eqref{sigma}, on the other hand, is the symplectic form (see~\cite[Section~4.3]{jet});
it is obtained alternatively by anti-symmetrizing the conservation law for~$I^\Omega_2(\u,\v)$
in the jets~$\u$ and~$\v$ (see~\cite[Corollary~3.10]{osi}).
Finally, the surface layer integral in~\eqref{sprod} is
obtained by symmetrizing~$I^\Omega_2(\u,\v)$ in its two arguments
(see~\cite[Theorem~1.1]{osi}).

In this paper, we always assume that the bilinear form defined by~\eqref{sigma}
is {\em{non-degenerate}}, and that~\eqref{sprod} is {\em{positive semi-definite}}.
These assumptions are sensible because they have
been justified in Minkowski space in~\cite{action}. However, a-priori
 the bilinear form~$\sigma^t_\rho$ may be degenerate (thus it would be more appropriate
to call it a ``presymplectic form''). In this case, our method is to choose~$\Jvary$ so small
that the restriction of~$\sigma^t_\rho$ to~$\J_{\rho, \sc} \times \J_{\rho, \sc}$
is non-degenerate. This procedure also justifies the name ``symplectic form.''
Particular examples where the symplectic form will be degenerate are systems involving
gauge symmetries. In these examples, the choice of~$\Jvary$ involves a gauge-fixing procedure
or the choice of a specific gauge.

We finally remark that in the smooth setting, the symplectic form is the
exterior derivative of $\gamma_\rho$ (for details see the proof of~\cite[Lemma~3.4]{jet}).
Moreover, the inner product~$(.,.)_\rho$ can be regarded as the symmetrized derivative of~$\gamma_\rho$.
We will come back to this point in more detail in Section~\ref{seccompconn}.

As shown in~\cite{jet, osi}, the above surface layer integrals satisfy conservation laws.
For self-consistency and for a better comparison with the nonlinear conservation law
in Section~\ref{secosinonlin}, we now derive these conservation laws again, but
with a different method where we differentiate the above formulas with respect to time.
\begin{Prp} \label{prplinconserve} For linearized solutions~$\u, \v \in \Jlin_{\rho, \sc}$,
the surface layer integrals in Definition~\ref{defosi} satisfy the identities
\begin{align}
\frac{d}{dt} \gamma^t_\rho(\v) &= -\int_{N^t} \nabla_\v \s\: d\mu_t(x) \label{cI1} \\
\frac{d}{dt} \sigma^t_\rho(\u, \v) &= 0 \label{cI2as} \\
\frac{d}{dt} (\u, \v)^t_\rho &= \int_{N^t} \Big( \nabla_\u \nabla_\v \s -2 \,\Delta_2[\u, \v] \Big)\: d\mu_t \:, \label{cI2symm}
\end{align}
where
\[ \Delta_2[\u,\v] := \frac{1}{2} \: \bigg( \int_{M} \big( \nabla_{1, \u} + \nabla_{2, \u} \big)
\big( \nabla_{1, \v} + \nabla_{2, \v} \big) \L(x,y)\: d\rho(y) - \nabla_\u \nabla_\v\: \s \bigg) \:. \]
\end{Prp}
\Proof Using~\eqref{mutdef}, we can rewrite the spacetime integrals
using Fubini's theorem as products of a time integral and a spatial integral.
Then time derivatives reduce to differentiating the upper or lower
limits of integration, like for example
\[ \frac{d}{dt} \int_{\Omega^t} (\cdots)\: d\rho(x) = 
\frac{d}{dt} \int_{-\infty}^t dt' \int_{N^{t'}} (\cdots)\: d\mu_{t'} = 
\int_{N^t} (\cdots)\: d\mu_t\:. \]
In this way, we obtain
\begin{align}
\frac{d}{dt} \gamma^t_\rho(\v) &= \int_{N^t} d\mu_t(x) \int_{M \setminus \Omega^t} d\rho(y)\:
\big( \nabla_{1,\v} - \nabla_{2,\v} \big) \L(x,y) \notag \\
&\qquad -\int_{\Omega^t} d\rho(x) \int_{N^t} d\mu_t(y)\:
\big( \nabla_{1,\v} - \nabla_{2,\v} \big) \L(x,y) \notag \\
&\!\overset{(*)}{=} \int_{N^t} d\mu_t(x) \int_M d\rho(y)\: \big( \nabla_{1,\v} - \nabla_{2,\v} \big) \L(x,y) \notag \\
&= \int_{N^t} d\mu_t(x) \Big(2 \nabla_\v \ell - \Delta \v - \nabla_\v \s \Big) 
= -\int_{N^t} \nabla_\v \s\: d\mu_t(x) \:, \label{gevolve}
\end{align}
where in~($*$) we used that the integrand is anti-symmetric in~$x$ and~$y$
and employed~\eqref{ldef} and~\eqref{eqlinlip2}.
Similarly,
\begin{align}
\frac{d}{dt} \sigma^t_\rho(\u, \v) &= \int_{N^t} d\mu_t(x)
\int_M d\rho(y)\: \big( \nabla_{1,\u} \nabla_{2,\v} - \nabla_{2,\u} \nabla_{1,\v} \big) \L(x,y) \notag \\
&= \int_{N^t} \Big( \la \u, \Delta \v \ra(x) - \la \v, \Delta \u \ra(x) \Big)\: d\mu_t(x) = 0 \\
\frac{d}{dt} (\u, \v)^t_\rho &= \int_{N^t} d\mu_t(x)
\int_M d\rho(y)\: \big( \nabla_{1,\u} \nabla_{1,\v} - \nabla_{2,\u} \nabla_{2,\v} \big) \L(x,y) \notag \\
&= -\int_{N^t} d\mu_t(x) \int_M d\rho(y)\: 
\big( \nabla_{1,\u} + \nabla_{2,\u} \big) \big( \nabla_{1,\v} + \nabla_{2,\v} \big) \L(x,y) \notag \\
&\quad\: + \int_{N^t} d\mu_t(x) \int_M d\rho(y)\: 
\nabla_{1,\u} \big( \nabla_{1,\v} + \nabla_{2,\v} \big) \L(x,y) \notag \\
&\quad\: + \int_{N^t} d\mu_t(x) \int_M d\rho(y)\: \nabla_{1,\v} \big( \nabla_{1,\u} + \nabla_{2,\u} \big) \L(x,y) \notag \\
&= \int_{N^t} \Big( -2 \,\Delta_2[\u, \v] + \la \u, \Delta \v \ra + \la \v, \Delta \u \ra + \nabla_\u \nabla_\v \s\Big)\: d\mu_t \:.
\label{sevolve}
\end{align}
This completes the proof.
\QED
We now explain why and in which sense the identities~\eqref{cI1}--\eqref{cI2symm}
can be understood as {\em{conservation laws}}. The relation~\eqref{cI2as}
implies that the symplectic form is conserved in time.
The relation~\eqref{cI1} shows that the same is true for~$\gamma^1_\rho(\v)$, provided that
the scalar component of~$\v$ is zero. Thus we get a conservation law for linearized solutions
with vanishing scalar component. The term~$\nabla_\u \nabla_\v \s$ in~\eqref{cI2symm}
involves again only the scalar components of the jets.
The term~$\Delta_2[\u, \v]$ in~\eqref{cI2symm}, on the other hand,
can be understood as an interaction term. Thus we get an exact conservation law only if
this interaction term vanishes. If the interaction term is non-zero, then~\eqref{cI2symm}
can still be interpreted as an approximate conservation law. This approximate conservation law is
indeed very useful for proving existence of solutions (for details see the energy estimates in~\cite[Section~3.2]{linhyp}).
We remark that the right side of~\eqref{cI2symm} can be rewritten using Green's operators;
see~\cite[Theorem~1.1]{osi}.

\subsection{The Cauchy Problem and Restrictions to Surface Layers} \label{seccauchy}
We now recall a few results of~\cite{linhyp} on the Cauchy problem for the linearized field
equations which will be needed later on.
We again assume that spacetime is globally hyperbolic (see~\cite[Definition~4.19]{linhyp})
and that it has a smooth manifold structure with global time function~$\scrt$ (see Definition~\ref{defsms}).
One of the main results in~\cite{linhyp} is to show that the Cauchy problem is well-posed.
However, the initial data must not be prescribed on a hypersurface, but instead
on a {\em{surface layer}}, i.e.\ in a spacetime strip which is extended in time on the Compton scale
(i.e.~$\Delta t \sim m^{-1}$). More precisely, given initial data~$\v_0 \in \J_{\rho, \sc}$,
there is a unique solution~$\v \in \Jlin_{\rho, \sc}$ which coincides with~$\v_0$ in the
surface layer at time~$t_0$ in the sense that~$\v-\v_0 \in \underline{\J_\rho}_{t_0}$
(for details see~\cite[Section~3.5]{linhyp}). It is convenient to always identify
the Cauchy data on a surface layer with the corresponding linearized solution.
We thus obtain a mapping
\beq \label{restrictionmap}
|^{t_0} \::\: \J_{\rho, \sc} \rightarrow \Jlin_{\rho, \sc} \:,\qquad \v_0 \mapsto \v_0|^{t_0} = \v \:,
\eeq
which we refer to as the {\em{restriction map}}.
The restriction map is very useful because it allows us to identify a nonlinear jet
at any time with a corresponding linearized solution, obtained by restricting the nonlinear
jet to the Cauchy surface layer and taking it as the initial data for the Cauchy problem.

Before going on, we briefly discuss and explain this construction.
One should keep in mind that in the Cauchy surface layer, the jets~$\v$ and~$\v_0$ do not need to coincide pointwise,
but only in a weak sense when tested in~$\Jtest$. This implies that identifying nonlinear jets
with linearized solutions using the restriction map may involve an error term which takes into account
corrections due to the microscopic spacetime structure.
Since the aim in this work is to get a connection to quantum field theory in Minkowski space,
we shall simply ignore these corrections. But they must clearly be taken into account when
quantum gravity effects are considered.

\section{Inner Solutions} \label{secinner}
\subsection{Definition and Basic Properties}
We again assume that spacetime has a smooth manifold structure and admits a global time function
(see Definition~\ref{defsms}).
Let~$v \in \Gamma(M, TM)$ be a vector field. Then
its {\em{divergence}} $\div v \in C^\infty(M, \R)$ may be defined by the relation
\[ \int_M \div v\: \eta(x)\: d\rho = -\int_M D_v \eta(x)\: d\rho(x) \:, \]
to be satisfied by all test functions~$\eta \in C^\infty_0(M, \R)$.
In a local chart~$(x,U)$, the divergence is computed by
\beq \label{defdiv}
\div v = \frac{1}{h}\: \partial_j \big( h\, v^j \big)
\eeq
(where, following the Einstein summation convention, we sum over~$j=0,\ldots,3$).

When integrating by parts using Gau{\ss}' theorem, we need to be able to make
sure that we do not get boundary values at infinity. To this end, it is convenient
to choose the Riemannian metric~$g_x$ introduced before~\eqref{vsprod}
to be compatible with the smooth manifold structure in the following sense.
We again assume that~\eqref{Tinclude} holds.

\begin{Def} \label{defadapted} The Riemannian metric~$g$ on~$\Gamma_x$ is {\bf{adapted at infinity}} if
there is a sequence~$(\eta_n)_{n \in \N}$ of compactly supported functions,
$\eta_n \in C^\infty_0(M, \R)$, with the following properties:
\bitem
\item[{\rm{(i)}}] The functions~$\eta_n$ are monotone increasing and exhaust~$M$
in the sense that for any compact set~$K \subset M$ there is~$N$ with~$\eta_n|_K \equiv 1$
for all~$n \geq N$.
\item[{\rm{(ii)}}] The derivatives tend uniformly to zero, i.e.
\[ \lim_{n \rightarrow \infty} \sup_{x \in M} \|D \eta_n\|_x = 0 \:, \]
where~$\|.\|_x$ is the norm on~$T_xM \subset \Gamma_x$ induced by the Riemannian metric~$g$.
\eitem
\end{Def} \noindent
In the later applications when~$M$ is topologically Minkowski space (see Section~\ref{seclinear}),
we can choose~$\eta_n$ as~$\eta(x/n)$ with a usual cutoff function~$\eta \in C^\infty_0(\R^4)$,
and~$g$ such that its restriction to~$TM$ is equivalent to the Euclidean metric.

\begin{Def} \label{definner} An {\bf{inner solution}} is a jet~$\v$ of the form
\[ \v = (\div v, v) \qquad \text{with} \qquad v \in \Gamma(M, TM) \:. \]
We make the following regularity and decay assumptions:
\bitem
\item[{\rm{(i)}}] The vector field~$v$
can be extended to a vector field~$\tilde{v} \in \Gamma(U, T\F)$ defined in a neighborhood~$U$ of~$M$
such that the directional derivative~$(D_{1,\tilde{v}} + D_{2,\tilde{v}}) \L(x,y)$
exists for all~$x \in U$ and~$y \in M$ and is integrable in~$y$, i.e.\
\[ \int_M \Big| \big(D_{1,\tilde{v}} + D_{2,\tilde{v}} \big) \L(x,y) \Big| \: d\rho(y) < \infty \qquad
\text{for all~$x \in U$}\:. \]
Moreover, the directional derivative~$D_{\tilde{v}} \ell(x)$ exists for all~$x \in U$ and is continuous in~$U$.
\item[{\rm{(ii)}}] The integral
\[ \int_M \L(x,y)\: \|\v(y)\|_y \:d\rho(y) \]
is finite and bounded locally uniformly in a neighborhood of~$M$
(where~$\|.\|_y$ is again the norm corresponding to the scalar product~\eqref{vsprod}
and the Riemannian metric adapted at infinity according to Definition~\ref{defadapted}).
\item[{\rm{(iii)}}] For any test jet~$\u \in \Jtest_\rho$, the
directional derivative~$D_v \u$ (computed in the same charts used for computing the higher derivatives in Definition~\ref{defJvary}) is again in~$\Jtest_\rho$.
\eitem
The vector space of all inner solutions is denoted by~$\Jin_\rho$.
The set of all inner solutions which are compactly supported on every set~$N^t$
is denoted by~$\Jin_{\rho, \sc}$.
\end{Def} \noindent
Note that~(i) implies that every inner solution is in~$\J^1_\rho \cap \Jdiff_\rho$
(see~\eqref{Jdiffdef} and Definition~\ref{defJvary}).

The name ``inner {\em{solution}}'' is justified by the following lemma:
\begin{Lemma} Every inner solution~$\v \in \Jin_\rho$ is a solution of the linearized field equations, i.e.\
\[ \la \u, \Delta \v \ra|_M = 0 \qquad \text{for all~$\u \in \Jtest_\rho$} \:. \]
\end{Lemma}
\Proof Applying the Gauss divergence theorem, one finds that for every~$f \in C^1_0(M, \R)$,
\[ \int_M \nabla_\v f\: d\rho =
\int_M \big( \div v\: f + D_v f \big)\: d\rho 
= \int_M \div \big( f v \big)\: d\rho = 0 \:. \]
Likewise, in the linearized field equations we may integrate by parts in~$y$.
In a formal computation, we obtain for any~$x \in U$,
\beq \label{pint}
\int_M \big( D_{1,\tilde{v}} + \nabla_{2,\v} \big) \L(x,y) \: d\rho(y) = 
D_{\tilde{v}} \big( \ell + \s \big)(x) \:.
\eeq
However, as the function~$\L(x,.)$ need not be compactly supported, we
need to insert the cutoff functions~$\eta_n$ in Definition~\ref{defadapted}.
Moreover, we need to be careful because the individual derivatives do not need to exist.

In order to prove~\eqref{pint}, making use of Definition~\ref{definner}~(i), we know
from Lebesgue's dominated convergence theorem that for any~$x \in U$,
\[ A(x) := \int_M \big( D_{1,\tilde{v}} + \nabla_{2,\tilde{\v}} \big) \L(x,y) \: d\rho(y) = 
\lim_{n \rightarrow \infty} \int_M \big( D_{1,\tilde{v}} + \nabla_{2,\tilde{\v}} \big) \L(x,y)\:
\eta_n(y)\: d\rho(y) \:. \]
Now we can integrate by parts to obtain
\beq \label{Aform}
A(x) =  \lim_{n \rightarrow \infty} \bigg( D_{\tilde{v}} \int_M \L(x,y)\: \eta_n(y)\: d\rho(y) 
- \int_M \L(x,y)\: \big( D_v \eta_n(y)\big)\: d\rho(y) \bigg)
\eeq
(here one needs to pull out the derivative~$D_{\tilde{v}}$ before the integral, because
the Lagrangian need not be differentiable; the integral, on the other hand, is well-defined because
the last integral is).
The last integral can be estimated by
\beq \label{intes}
\bigg| \int_M \L(x,y) \: \big( D_v \eta_n(y) \big)\: d\rho(y) \bigg|
\leq \sup_M \|D_v \eta_n\| \int_M \|v(y)\|_y \:\L(x,y) \: d\rho(y) \:.
\eeq
According to Definition~\ref{definner}~(ii), the obtained integral is bounded locally
uniformly in~$x$. Using Definition~\ref{defadapted}~(ii), we conclude that
the last integral in~\eqref{Aform} tends to zero as~$n \rightarrow \infty$, locally uniformly in~$x$.

As a consequence, also the first integral in~\eqref{Aform} converges as~$n \rightarrow 0$, locally uniformly in~$x$.
In order to prove~\eqref{pint}, it remains to show that this limit is given by
\beq \label{limn}
\lim_{n \rightarrow \infty} \bigg( D_{\tilde{v}} \int_M \L(x,y)\: \eta_n(y)\: d\rho(y) \bigg) =
D_{\tilde{v}} \big( \ell + \s \big)(x) \:.
\eeq
Assume conversely that this equation does not hold for all~$x \in U$. Then by continuity
(note that the left side of~\eqref{limn} is continuous as a locally uniform limit of continuous
function, as is the right side by Definition~\ref{definner}~(i)), the equation~\eqref{limn} is violated in an open set.
Therefore, we may
choose a path~$\gamma : [t_0, t_1] \rightarrow U$ along the integral curves of~$\tilde{v}$ such that
\[ \int_{t_0}^{t_1} \lim_{n \rightarrow \infty} \bigg( D_{\tilde{v}} \int_M \L\big(\gamma(t),y \big)\: \big(1-\eta_n(y)\big)\: d\rho(y) \bigg)\: dt
\neq 0 \:. \]
Due to the locally uniform convergence, we may interchange the integral and the limit to conclude that
\[ \lim_{n \rightarrow \infty} \int_M \L(x,y)\: \big(1-\eta_n(y)\big)\: d\rho(y) \Big|^{x=\gamma(t_1)}_{x=\gamma(t_0)} 
\neq 0 \:. \]
On the other hand, using assumption~(iv) on page~\pageref{Cond4}, the limits on the left
vanish using Lebesgue's dominated convergence theorem. This is a contradiction.
Hence~\eqref{limn} holds. This concludes the proof of~\eqref{pint}.

We rewrite~\eqref{pint} as
\[ \Delta \tilde{\v}(x) = \nabla_{\tilde{\v}} \ell(x) \qquad \text{for all~$x \in U$} \]
(where the scalar component of~$\v$ can be extended to~$U$ arbitrarily).
The next and final step is to show that for any~$\u \in \Jtest_\rho$ and~$x \in M$, the jet derivative~$\nabla_\u$
of this equation exists and vanishes. To this end, we write the jet derivative of the right side as
\[ \nabla^2\ell|_x(\u,\v) = \nabla_{\v(x)} \big( \nabla_\u \ell(x)\big) - \nabla_{D_v \u} \ell(x) \]
(where the first summand on the right is an iterated directional derivative).
The last summand vanishes because of the weak EL equations,
using that~$D_v \u \in \Jtest_\rho$ (see Definition~\ref{definner}~(iii)).
In order to treat the first summand, we note that the function~$\nabla_\u \ell$ vanishes identically
on~$M$ by the weak EL equations. Therefore, this function is differentiable in the direction
of every vector field on~$M$, and this directional derivative is zero.
This concludes the proof.
\QED

Inner solutions have the nice property that surface layer integrals simplify to
standard surface integrals, as is exemplified in the following proposition for the
conserved one-form and the symplectic form in Definition~\ref{defosi}.
\begin{Def} Let~$\v = (\div v, v) \in \Jin_\rho$ be an inner solution and~$\Omega \subset M$ 
closed with smooth boundary~$\partial \Omega$.
On the boundary, we define the measure~$d\mu(\v,x)$ as the contraction of the volume form on~$M$
with~$v$, i.e.\ in local charts
\beq \label{mudef}
d\mu(\v,x) = h\: \epsilon_{ijkl} \:v^i\: dx^j dx^k dx^l \:,
\eeq
where~$\epsilon_{ijkl}$ is the totally anti-symmetric Levi-Civita symbol
(normalized by~$\epsilon_{0123}=1$).
\end{Def} \noindent
We also use the notation
\[ \mu(\v,N_t) =\int_{N_t} d\mu(\v,x) \:. \]
\begin{Prp} \label{prpintpart}
Let~$\v \in \Jin_{\rho, \sc}$ be a spatially compact inner solution
and~$\u \in \Jlin_\rho$ a linearized solution. Then for any time~$t$,
\begin{gather}
\gamma^t_\rho(\v) = \s \,\mu(\v,N^t) \label{innerflux} \\
\sigma^t_\rho(\u, \v) = 0 \:. \label{innersigma}
\end{gather}
\end{Prp}
\Proof In~\eqref{gamma} we integrate by parts with the help of the Gau{\ss} divergence theorem.
We thus obtain
\begin{align*}
\gamma^t_\rho(\v)
&= \int_{N^t} d\mu(\v,x) \int_{M \setminus \Omega^t} d\rho(y)\:
\L(x,y) + \int_{\Omega^t} d\rho(x) \int_{N^t} d\mu(\v,y)\: \L(x,y) \\
&= \int_{N^t} d\mu(\v,x) \int_M d\rho(y)\: \L(x,y)
= \s \int_{N^t} d\mu(\v,x) = \s\, \mu \big(\v, N^t \big) \:,
\end{align*}
where in the last line we used the symmetry of~$\L$ and employed the EL equations.
This gives~\eqref{innerflux}.

In order to derive~\eqref{innersigma}, we integrate by parts in~\eqref{sigma},
\begin{align*}
\sigma^t_\rho(\v, \u) &= \int_{N^t} d\mu(\v,x) \int_{M \setminus \Omega^t} d\rho(y)\: \nabla_{2,\u} \L(x,y)
+ \int_{\Omega^t} d\rho(x) \int_{N^t} d\mu(\v,y)\: \nabla_{1,\u} \L(x,y) \\
&= \int_{N^t} d\mu(\v,x) \int_M d\rho(y)\: \nabla_{2,\u} \L(x,y) \:.
\end{align*}
Adding the weak EL equations~$\nabla_\u \ell(x)=0$, we obtain
\begin{align*}
\sigma^t_\rho(\v, \u) 
&= \int_{N^t} d\mu(\v,x) \bigg( \int_M \big(\nabla_{1,\u} + \nabla_{2,\u} \big) \L(x,y)\:d\rho(y)
- \big(\nabla_\u \,\s\big)(x) \bigg) \\
&= \int_{N^t} (\Delta \u)(x)\: d\mu(\v,x) = 0 \:,
\end{align*}
giving the result.
\QED
We next show that inner solutions can always be used for testing:
\begin{Prp} \label{prptest} Let~$v$ be a solution of the linearized field equations~\eqref{eqlinlip}.
Then the linearized field equations are also satisfied when testing in~$\Jin_\rho$, i.e.\
\beq \label{eqin}
\la \u, \Delta \v \ra|_M = 0 \qquad \text{for all~$\u \in \Jin_\rho$} \:.
\eeq
\end{Prp}
\Proof Let~$\u = (\div u, u)$ be an inner solution.
Using that the scalar component of the test jets form a subspace~\eqref{Gammatest}
are nowhere trivial~\eqref{Cnontriv}, the linearized field equations~\eqref{eqlinlip} imply
that
\[ \int_M \big( \nabla_{1, \v} + \nabla_{2, \v} \big) \L(x,y)\: d\rho(y) - \big( \nabla_\v \,\s \big)(x)=0
\qquad \text{for all~$x \in M$}\:. \]
Hence all the derivatives in the direction of the vector field~$u$ vanish.
This gives the result.
\QED

Let us explain the significance of above constructions.
Inner solutions can be regarded as infinitesimal generators of transformations of~$M$
which leave the measure~$\rho$ unchanged. Therefore, inner solutions do not change the
causal fermion system, but merely describe symmetry transformations of the universal measure.
In view of Proposition~\ref{prptest}, it is a good idea to enlarge~$\Jtest$
such as to include all inner solutions (note that every inner solution
is in~$\Jdiff$ as defined in~\eqref{Jdiffdef}, simply because the function~$\ell$
vanishes identically on~$M$, so that its derivative in~\eqref{Gdiffdef} exists and vanishes
for every inner solution). With this in mind, in what follows we always assume that
\beq \label{Jinsub}
\Jin_\rho \subset \Jtest_\rho \:.
\eeq

\subsection{Construction of Spatially Compact Inner Solutions}
In this section we use methods of hyperbolic partial differential equations in order to
construct useful classes of inner solutions. We first show that the
scalar component of an inner solution can be chosen arbitrarily:

\begin{Prp} \label{prpinnera}
Let~$a \in \Cisc(\scrM, \R)$ be a smooth function with spatially compact support.
Then there is a spatially compact vector field with
\[ \div v = a \:. \]
In the case that~$a$ is compactly supported, the vector field~$v$
can be chosen to be supported in the future of~$a$, in the sense that
for all~$t \in \R$ the implication
\[ a|_{\Omega^t} \equiv 0 \quad \Longrightarrow \quad v|_{\Omega^t} \equiv 0 \]
holds (with~$\Omega^t$ as defined in~\eqref{Omegatdef}).
\end{Prp}
\Proof Our task is to solve the equation~$\div v = a$, which can be written equivalently as
\beq \label{diveq}
\partial_j \big( h\, v^j \big) = h a \:.
\eeq
We first consider the case that~$a$ has compact support.
In order to solve the partial differential equation~\eqref{diveq}, it is useful to choose a Lorentzian
metric~$g$. The choice of the metric is irrelevant, and the arbitrariness in choosing the metric
corresponds to the fact that~\eqref{diveq} is an underdetermined equation which admits
many different solutions. For simplicity, we choose~$g$ as the Minkowski metric.
Let~$\Box$ be the wave operator in Minkowski space. Using for example retarded Green's
operators, there is a solution~$\phi \in \Cisc(\scrM, \R)$ with~$\Box \phi = h a$.
Then the vector field
\beq \label{vconstruct}
v^j := \frac{1}{h} \: g^{jk}\: \partial_k \phi \:.
\eeq
satisfies~\eqref{diveq}.

In the case that the function~$a$ merely has spatially compact support, we decompose~$a$ as
\[ a =a_+ + a_- \:, \]
where~$a_+$ is supported in the set~$\{t > 0\}$ and~$a_-$ is supported in the set~$\{t < 1 \}$.
Denoting the advanced and retarded Green's operators of the scalar wave equation in Minkowski
space by~$S^\vee$ and~$S^\wedge$, respectively, the function
\[ \phi := S^\wedge \big( h a_+ \big) + S^\vee\big( h a_- \big) \]
is a well-defined solution of the equation~$\Box \phi = h a$ which is smooth and has spatially compact
support. Therefore, we can again define the vector field~$v$ by~\eqref{vconstruct}.
This gives the result.
\QED
In what follows, we always assume that this vector field satisfies all the regularity and decay assumptions
in Definition~\ref{definner}. We then obtain a corresponding inner solution
\[ \v := (a, v) \in \Jin_{\rho, \sc} \:. \]

According to Lemma~\ref{prpintpart}, for a spatially compact
inner solution~$\v \in \Jin_{\rho, \sc}$
the surface layer inner product~$\gamma^t_\rho(\v)$ in~\eqref{gamma} reduces to the flux of~$v$ through
the surface~$N^t$.
According to the Gauss divergence theorem, this flux integral is time independent
provided that~$v$ is divergence-free. Keeping in mind that the divergence of~$v$ is precisely the
scalar component of the resulting inner solution, we get direct agreement with the
conservation law~\eqref{cI1}. We now show that the
flux integral~\eqref{innerflux} can be arranged to have an arbitrary value.
\begin{Prp} \label{prpinnerflux}
Given~$c \in \R$, there is a spatially compact inner solution~$\w =(0, w) \in \Jin_{\rho,\sc}$ without
scalar component with~$\gamma^t_\rho(\w)=c$.
\end{Prp}
\Proof As in the proof of Proposition~\ref{prpinnera} we consider the scalar wave equation.
We choose a compactly supported smooth vector field~$v$ on~$N^t$
such that~$\gamma^t_\rho(\v)=c$. Next, we let~$\phi$ be the solution of the Cauchy problem
\[ \Box \phi = 0 \:,\qquad \phi|_{N^t}=0,\; \frac{1}{h} \: g^{jk}\: \partial_k \phi = v^j \:. \]
Due to finite propagation speed, this solution is spatially compact.
Moreover, the vector field~$w$ with components
\[ w^j := \frac{1}{h} \: g^{jk}\: \partial_k \phi \]
is divergence-free and coincides on~$N^t$ with~$v$.
\QED

\subsection{Application: Linearized Solutions without Scalar Components}
As an application of the above results, we now construct a space of linear solutions
with particularly nice properties. Indeed, combining Propositions~\ref{prpinnera} and~\ref{prpinnerflux}
with the conservation law of~\eqref{cI1}
immediately gives the following result:
\begin{Corollary} \label{corvmu}
For any solution~$\v \in \Jlin_{\rho, \sc}$ there is an inner solution~$\u \in \Jin_{\rho, \sc}$
such that
\bitem
\item[\rm{(i)}] The jet~$\v-\u$ has no scalar component.
\item[\rm{(ii)}] $\gamma^t_\rho(\v-\u)=0$ for all~$t$.
\eitem
\end{Corollary}
Clearly, the jet~$\u$ is not unique, because the vector field~$u$ can be modified by a divergence-free
vector field with no flux through~$N^t$. But, similar to a gauge freedom,
this arbitrariness has no physical significance.

Choosing for every basis vector~$\v \in \Jlin_{\rho, \sc}$ a corresponding jet~$\u \in \Jin_{\rho, \sc}$,
writing~$\v-\u=(0, w)$ we obtain a vector field on~$T\F$ along~$M$.
All the resulting vector fields span a vector space which we denote by~$\Glin_\sc$.
In this way, we obtain a linear mapping~$\mathscr{V} \::\: \Jlin_\sc \rightarrow \Glin_\sc$
with the property that for every~$w \in \Glin_\sc$, the corresponding jet~$\w=(0,w)$
is a linearized solution with vanishing inner flux, i.e.\ $\gamma^t_\rho(\w)=0$.
Moreover, from~\ref{prpintpart} we know that the symplectic form is not
affected by this construction, i.e.\ $\sigma^t_\Omega(\u, \v)=
\sigma^t_\Omega((0, \mathscr{V} \u), (0, \mathscr{V} \v))$.
For ease in notation, in what follows we implicitly identify the vector fields with
corresponding jets with zero scalar components. For example, we simply
write~$\gamma^t_\rho (\mathscr{V} \u)$ and similar for the symplectic form.
In view of~\eqref{Jinsub}, the jets in~$\Glin_\sc$ again consist of test jets.

The construction which we just carried out for linearized solutions can be
performed similarly for jets in the image of the causal Green's operators.
According to Proposition~\ref{prpinnera}, this can be arranged while preserving
the causal support properties of these jets
(more precisely, by scaling the Minkowski metric~$g$ in~\eqref{vconstruct} appropriately,
one can arrange that the light cones of~$g$ lie inside the causal cones of the Green's operators).
With a slight abuse of notation, we denote these modified operators again
by~$S^\vee$ and~$S^\wedge$.
The result of this construction is summarized as follows.
\begin{Corollary} \label{cornoscalar}
By adding  suitable inner solutions to the jets in~$\Jvary_{\rho,0}$
and~$\J_{\rho, \sc}$, we obtain vector spaces of vector fields
\[ \Gvary_{\rho,0}, \:\Gamma_{\rho, \sc} \;\subset\; \Gtest \]
together with mappings~$\Delta$ and~$G$ related to each other by the exact sequence
\beq \label{exact2}
0 \rightarrow \Gvary_{\rho,0} \overset{\Delta}{\longrightarrow} \J^*_{\rho,0}
\overset{G}{\longrightarrow} \Gamma_{\rho,\sc}
\overset{\Delta}{\longrightarrow} \J^*_{\rho,\sc} \rightarrow 0 \:.
\eeq
Introducing the subspace of solutions by
\[ \Glin_{\rho,\sc} := G \big( \J^*_{\rho,0} \big) \subset \Gamma_{\rho,\sc} \:, \]
the conserved one-form vanishes on this subspace,
\beq \label{noIt}
\gamma^t_\rho(v) = 0 \qquad \text{for all~$v \in \Glin_{\rho,\sc}$ and all~$t \in \R$}\:.
\eeq
The causal Green's operators in~\eqref{Scausal}
and the $p^\text{th}$ order perturbation map in~\eqref{Pp} become mappings
\begin{align}
S^\vee, S^\wedge &\::\: \;\;\;\J^*_{\rho,0}\;\;\; \rightarrow \Gamma_{\rho, \sc} \label{Scausal2} \\
\Pert_\rho^{(p)} &\::\: \big( \Glin_{\rho, \sc} \big)^p \rightarrow
\Gamma_{\rho, \sc} \:. \label{Pp2}
\end{align}
\end{Corollary}
In what follows, we shall work exclusively with the jet spaces in~\eqref{exact2}
as well as with the spatially compact inner solutions~$\Jin_{\rho, \sc}$.

\section{A Conservation Law for a Nonlinear Surface Layer Integral} \label{secosinonlin}
We now introduce another conservation law for a surface layer integral.
Instead of working with linearized solutions, we directly compare the perturbed measure~$\tilde{\rho}$,
which takes into account the nonlinear interaction, with the vacuum measure~$\rho$. 
We again assume that the measure~$\tilde{\rho}$ is of the form~\eqref{rhoFf}.
For technical simplicity and for notational consistency with~\cite{noether, jet, osi}, we
first define the surface layer integral for a compact subset~$\Omega \subset M$.
Later on, the surface layer integral will also be used as in Definition~\ref{defosi}
for the set~$\Omega^t$ in~\eqref{Omegatdef}.

\begin{Def} \label{defosinl}
Given a compact subset~$\Omega \subset M$, the {\bf{nonlinear surface layer integral}}
$\gamma^\Omega(\tilde{\rho}, \rho)$ is defined by
\beq \label{osinl}
\gamma^\Omega(\tilde{\rho}, \rho) =
\int_{\tilde{\Omega}} d\tilde{\rho}(x) \int_{M \setminus \Omega} d\rho(y)\: \L(x,y)
- \int_{\Omega} d\rho(x) \int_{\tilde{M} \setminus \tilde{\Omega}}  d\tilde{\rho}(y)\: \L(x,y) \:,
\eeq
where~$\tilde{M}:= \supp \tilde{\rho}$ and~$\tilde{\Omega}:=F(\Omega)$.
\end{Def}
Using the definition of the push-forward measure, the nonlinear surface layer integral can be
written alternatively as
\beq \label{osinl2}
\gamma^\Omega(\tilde{\rho}, \rho) =
\int_{\Omega} d\rho(x) \int_{M \setminus \Omega} d\rho(y)\: 
\Big( f(x)\, \L\big(F(x),y \big) -  \L\big(x, F(y) \big)\, f(y) \Big)\:.
\eeq
Clearly, this surface layer integral is closely related to the surface layer integrals in Definition~\ref{defosi}.
Indeed, expanding the integrand in~\eqref{osinl2} linearly, one gets precisely the integrand in~\eqref{gamma}.
Expanding to second order, one gets the integrand in~\eqref{sprod} divided by two.
Therefore, the nonlinear surface layer integral can be regarded as a generalization of~\eqref{gamma}
and~\eqref{sprod} which takes into account higher order corrections.

The derivation of the corresponding conservation law follows
the idea first given in~\cite{noether}. In analogy to~\eqref{ldef} we set
\beq \label{tildeldef}
\tilde{\ell}(x) := \int_\F \L(x,y)\: d\tilde{\rho}(y) - \s
\eeq

\begin{Thm} \label{thmosinlvol}
The nonlinear surface layer integral can be written as the volume term
\beq
\gamma^\Omega(\tilde{\rho}, \rho) = \int_\Omega 
\bigg( \Big( f(x)\, \ell\big( F(x) \big) - \tilde{\ell}(x) \Big) 
+ \s \:\big(f(x) - 1 \big) \bigg) \: d\rho(x)
\label{osinlvol} \:.
\eeq
\end{Thm}
\Proof Using the antisymmetry of the integrand in~\eqref{osinl2}, we obtain
\begin{align*}
\gamma^\Omega(\tilde{\rho}, \rho) &= \int_{\Omega} d\rho(x) \int_M d\rho(y)\: 
\Big( f(x)\, \L\big(F(x),y \big) -  \L\big(x, F(y) \big)\, f(y) \Big) \\
&= \int_{\Omega} d\rho(x) \Big( f(x)\, \big( \ell\big( F(x) \big) +\s \big) - \big( \tilde{\ell}(x) +\s) \Big) \:,
\end{align*}
where in the last line we applied~\eqref{ldef} and~\eqref{tildeldef}.
\QED

We now prove that the volume term on the right side of~\eqref{osinlvol} can be arranged
to vanish with the help of inner solutions, order by order in perturbation theory.
The general reason why this method works is clarified in Appendix~\ref{appconserve}
with a non-perturbative argument.

\begin{Thm} \label{thmosinoconserve}
In a perturbative treatment, by adding to every order~$p$
a suitable spatially compact inner solution~$\v^{(p)} \in \Jin_\sc$,
one can arrange that~$\gamma^\Omega(\tilde{\rho}, \rho)$ vanishes
for every compact~$\Omega \subset M$.
\end{Thm}
\Proof Our strategy is to arrange that the integrand in~\eqref{osinlvol}
vanishes identically. Clearly, expanding the integrand to higher order in perturbation theory
gets very complicated. Here it suffices to note that, since the interaction takes place in a compact
set and we have finite propagation speed, we get a function~$g$ of spatially compact support.
Our goal is to show that this function can be compensated by the linear contribution
of the spatially compact inner solution~$\v^{(p)} \in \Jin_\sc$. This linear contribution is computed by
\begin{align*}
&f(x)\, \ell\big( F(x) \big) - \tilde{\ell}(x) 
+ \s \:\big(f(x) - 1 \big) \\
&= \int_M d\rho(y)\: 
\Big( f(x)\, \L\big(F(x),y \big) -  \L\big(x, F(y) \big)\, f(y) \Big) \\
&\asymp \int_M d\rho(y)\: \big( \nabla_{1, \v^{(p)}} - \nabla_{2, \v^{(p)}} \big) \L(x,y) \\
&= 2 \,\nabla_{\v^{(p)}} \big( \ell + \s \big)(x) - \big( \Delta \v^{(p)} - \nabla_{\v^{(p)}} \,\s \big)(x) 
= \big(\nabla_{\v^{(p)}} \,\s\big)(x) \:,
\end{align*}
where in the last step we used that~$\rho$ satisfies the weak EL equations
and that~$\v^{(p)}$ is a solution of the linearized field equations.
As shown in Proposition~\ref{prpinnera}, the scalar component of~$\v^{(p)}$
can be arranged to be~$-g/\s$. This concludes the proof.
\QED

In what follows, we shall always work with the modified perturbation expansion
where~$\gamma^\Omega(\tilde{\rho}, \rho)$ vanishes for every compact~$\Omega$.
For clarity, we write the inner solution separately. Thus we write
the resulting perturbation expansion in modification of~\eqref{Pertdef},
\eqref{Pertexpand} and \eqref{Pp} as
\beq \label{Pert2}
\begin{split}
\Pert_\rho + \mathfrak{N}_\rho &: U \subset \Glin_{\rho, \sc} \rightarrow \calB \:,
\qquad
\big( \Pert_\rho + \mathfrak{N}_\rho \big)(\lambda w) = \sum_{p=1}^\infty \lambda^p\:
\big( \Pert_\rho^{(p)} + \mathfrak{N}_\rho^{(p)} \big)
\end{split}
\eeq
with
\beq \label{PNpert}
\Pert_\rho^{(p)} \::\: \big( \Glin_{\rho, \sc} \big)^p \rightarrow \Gamma_{\rho, \sc}
\qquad \text{and} \qquad
\mathfrak{N}_\rho^{(p)} \::\: \big( \Glin_{\rho, \sc} \big)^p \rightarrow \Jin_{\rho, \sc} \:.
\eeq
We often write the $p^\text{th}$ order jets as
\beq \label{wnp}
w^{(p)} = \Pert_\rho^{(p)}(w, \ldots, w) \in \Gamma_{\rho, \sc} \qquad \text{and} \qquad
\mathfrak{n}^{(p)} = \mathfrak{N}_\rho^{(p)}(w, \ldots, w) \in \Jin_{\rho, \sc} \:.
\eeq
Since we work with spatially compact jets throughout,
the corresponding surface layer integral at time~$t$ introduced in analogy to Definition~\ref{defosi} by
\beq
\gamma^t(\tilde{\rho}, \rho) := \int_{\tilde{\Omega}_t} d\tilde{\rho}(x) \int_{M \setminus \Omega^t} d\rho(y)\: \L(x,y)
- \int_{\Omega^t} d\rho(x) \int_{\tilde{M} \setminus \tilde{\Omega}_t}  d\tilde{\rho}(y)\: \L(x,y) \label{Inonlin}
\eeq
is well-defined to every order in perturbation theory and is time independent.

\section{The Approximation of Small Inner Solutions} \label{secinnersmall}
In the above constructions, inner solutions were used
twice: in order to arrange that the linearized solutions have no scalar
component (see Corollary~\ref{cornoscalar} and the preceding construction)
and in order to arrange the conservation of the nonlinear surface layer integral
(see Theorem~\ref{thmosinoconserve}).
The inner solutions have the interpretation as describing an infinitesimal diffeomorphism
of spacetime needed in order to compensate the volume change induced by the interaction.
In applications to a physical scattering process in Minkowski space, this volume change and consequently also the
inner solutions are extremely small (for details see Appendix~\ref{secapproxinner}).
With this in mind, in what follows we make the following approximation:
\begin{Def} \label{defsmallinner}
The {\bf{approximation of small inner solutions}} is the simplification
where the inner solutions are taken into account only linearly. 
Thus all contributions to the interaction and to surface layer integrals which
involve one inner solution~$\v \in \Jin_{\rho, \sc}$ and another jet in~$\Jin_{\rho, \sc} \cup \Gamma_{\rho, \sc}$
are neglected.
\end{Def}

\section{Description of a Scattering Process} \label{secscatter}
We now explain how to describe a physical scattering process.
We have the situation in mind that the interaction takes place in a finite time interval,
whereas before and after this time interval, the dynamics is linear.
Moreover, the interaction should take place in a bounded region space.
In order to model this situation, we first explain how 
linear systems are described mathematically (Section~\ref{seclinear}).
A scattering process will then be modelled by a measure which at
large positive and large negative times behaves like a linear system
(Section~\ref{secscatters}). We finally analyze the question how the jet spaces
can be endowed with almost-complex and complex
structures (Section~\ref{seccomplexlinear}--\ref{secwhencomplex}).

\subsection{Linear Systems in Minkowski Space} \label{seclinear}
Let~$\rho \in \calB$ be a critical measure.
Describing the system as a {\em{linear system}} is the
approximation where all second and higher orders in the perturbation expansion are neglected, i.e.\
in suitable charts,
\beq \label{onlyfirst}
\Pert_\rho = \Pert_\rho^{(1)} \::\: U \subset \Glin_{\rho, \sc}  \rightarrow \calB \:.
\eeq
In other words, $\calB$ is identified locally with~$\Jlin_{\rho, \sc}$.
This gives~$\calB$ in a neighborhood of~$\rho$ the structure of a vector space.
This vector space structure also gives rise to a canonical connection~$\nabla$ on~$\calB$.

Combining this assumptions with the previous constructions
greatly simplify the conservation laws
for the surface layer integrals in Section~\ref{secosi}. 
Indeed, since the jets have no scalar components (see Corollary~\ref{cornoscalar}),
both~\eqref{cI1} and the first term in the integrand in~\eqref{cI2symm} vanish.
Moreover, the fact that the higher orders vanish~\eqref{onlyfirst} implies
that also the second term in the integrand in~\eqref{cI2symm} vanishes.
Therefore, the surface layer integrals~$\gamma^t_\rho$, $\sigma^t_\rho$ and~$(.,.)^t_\rho$
in Definition~\ref{defosi} are all {\em{conserved}} in the sense that they are time independent.

In order to make the setting more concrete, we now assume that
the measure~$\rho$ describes a linear perturbation of {\em{Minkowski space}}.
To this end, we let~$\rho_{\vac}$ be a critical measure formed of regularized
Dirac seas in Minkowski space (for example as constructed in~\cite[Section~1.2]{cfs}).
We always identify points of~$M_\vac := \supp \rho_\vac$ with corresponding points
of Minkowski space~$\scrM$.
Since the system is linear, we can write the interacting measure~$\rho$ in suitable charts as
\[ \rho = F_* (f \rho_\vac) \qquad \text{with} \qquad F(x) = x + w(x) + n(x) \:, \]
where~$w \in \Glin_{\rho_\vac, \sc}$ is a linearized solution
and~$\mathfrak{n} = (f=\div n, n) \in \Jin_{\rho_\vac, \sc}$
is an inner solution. Linearizing the short notation~\eqref{shortnotation},
we simply write
\beq \label{linearMink}
\rho = w + \n \:,
\eeq
where now and in what follows we always perturb the {\em{vacuum}} measure~$\rho_{\vac}$.
Using the limiting case of small inner solutions (see Definition~\ref{defsmallinner}),
the surface layer integrals of Definitions~\ref{defosi} and~\ref{defosinl}
can be written for~$u, v \in \Glin_{\rho, 0}$ as follows,
\begin{align}
\gamma^t_{\rho}({u}) &= (w, {u}) + \sigma^t(w, {u}) \label{cons1} \\
\sigma^t_{\rho}({u}, {v}) &= \sigma^t({u}, {v}) \label{cons21} \\
({u}, {v})^t_{\rho} &= ({u}, {v})^t \:, \label{cons22} \\
\gamma^t(\rho, \rho_\vac) &= \gamma^t_{\rho_\vac}(\mathfrak{n}) + \frac{1}{2}\: (w, w)^t \:, \label{cons3}
\end{align}
where the bilinear forms~$(.,.)$ and~$\sigma(.,.)$ {\em{without}} index
always refer to the vacuum measure~$\rho_\text{vac}$.
In Proposition~\ref{prpintpart}, the term~$\gamma^t_{\rho_\vac}(\mathfrak{n})$ was computed
to be the flux~\eqref{innerflux} of the vector field~$n$ through the hypersurface~$N^t$.
We refer to this contribution as the
\beq \label{innerfluxdef}
\text{\em{inner flux}} \qquad \gamma^t_{\rho_\vac}(\mathfrak{n}) = \s \,\mu(\v,N^t) \:.
\eeq
All the above surface layer integrals are conserved.

We remark that, as shown in~\cite{action}, the symplectic form and the surface layer inner product
are non-trivial; they diverge in the limit~$\delta \searrow 0$ of the order~$\sim \delta^{-4}$
(see~\cite[eqns~(1.3)--(1.6)]{action}; here~$\delta$ denotes a length scale
of the ultraviolet regularization).
Moreover, the calculations in~\cite[Section~5]{noether} show that
the one-form~$\gamma^t_{\rho_\vac}$ vanishes to the order~$\sim \delta^{-4}$.

\subsection{Scattering Systems in Minkowski Space} \label{secscatters}
\begin{figure}
%
\psscalebox{1.0 1.0} 
{
\begin{pspicture}(-0.7,-2.2)(18.07,2.2)
\definecolor{colour0}{rgb}{0.8,0.8,0.8}
\definecolor{colour1}{rgb}{0.6,0.6,0.6}
\rput[bl](10.836667,-1.4633334){\normalsize{$x$}}
\rput[bl](7.5916667,-1.9216666){\normalsize{$l = - \zeta(s)$}}
\psframe[linecolor=colour0, linewidth=0.02, fillstyle=solid,fillcolor=colour0, dimen=outer](12.2,2.2)(7.49,-2.2)
\rput[bl](11.72,1.76){\normalsize{$\tilde{M}$}}
\psline[linecolor=black, linewidth=0.02, linestyle=dashed, dash=0.17638889cm 0.10583334cm](7.51,-1.01)(12.31,-1.01)
\psline[linecolor=black, linewidth=0.02, linestyle=dashed, dash=0.17638889cm 0.10583334cm](7.51,0.99)(12.31,0.99)
\rput[bl](12.39,0.88){\normalsize{$t_\text{\rm{out}}$}}
\rput[bl](12.4,-1.12){\normalsize{$t_\text{\rm{in}}$}}
\psellipse[linecolor=colour1, linewidth=0.02, linestyle=dashed, dash=0.17638889cm 0.10583334cm, fillstyle=solid,fillcolor=colour1, dimen=outer](9.945,0.05)(1.645,0.48)
\rput[bl](8.69,-0.08){\normalsize{scattering region}}
\psframe[linecolor=colour0, linewidth=0.02, fillstyle=solid,fillcolor=colour0, dimen=outer](4.71,-1.0)(0.0,-2.19)
\psframe[linecolor=colour0, linewidth=0.02, fillstyle=solid,fillcolor=colour0, dimen=outer](4.72,2.2)(0.01,1.01)
\rput[bl](3.82,1.77){\normalsize{$M_\text{out}$}}
\rput[bl](3.93,-1.44){\normalsize{$M_\text{in}$}}
\psbezier[linecolor=black, linewidth=0.04, arrowsize=0.05291667cm 4.0,arrowlength=1.4,arrowinset=0.0]{->}(5.03,1.51)(5.59,1.98)(6.83,1.75)(7.32,1.48)
\psbezier[linecolor=black, linewidth=0.04, arrowsize=0.05291667cm 4.0,arrowlength=1.4,arrowinset=0.0]{->}(5.04,-1.6)(5.6,-1.13)(6.84,-1.36)(7.33,-1.63)
\rput[bl](5.84,1.25){\normalsize{$\iota_\text{out}$}}
\rput[bl](5.83,-1.84){\normalsize{$\iota_\text{in}$}}
\rput[bl](0.98,1.58){\normalsize{$w_\text{out} + \n_\text{out}$}}
\rput[bl](1.04,-1.71){\normalsize{$w_\text{in}+\n_\text{in}$}}
\rput[bl](8.43,1.54){\normalsize{$\Pert(w) + \mathfrak{N}(w)$}}
\rput[bl](9.31,-1.71){\normalsize{$w$}}
\end{pspicture}
}
\caption{A scattering system with retarded perturbation expansion.}
\label{figscatter}
\end{figure}
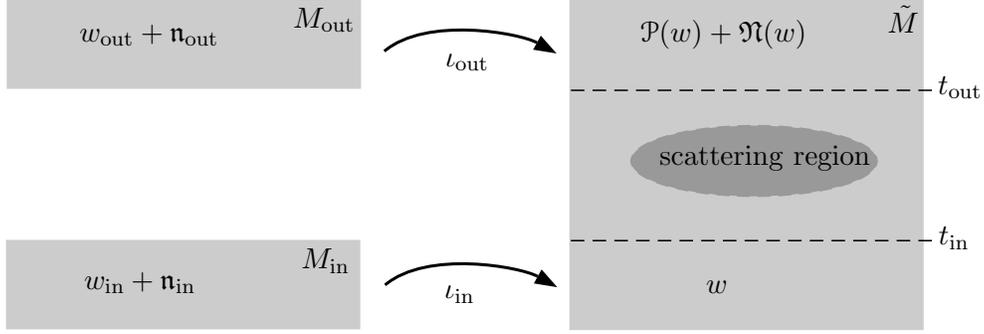

A {\em{scattering system}} is defined as an interacting system~$\tilde{\rho}$ which
asymptotically for large negative and for large positive times
goes over to linear system~$\rho_\text{in}$ and~$\rho_\text{out}$, respectively
(see Figure~\ref{figscatter}).
Thus we let~$\tilde{\rho}$ be again a critical measure which has a smooth
manifold structure and admits a global time function (see Definition~\ref{defsms}).
For simplicity, we assume that this smooth manifold is~$\R^4$.
Next, we assume that there are two linear
systems~$\rho_\text{in}$ and~$\rho_\text{out}$ as well as injections
\beq \label{iotainout}
\iota_\text{in} : M_\text{in} \hookrightarrow \tilde{M} \qquad \text{and} \qquad
\iota_\text{out} : M_\text{out} \hookrightarrow \tilde{M} \:.
\eeq
and that the images~$\iota_\text{in}(M_\text{in})$
and~$\iota_\text{out}(M_\text{out})$ contain the asymptotic past and future
of~$\tilde{M}$, respectively. Furthermore, we assume that the mappings~$\iota_\text{in}$
and~$\iota_\text{out}$ are close to the identity.
For brevity, we do not quantify what ``close to the identity'' means.
For our purposes, it suffices to have the intuitive picture that~$\rho_\text{in}$ and~$\rho_\text{out}$
should be ``good approximations'' to~$\tilde{\rho}$ in the respective asymptotic ends.

The linear systems describing the asymptotic ends can be
described again in the form~\eqref{linearMink}, where for clarity we
add subscripts ``in'' and ``out,'' i.e.\
\beq \label{restrict}
\begin{split}
\rho_\text{in} &= \big( w_\text{in}(x) + \n_\text{in}(x) \big) \: \Theta\big( t_\text{in}-\scrt(x) \big) \\
\rho_\text{out} &= \big( w_\text{out}(x) + \n_\text{out}(x) \big) \: \Theta\big( \scrt(x)-t_\text{out} \big)
\end{split}
\eeq
(this means that both~$\iota_\text{in}$ and~$\iota_\text{out}$
in~\eqref{iotainout} simplify to the inclusion map).
Moreover, we assume that~$\tilde{\rho}$ can be obtained from~$\rho_{\vac}$
perturbatively, i.e.\
\beq \label{pertansatz}
\tilde{\rho} = \big(\Pert + \mathfrak{N} \big)(w)+ \mathfrak{n}
\eeq
for a linearized solution~$w \in \Glin_{\rho_\vac, 0}$ and an inner solution~${\mathfrak{n}}$
in the Minkowski vacuum.
For ease in notation, $\Pert$ without an index always refers
to a perturbation of the vacuum measure~$\rho_\vac$.
We also note that the reason for simply adding the inner solution in~\eqref{pertansatz}
is that, in view of the limiting case of small inner solutions (Definition~\ref{defsmallinner}),
the inner solution enters the perturbation expansion only linearly.

The Heaviside functions in~\eqref{restrict} require a brief
explanation. At first sight, the multiplication might seem problematic because multiplying
a critical measure by a characteristic function does not yield a critical measure.
However, the EL equations will be violated only in a boundary layer around
the surfaces~$t=\tin$ and~$t=\tout$, respectively. In order not to distract from
the main construction, we here simply disregard such boundary effects.

In order to simplify the situation further, we shall restrict attention the situation
in which the perturbation expansion is performed purely with {\em{retarded}} Green's operators.
This means that the interaction changes the system only towards the future.
As a consequence, the linearized solution~$w$ coincides with the incoming jet~$w_\text{in}$.
Moreover, for simplicity we set the incoming inner solution~$\mathfrak{n}_\text{in}=0$ to zero
(this is no loss in generality because, in the approximation of small inner solutions,
the linear solutions are treated linearly, so that an incoming inner solution changes the
inner flux only by an additive constant).
Then the outgoing jets are obtained as the sum of all the jets of the perturbation
expansion. Thus, to summarize,
\begin{align}
\big( w_\text{in} + {\mathfrak{n}}_\text{in} \big) \big|_{\{t<\tin\}}  =
w_\text{in} \big|_{\{t<\tin\}} &= w = \Pert(w) \big|_{\{t<\tin\}} \label{wasyin} \\
\big( w_\text{out} + {\mathfrak{n}}_\text{out} \big) \big|_{\{t>\tout\}} &= 
\big(\Pert + \mathfrak{N} \big)(w) \big|_{\{t>\tout\}} \label{wasydef}
\end{align}
(where we again use the notation~\eqref{Pertexpand} and work in charts on~$\F$).
For clarity, we point out that working with a retarded time evolution merely is a technical simplification
which makes it possible to identify the incoming jets with the linear perturbations.
But one could work just as well with other choices of Green's operators without changing any of our results.

As pointed out above, the surface layer integrals of Definitions~\ref{defosi} and~\ref{defosinl}
are conserved both in the asymptotic future and past.
But are they also conserved in the interaction region? In other words, do they
coincide for the in- and outgoing jets? 
This is a subtle point, and for clarity we collect our previous results as separate corollaries.
\begin{Corollary} \label{corconsscatter} Let~$\tilde{\rho}$ be a critical measure describing a scattering system
with incoming and outgoing measures~$\rho_\text{\rm{in}}$ and~$\rho_\text{\rm{out}}$.
Moreover, let~$\tilde{u}, \tilde{v} \in \Glin_{\tilde{\rho}, \sc}$ be linearized solutions. Then
\begin{align}
\gamma^t_{\tilde{\rho}} \big(\tilde{u} \big) \big|_{t_\text{\rm{in}}}^{t_\text{\rm{out}}} &= 0 \label{cI1nl} \\
\sigma^t_{\tilde{\rho}} \big( \tilde{u}, \tilde{v} \big) \big|_{t_\text{\rm{in}}}^{t_\text{\rm{out}}}
&= 0 \label{cI2asnl} \\
\big(\tilde{u}, \tilde{v} \big)^t_{\tilde{\rho}} \big|_{t_\text{\rm{in}}}^{t_\text{\rm{out}}} &=
-2 \int_{M} \Delta_2[\tilde{u}, \tilde{v}]\: d\tilde{\rho}\:, \label{cI2symmnl} \:.
\end{align}
\end{Corollary}
\Proof Follows immediately from Proposition~\ref{prplinconserve}.
\QED

\begin{Corollary} \label{cornl} Let~$\tilde{\rho}$ be a critical measure describing a scattering system
with incoming and outgoing measures~$\rho_\text{\rm{in}}$ and~$\rho_\text{\rm{out}}$. Then
the nonlinear surface layer integral is conserved,
\beq \gamma^t(\tilde{\rho}, \rho_\vac) \big|_{t_\text{\rm{in}}}^{t_\text{\rm{out}}} = 0\:. \label{cosinl}
\eeq
Moreover, it can be computed at times~$t=t_\text{\rm{in}}$ and~$t=t_\text{\rm{out}}$ by
\beq \label{nl}
\gamma^t \big(\tilde{\rho}, \rho_\vac \big) = \s \,\mu\big( \mathfrak{N}(w) ,N^t \big) + \frac{1}{2}\: \big(\Pert(w), \Pert(w) \big)^t \:,
\eeq
where~$\mu$ is the measure~\eqref{mudef}.
\end{Corollary}
\Proof Follows immediately from Theorem~\ref{thmosinoconserve} using the asymptotic
form of the solutions~\eqref{wasyin} and~\eqref{wasydef} together with the formulas
in the asymptotic regions~\eqref{cons3} and~\eqref{innerfluxdef}.
\QED

To avoid confusion, we point out that the jets~$\tilde{u}$ and~$\tilde{v}$ 
in Corollary~\ref{corconsscatter} are linearized solutions
with respect to the {\em{interacting}} measure~$\tilde{\rho}$, but {\em{not}} the vacuum measure~$\rho_\vac$.
This means that, in order to compute these jets, one again needs to invoke a perturbation expansion.
More precisely, using~\eqref{DP}, $\tilde{u}$ can be written as
\beq \label{tildeu}
\tilde{u} = D\Pert|_w \, u \qquad \text{with~$u \in \Glin_{\rho, \sc}$}\:.
\eeq
According to~\eqref{Pp2}, in the perturbation expansion we always
add suitable inner solutions such that the jet~$\tilde{\u}$ has no scalar component.
As a consequence, the conserved one-form~$\gamma^t_\rho$ is indeed conserved.
The symplectic form~\eqref{cI2asnl} and the nonlinear surface layer integral~\eqref{cosinl}
are conserved as well. The surface layer inner product~\eqref{cI2symmnl}, however, in
general is {\em{not}} conserved in an interacting system.

\subsection{The Complex Structure of Linear Systems} \label{seccomplexlinear}
Our next goal is to endow the linearized solutions with a complex structure.
For clarity, we first give the construction for {\em{linear systems}}.
In this case, both the symplectic form and the surface layer inner product are conserved,
and using the formulas~\eqref{cons21} and~\eqref{cons22},
we obtain the conservation laws
\[ \sigma^t(u,v) \big|_{t_\text{\rm{in}}}^{t_\text{\rm{out}}} = 0 \quad \text{and} \quad
\big(u, v \big)^t \big|_{t_\text{\rm{in}}}^{t_\text{\rm{out}}} = 0 
\qquad \text{for all~$u, v \in \Glin_{\rho, \sc}$} \:, \]
where the bilinear forms are given by~\eqref{sigma} and~\eqref{sprod}
with~$\rho=\rho_\vac$ (and we again omit the subscript~$\rho_\vac$).

For clarity, we first give the basic construction and discuss the involved
assumptions afterward (after~\eqref{Idef} below).
We assume that~$(.,.)$ is positive semi-definite.
Then dividing out the null space and forming the completion, we obtain
a real Hilbert space denoted by~$\h^\R$.
Next, we assume that~$\sigma$ is a bounded bilinear functional on this Hilbert space.
Then we can represent it relative to the scalar product by
\beq \label{Sdef}
\sigma(u, v) = (u,\, \mathscr{T}\, v) \:,
\eeq
where~$\mathscr{T}$ is a uniquely determined bounded operator on~$\h^\R$.
Since the symplectic form is anti-symmetric and the scalar product is symmetric, it is obvious that
\[ \mathscr{T}^* = -\mathscr{T} \]
(where the adjoint is taken with respect to the scalar product~$(.,.)$).
Finally, we assume that~$\mathscr{T}$ is invertible.
Then setting
\beq \label{Idef}
J := -(-\mathscr{T}^2)^{-\frac{1}{2}}\: \mathscr{T}
\eeq
defines a complex structure on the real Hilbert space~$\h^\R$.

The above assumptions are justified by the fact that they are satisfied
for the surface layer integrals in Minkowski space.
Indeed, as shown in~\cite{action}, the bilinear form~$(.,.)$ is positive
semi-definite. As explained at the end of Section~\ref{secosi}, we choose
the jet space~$\Jtest$ such that~$\sigma$ is non-degenerate.
By choosing~$\Jtest$ even smaller if necessary (for example by restricting
attention to the bosonic and fermionic jets as analyzed in~\cite{action}),
we can arrange that the symplectic form is bounded relative to the scalar product~\eqref{Sdef}
and that the resulting bounded operator~$\mathscr{T}$ is invertible.

We next complexify the vector space~$\Glin_\sc$ and denote
its complexification by~$\Gamma^\C$. We also extend~$J$ to a complex-linear operator
on~$\Gamma^\C$. The fact that~$J^*=-J$ and~$J^2=-\1$ implies that~$J$ has
the eigenvalues~$\pm i$. Consequently, $\Gamma^\C$ splits into a direct sum of the
corresponding eigenspaces, which we refer to as the {\em{holomorphic}} and 
{\em{anti-holomorphic subspaces}}, i.e.\
\[ 
\Gamma^\C = \Gamma^\hol \oplus \Gamma^\ah \qquad \text{with} \qquad
\Gamma^\hol := \chi^\hol \:\Gamma^\C \:,\;\;\;
\Gamma^\ah := \chi^\ah \:\Gamma^\C \:, \]
where we set
\beq \label{chiholdef}
\chi^\hol = \frac{1}{2}\: (\1 - i J) \qquad \text{and} \qquad
\chi^\ah = \frac{1}{2}\: (\1 + i J) \:.
\eeq
We also complexify the inner product~$(.,.)$ and the symplectic form to {\em{sesquilinear}}
forms on~$\Gamma^\C$ (i.e.\ anti-linear in the first and linear in the second argument).
Moreover, we introduce a positive semi-definite inner product~$(.|.)$ by
\[ 
(.|.) = (\,.\,,(-\mathscr{T}^2)^\frac{1}{2}\, .\,) = \sigma( \,.\,, J \,.\, ) \::\: \Gamma^\C \times \Gamma^\C \rightarrow \C \:. \]
This positive semi-definite inner product product gives rise to a Hilbert space structure.
In order to work out the similarities and differences to quantum theory, it is best to
form the Hilbert space as the completion of the holomorphic subspace, i.e.\
\beq \label{Hprod}
\h := \overline{\Gamma^\hol}^{(.|.)} \:.
\eeq
We denote the induced scalar product on~$\h$ by~$\la.|. \ra$. Then~$(\h, \la.|.\ra)$
is a complex Hilbert space. It has the useful property that
\beq \label{imsigma}
\im \la u| v \ra = \im \sigma(u, J v) = \re \sigma(u, v) \:.
\eeq

\subsection{Complex Connections and the Holomorphic Perturbation Expansion} \label{seccompconn}
We now return to the scattering system described by the measure~$\tilde{\rho}$
as introduced in Section~\ref{secscatters}.
Since this system goes over to linear systems asymptotically,
we can use the construction of the previous section to obtain complex structures
for the incoming and outgoing linearized solutions.
The main complication is that, in view of~\eqref{cI2symmnl},
the surface layer inner product in general is not conserved.
As a consequence, also the operator~$\mathscr{T}$ as defined by~\eqref{Sdef}
in the two asymptotic regions will in general be different.
The same will be true for the resulting complex structures.

In order to get a better understanding of this fact, it is helpful
to consider the smooth setting with differential geometric notions.
Then, as mentioned at the end of Section~\ref{secosi}, the {\em{symplectic form}}
can be regarded as the exterior derivative of the conserved one-form, i.e.\
\beq \label{dgshort}
\sigma^t_{\tilde{\rho}} = d \gamma^t_{\tilde{\rho}} \:.
\eeq
In this way, the conservation law for~$\sigma^t_{\tilde{\rho}}$ follows immediately
from that for~$\gamma^t_{\tilde{\rho}}$. For clarity, we write~\eqref{dgshort}
in more detail as
\beq \label{dgamma}
\sigma^t_{\tilde{\rho}} \big(\tilde{u},\tilde{v} \big) = \big( D_{\tilde{u}} \gamma^t_{\tilde{\rho}}\big) \big( \tilde{v} \big) - \big( D_{\tilde{v}} \gamma^t_{\tilde{\rho}} \big)\big( \tilde{u} \big) \:,
\eeq
where~$D$ denotes the partial derivatives performed in any chart of the nonlinear
solution space~$\calB$. Considering the perturbation map as a chart and using~\eqref{tildeu},
we can write~\eqref{dgamma} in even more detail as follows,
\begin{align*}
\gamma_{\Pert(w)}\big( D\Pert|_w \,.\, \big) \:&:\: \Glin_{\sc} \rightarrow \R \\
D \Big(\gamma_{\Pert(w)}\big( D\Pert|_w \,.\, \big)\Big) \:&:\: \Glin_{\sc} \times \Glin_\sc \rightarrow \R \\
\sigma_{\Pert(w)} \big( D\Pert|_w\, u, D\Pert|_w\, v \big) &=
D \Big(\gamma_{\Pert(w)}\big( D\Pert|_w \,.\, \big)\Big) (u,v)
- D \Big(\gamma_{\Pert(w)}\big( D\Pert|_w \,.\, \big)\Big) (v,u) \:.
\end{align*}
However, since this notation is rather cumbersome, in what follows we prefer the shorter notation~\eqref{dgamma}.
The {\em{surface layer inner product}}, on the other hand,
can be regarded as a symmetrized derivative~$\gamma^t_{\tilde{\rho}}$.
In order to give such symmetric derivatives a differential geometric meaning, one must
work with covariant derivatives. Therefore, we write the surface layer inner product at time~$t$
in analogy to~\eqref{dgamma} as
\begin{align}
(\tilde{u}, \tilde{v})^t_{\tilde{\rho}} &= \big( \nabla^t_{\tilde{u}} \gamma_{\tilde{\rho}}\big)\big( \tilde{v} \big) + \big(\nabla^t_{\tilde{v}} \gamma_{\tilde{\rho}}\big)\big( \tilde{u} \big) \label{invariant} \\
&= \big(D_{\tilde{u}} \gamma_{\tilde{\rho}}\big)\big( \tilde{v} \big) + \big(D_{\tilde{v}} \gamma_{\tilde{\rho}}\big)\big( \tilde{u} \big) - \gamma_{\tilde{\rho}} \big( \Gamma^t (\tilde{u}, \tilde{v} ) \big) \:, \label{chart}
\end{align}
where in the last line we wrote the covariant derivative in the chart~$\Pert$
with ``Christoffel symbols''~$\Gamma$.
Since the resulting bilinear form should be symmetric, the connection must
be {\em{torsion-free}}, i.e.
\[ \Gamma^t(\tilde{u}, \tilde{v}) = \Gamma^t( \tilde{v}, \tilde{u}) \qquad \text{for all~$\tilde{u},
\tilde{v} \in \Glin_{\tilde{\rho}, \sc}$}\:. \]
As is the case in the classical differential geometric setting, the equation~\eqref{invariant} is invariant
and thus does not depend on the choice of charts or Green's operators.
The representation~\eqref{chart}, however, does depend on the chart.
For example, writing it for the perturbation map with advanced
Green's operators would give rise to different Christoffel symbols.
We also point out that the index~$t$ indicates that the connection is time-dependent.
In this way, the fact that the surface layer inner product is not conserved corresponds to
the fact that we have no distinguished connection on~$\calB$.

Clearly, this procedure raises the question whether there is a canonical way to choose the connection.
Before analyzing this question in detail in Sections~\ref{secalmostcomplex}
and~\ref{secwhencomplex}, we now give a few further constructions.
We thus assume that a connection~$\nabla$ on~$\calB$ is given
(for example the connection~$\nabla^t$ above). We denote the corresponding
surface layer inner product by
\[ (\tilde{u}, \tilde{v})_{\tilde{\rho}} = \big(\nabla_{\tilde{u}} \gamma_{\tilde{\rho}}\big)\big( \tilde{v} \big)
+ \big(\nabla_{\tilde{v}} \gamma_{\tilde{\rho}}\big)\big( \tilde{u} \big) \:. \]
Modifying the construction for linear systems~\eqref{Sdef} and~\eqref{Idef},
\beq \label{sigtilde}
\sigma(\tilde{u}, \tilde{v}) = (\tilde{u},\, \tilde{\mathscr{T}}\, \tilde{v})_{\tilde{\rho}} 
\qquad \text{and} \qquad \tilde{J} = \big(-\tilde{\mathscr{T}}^2 \big)^{-\frac{1}{2}}\: \tilde{\mathscr{T}} \:,
\eeq
we obtain an {\em{almost-complex structure}} on~$\Jlin_{\tilde{\rho}}$.
We again complexify the vector space $\Jlin_{\tilde{\rho}, \sc}$ and denote
its complexification by~$\Gamma_{\tilde{\rho}}^\C$.
It splits into a direct sum of the holomorphic and anti-holomorphic subspaces, i.e.\
\[ \Gamma_{\tilde{\rho}}^\C = \Gamma_{\tilde{\rho}}^\hol \oplus \Gamma_{\tilde{\rho}}^\ah \qquad \text{with} \qquad
\Gamma_{\tilde{\rho}}^\hol := \tilde{\chi}^\hol \:\Gamma_{\tilde{\rho}}^\C \:,\;\;\;
\Gamma_{\tilde{\rho}}^\ah := \tilde{\chi}^\ah \:\Gamma_{\tilde{\rho}}^\C \:, \]
where we set
\[ 
\tilde{\chi}^\hol = \frac{1}{2}\: (\1 - i \tilde{J}) \qquad \text{and} \qquad
\tilde{\chi}^\ah = \frac{1}{2}\: (\1 + i \tilde{J}) \:. \]
We also complexify the scalar product~$(.,.)_{\tilde{\rho}}$ to a sesquilinear form denoted by
\[ 
(.|.)_{\tilde{\rho}} \::\: \Gamma_{\tilde{\rho}}^\C \times \Gamma_{\tilde{\rho}}^\C \rightarrow \C \:. \]
Here we need to assume that~$(.|.)_{\tilde{\rho}}$ is positive semi-definite, and that
the resulting operator~$\tilde{\mathscr{T}}$ is bounded and invertible.
This poses implicit conditions on the admissible choices of the connection~$\nabla$.

We point out that the operator~$\tilde{\mathscr{T}}$ is defined independently of the choice of
surface layers. It can be computed in both asymptotic regions. For clarity, we denote
these operators by~$\mathscr{T}_\text{in}$ and~$\mathscr{T}_\text{out}$, respectively.
The fact that these operators are defined invariantly means that they are compatible
with the linearized time evolution, i.e.
\beq \label{Sinout}
\mathscr{T}_\text{out} = U \, \mathscr{T}_\text{in} \, U^{-1} \qquad \text{where} \qquad
U := D \Pert|_w \::\: u \rightarrow \tilde{u} \:.
\eeq
However, one must keep in mind that the scalar product~$(.,.)_{\tilde{\rho}}$,
and therefore also the operator~$\tilde{\mathscr{T}}$, have different forms in the asymptotic regions.
Indeed, using the formulas for the surface layer inner products
for linear systems in~\eqref{cons1}--\eqref{cons22}
in~\eqref{chart} one sees that
\begin{align}
(\tilde{u}, \tilde{v})^{t_\text{in}}_{\tilde{\rho}} &= (u, v ) - \big(w, \Gamma(u,v) \big) \label{spin} \\
(\tilde{u}, \tilde{v})^{t_\text{out}}_{\tilde{\rho}} &=
\big(\tilde{u}, \tilde{v} \big) + \big(\Pert(w), D^2\Pert|_w \,(u,v) \big)
- \big(\Pert(w), D\Pert|_w \,\Gamma(u,v) \big) \:. \label{spout}
\end{align}
According to~\eqref{sigtilde}, this also modifies the form of~$\tilde{\mathscr{T}}$
(note that, according to~\eqref{cI2asnl}, the symplectic form~$\sigma$ has the same form
in both asymptotic regions).
In particular, the scalar product in the outgoing region, and consequently also~$\mathscr{T}_\text{out}$,
are not computable from the knowledge of the outgoing linearized solutions~$\tilde{u}$ and~$\tilde{v}$ alone.
Instead, one must know the history of the scattering process.
A more geometric way of understanding this fact is that the transformation law of the Christoffel symbols
depends on the scattering process. This becomes clearer if one writes~\eqref{spout} as
\[ (\tilde{u}, \tilde{v})^{t_\text{out}}_{\tilde{\rho}} =
\big(\tilde{u}, \tilde{v} \big) - \big(\Pert(w), \tilde{\Gamma}(\tilde{u},\tilde{v}) \big) \]
with the transformed Christoffel symbols
\[ \tilde{\Gamma}(\tilde{u},\tilde{v}) := D\Pert|_w \,\Gamma(u,v) - D^2\Pert|_w \,(u,v) \:, \]
showing that the interaction as described by~$D^2\Pert|_w$ enters the transformation of the
Christoffel symbols.

Next, it is instructive to write~\eqref{Sinout} as
\[ D \Pert|_w\, \mathscr{T}_\text{in} = \mathscr{T}_\text{out} \, D \Pert|_w \:. \]
Applying the functional calculus, we obtain a similar relation for the operators~$\Gamma_\text{in}$
and~$\Gamma_\text{out}$. We thus obtain
\beq \label{DPert}
D \Pert|_w\, \chi^\hol_\text{in} = \chi^\hol_\text{out} \,D \Pert|_w \:.
\eeq
This means that the linearized time evolution preserves the complex structure.
The equation can be interpreted in analogy to the Cauchy-Riemann equation
as stating that the derivative of~$\Pert$ maps the holomorphic subspaces to each other.
Unfortunately, the last equation is of no use for the perturbative treatment,
because expanding~\eqref{DPert} in powers of the coupling constant~$\lambda$, 
the operators~$\chi^\hol_\text{in}$ and~$\chi^\hol_\text{out}$
also need to be expanded, leading to a complicated mixing of the holomorphic and
anti-holomorphic components.
This complication can be avoided if the almost-complex structure
can be integrated to give rise to a complex structure. This motivates the following
definition.

\begin{Def} \label{defcc}
$\nabla$ is a {\bf{holomorphic connection}} if the almost-complex structure~$\tilde{J}$
defined in~\eqref{sigtilde} is a complex structure.
\end{Def}

We finally explain the implication of a holomorphic connection.
Thus suppose that~$\calB$ admits a holomorphic connection (the problem of existence
will be considered in Section~\ref{secwhencomplex} below).
Then, as in complex geometry, one can choose holomorphic and anti-holomorphic
coordinates. Working in such a complex chart, the operator~$J$ reduces to complex conjugation.
This means in our language that there is a chart~$\Pert : \Glin_\sc
\rightarrow \calB$ (no longer retarded, but involving a specific combination
of different Green's operators) such that the operator~$J$
is constant, i.e.
\[ J = J_\text{in} = J_\text{out} \:. \]
As a consequence, in~\eqref{DPert} one can omit the indices ``in'' and ``out,''
\beq \label{linhol}
D \Pert|_w\, \chi^\hol = \chi^\hol \,D \Pert|_w \:.
\eeq
In contrast to~\eqref{DPert}, this equation can be evaluated order by order
in perturbation theory to obtain the following result:

\begin{Thm} {\bf{(holomorphic perturbation expansion)}} \label{thmholpert}
Suppose that~$\Pert$ is a perturbation expansion compatible with
a complex structure induced by a holomorphic connection on~$\calB$. Then $\Pert$
preserves the complex structure to every order in perturbation theory, i.e.\
for all~$p \in \N$ and all~$w \in \Glin_\sc$,
\[ \chi^\hol \,\Pert^{(p)}(w,\ldots,w) = 
\chi^\hol \,\Pert^{(p)} \big( \chi^\hol \,w,\ldots,\chi^\hol\, w \big) \:. \]
\end{Thm} 
\Proof Multiplying~\eqref{linhol} by~$\chi^\hol$ and using that~$\chi^\hol$ is idempotent, we obtain
\[ \chi^\hol \,D \Pert|_w\, u = \chi^\hol \,D \Pert|_w\, (\chi^\hol \,u) \:. \]
Substituting the perturbation series~\eqref{dPertexpand}, the contribution~$\sim \lambda^{p-1}$ gives
\beq \label{toexpand}
\chi^\hol\, \Pert^{(p)} \big(\! \underbrace{w,\ldots, w}_{\text{$p-1$ factors}} \!, u \big) = 
\chi^\hol\, \Pert^{(p)} \big(\! \underbrace{w,\ldots, w}_{\text{$p-1$ factors}}\!, \chi^\hol\, u \big) \:.
\eeq
We now set~$u=w$ and choose~$w$ as
\beq \label{wchoose}
w = \cos(\alpha)\: v + \sin(\alpha)\: J v
= e^{i \alpha}\: v^\hol + e^{-i \alpha}\: v^\ah \:,
\eeq
where~$v \in \Glin_\sc$ and~$v^\hol := \chi^\hol v$, $v^\ah := \chi^\ah v$.
Using that~$\Pert^{(p)}$ is multilinear and symmetric, expanding~\eqref{toexpand} gives
\begin{align*}
\sum_{q=0}^p & \begin{pmatrix} p \\ q \end{pmatrix} e^{i (p-2q) \alpha}\:
\chi^\hol\, \Pert^{(p)} \big( \underbrace{v^\ah ,\ldots, v^\ah}_{\text{$q$ factors}},
\underbrace{v^\hol ,\ldots, v^\hol}_{\text{$p-q$ factors}} \big) \\
&= \sum_{q=0}^{p-1} \begin{pmatrix} p-1 \\ q \end{pmatrix} e^{i (p-2q) \alpha}\:
\chi^\hol\, \Pert^{(p)} \big( \underbrace{v^\ah ,\ldots, v^\ah}_{\text{$q$ factors}},
\underbrace{v^\hol ,\ldots, v^\hol}_{\text{$p-q$ factors}} \big) \:.
\end{align*}
Since~$\alpha$ can be chosen arbitrarily, the contributions must
vanish to every order~$q$.
Since the combinatorial factors on the left and right are different unless~$q=0$, it follows that
\[ \chi^\hol\, \Pert^{(p)} \big( \underbrace{v^\ah ,\ldots, v^\ah}_{\text{$q$ factors}},
\underbrace{v^\hol ,\ldots, v^\hol}_{\text{$p-q$ factors}} \big) = 0
\qquad \text{for~$q=1,\ldots, p$}\:. \]
This gives the result.
\QED
Stated in words, this result means that the holomorphic component~$\chi^\hol \Pert$
of the perturbation map to every order in perturbation theory depends only on the
holomorphic jets. This explains the name ``holomorphic perturbation expansion.''
Clearly, this theorem holds similarly for the anti-holomorphic component.
The anti-holomorphic component can be obtained from the holomorphic component
by taking the complex conjugate or, equivalently, by the replacement~$J \rightarrow -J$.

\subsection{A Canonical Almost-Complex Structure with Interaction} \label{secalmostcomplex}
We now come to the question of how to choose the connection~$\nabla$ in~\eqref{invariant}.
Equivalently, we can ask how to choose the operator~$\tilde{\mathscr{T}}$, because given~$\tilde{\mathscr{T}}$
we can use~\eqref{sigtilde} to {\em{define}} the scalar product~$(.,.)_{\tilde{\rho}}$, i.e.
\beq \label{prodpos}
(\tilde{u}, \tilde{v})_{\tilde{\rho}} := \sigma(\tilde{u}, \,\tilde{\mathscr{T}}^{-1}\, \tilde{v}) \:,
\eeq
which in turn determines the connection~$\nabla$ via~\eqref{invariant}.

The operator~$\tilde{\mathscr{T}}$ should be determined by its properties, which we now collect.
Since the complex structure at time~$t$ should depend only on the state at time~$t$,
but should be independent of the history of the physical system, the operator~$\tilde{\mathscr{T}}$ must
have the same form in the two asymptotic regions, i.e.\ $\mathscr{T}_\text{in}=\mathscr{T}_\text{out}$
(where we consider both~$\rho_\text{in}$ and~$\rho_\text{out}$ as linear perturbations
of the same vacuum measure~$\rho_\vac$). Using~\eqref{Sinout}, we can write this
condition as
\beq \label{commute}
\tilde{\mathscr{T}} \,U = U\, \tilde{\mathscr{T}} \:.
\eeq
Thus we seek for an operator~$\tilde{\mathscr{T}}$ which commutes with the linearized time
evolution~$U$. Moreover, the operator~$\tilde{\mathscr{T}}$ must be chosen such that it is invertible
and such that the inner product defined by~\eqref{prodpos} is positive semi-definite.
The question is whether an operator~$\mathscr{T}$ with the above properties exists and, if yes, if
it is unique. 

For clarity and technical simplicity, we begin with the case that~$\Glin_{\tilde{\rho}, \sc}$ is {\em{finite-dimensional}}
and treat the infinite-dimensional situation afterward.
Then on the complexification~$\Gamma^\C_{\tilde{\rho}}$, the symplectic form gives rise to an indefinite inner product,
\beq \label{Kreindef}
\bra .|. \ket \::\: \Gamma^\C_{\tilde{\rho}} \times \Gamma^\C_{\tilde{\rho}} \rightarrow \C \:,\qquad
\bra u|v \ket = \im \sigma(\overline{u}, v)
\eeq
(the bar indicates that we extend~$\sigma$ to a {\em{sesqui}}linear form on~$\Gamma^\C_{\tilde{\rho}}$).
The fact that~$U$ is a symplectomorphism implies that~$U$ is a unitary
operator on the indefinite inner product space~$(\Gamma^\C_{\tilde{\rho}}, \bra .|. \ket)$.
The relation~\eqref{commute} implies that the operators~$\tilde{\mathscr{T}}$ and~$U$
must have the same invariant subspaces. Moreover, the positivity requirement on the
inner product~\eqref{prodpos}
yields that the invariant subspaces of~$U$ must be definite eigenspaces,
and that the corresponding eigenvalues of the operator~$-i \tilde{\mathscr{T}}$ must be positive
if the eigenspace is positive definite, whereas they must be negative if the eigenspace
is negative definite.
We conclude that in the formulation with indefinite inner product spaces,
the above questions can be rephrased as follows:
\begin{Prp} \label{prpindef} Assume that~$\Glin_{\tilde{\rho}}$ is finite-dimensional.
There is an operator~$\tilde{\mathscr{T}}$ satisfying~\eqref{commute}
with the property that the inner product~$(.,.)_{\tilde{\rho}}$
defined by~\eqref{prodpos} is positive definite if and only if
the operator~$U$ on~$(\Gamma^\C_{\tilde{\rho}}, \bra .|. \ket)$ is diagonalizable and has
a pseudo-orthonormal eigenvector basis, i.e.
\beq \label{Unotation}
U = \sum_{\ell=1}^L s_\ell \,\lambda_\ell \: |\phi_\ell \ket \bra \phi_\ell| 
\qquad \text{with} \qquad s_\ell := \bra \phi_\ell | \phi_\ell \ket \in \{\pm 1\} \:.
\eeq
The operator~$-i \tilde{\mathscr{T}}$ can be chosen as any invertible symmetric operator 
on~$(\Gamma^\C_{\tilde{\rho}}, \bra .|. \ket)$ which commutes with~$U$
and whose positive and negative eigenvalues correspond to positive and
negative definite eigenspaces, respectively.
\end{Prp}
Before going on, we remark that in the non-interacting situation, the spectral
decomposition~\eqref{Unotation} can be understood as follows.
In this situation, it was shown in~\cite{action} that the positive and negative
definite subspaces of~$(\Gamma^\C_{\tilde{\rho}}, \bra.,. \ket)$ reproduce the usual frequency splitting.
Moreover, in this setting the time evolution operator can be written as~$U = e^{-i (\tout-\tin) H}$
with a Hamiltonian~$H$, whose positive and negative spectral subspaces
are the subspaces of positive and negative frequencies, respectively.
Therefore, the Hamiltonian has definite eigenspaces.
Applying the functional calculus, we conclude that also the operator~$U$ is diagonalizable
and has definite invariant subspaces, giving~\eqref{Unotation}.

Applying the functional calculus~\eqref{sigtilde} to the operator~$-i \tilde{\mathscr{T}}$,
the positive eigenvalues become plus one, whereas the negative
eigenvalues becomes minus one. We thus obtain a unique operator~$J$:
\begin{Prp} \label{prpalmost}
Under the assumptions of Proposition~\ref{prpindef}, there is a unique almost-complex
structure given by
\beq \label{Jform}
J = i \sum_{\ell=1}^L |\phi_\ell \ket \bra \phi_\ell| \:.
\eeq
\end{Prp}

We finally explain how our findings can be generalized to the {\em{infinite-dimensional}} setting.
In this case, the indefinite inner product~\eqref{Kreindef} gives rise to the
structure of a Krein space~$(\K, \bra .|. \ket)$ (see for example~\cite{bognar};
as the scalar product generating the Krein space topology
one can simply take the surface layer scalar product in~\eqref{sprod}).
The linearized time evolution operator~$U$ is a unitary operator on this Krein space.
The conditions specified in Proposition~\ref{prpindef} are generalized by the condition
that the Krein space should have an orthogonal decomposition into two invariant subspaces of~$U$,
\beq \label{Kdecomp}
\K = \K_+ \oplus \K_- \:,
\eeq
where~$\K_+$ is a positive and~$\K_-$ a negative definite subspace of~$\K$.
The operator~$J$, \eqref{Jform}, generalizes to
\[ J = \big(i \1_{\K_+} \big) \oplus \big(-i \1_{\K_-} \big) \:. \]
Keeping in mind that a unitary operator on a Krein space does not need to have a spectral decomposition,
the decomposition into indefinite invariant subspaces~\eqref{Kdecomp} poses a strong constraint for the existence of
a canonical almost-complex structure.

\subsection{Conditions for a Canonical Complex Structure} \label{secwhencomplex}
We now explore if the canonical almost-complex structure introduced in the previous section
gives rise to a complex structure. We again begin in the {\em{finite-dimensional}} setting.
We assume that the conditions in Proposition~\ref{prpindef} are satisfied.
In order to further simplify the setting, we strengthen these conditions
by assuming that all eigenspaces of~$U$ are definite
(these assumptions will be discussed below). We choose contours~$\Gamma_+$
and~$\Gamma_-$ which enclose the eigenvalues corresponding to the positive definite
respectively negative definite eigenspaces in counter-clockwise orientation.
Then the operators
\[ \Pi_\pm := -\frac{1}{2 \pi i} \int_{\Gamma_\pm} (U - \lambda)^{-1}\: d\lambda \]
are projection operators in~$(\K, \bra.|.\ket)$ onto the invariant
definite subspaces~$\K_\pm$ of~$U$. The operator~$J$ in~\eqref{Jform} can be written as
\[ J = i \,\Pi_+ - i \,\Pi_- \:. \]

\begin{Prp} \label{prpcomplex} Assume that~$\Glin_{\tilde{\rho}, \sc}$ is finite-dimensional
and that all eigenspaces of~$U$ are definite.
Then the almost-complex structure of Proposition~\ref{prpalmost} gives rise to
a complex structure if and only if for all~$w \in \Glin_{\tilde{\rho}, \sc}$
the following implication holds:
\beq \label{commcond}
\lambda_\ell \neq \lambda_{\ell'} \quad \text{and} \quad s_\ell, s_{\ell'} > 0 
\qquad \Longrightarrow \qquad \Pi_- D^2\Pert|_w(\phi_\ell, \phi_{\ell'}) = 0 \:.
\eeq
Here we again used the notation~\eqref{Unotation}, and~$D^2 \Pert|_w$ is the quadratic
correction to the linearized dynamics from~$\tin$ to~$\tout$.
\end{Prp}
\Proof The subspaces~$\Gamma^\hol_{\tilde{\rho}} \subset \Gamma^\C_{\tilde{\rho}}$ define a distribution on~$\calB$.
Our goal is to verify whether this distribution is integrable.
This is the case if and only if for any holomorphic sections~$u^\hol$ and~$v^\hol$ also their
commutator is holomorphic.

We first simplify this condition by showing that for any~$u, v \in \Gamma^\hol_{\tilde{\rho}}$,
it suffices to check the condition~$[u^\hol, v^\hol](w) \in \Gamma^\hol_{\tilde{\rho}}$
for arbitrarily chosen sections~$u^\hol$ and~$v^\hol$ with~$u^\hol(w)=u$
and~$v^\hol(w)=v$.
Indeed, other  holomorphic sections~$\hat{u}^\hol$ and~$\hat{v}^\hol$
with~$\hat{u}^\hol(w)=u$ and~$\hat{v}^\hol(w)=v$ can be written as
\[ \hat{u}^\hol = u^\hol + f\, \Delta u^\hol \qquad \text{and} \qquad
\hat{v}^\hol = v^\hol + g\, \Delta v^\hol \]
with two holomorphic sections~$\Delta u^\hol$ and~$\Delta v^\hol$
and two scalar functions~$f$ and~$g$ which vanish at~$w$.
A direct computation shows that the commutators~$[\hat{u}^\hol,\hat{v}^\hol]$
and~$[u^\hol,v^\hol]$ differ at~$w$ by a vector in~$\Gamma^\hol_{\tilde{\rho}}$
(in fact, this vector is a linear combination of~$\Delta u^\hol$ and~$\Delta v^\hol$).
Therefore, the condition~$[u^\hol, v^\hol](w) \in \Gamma^\hol_{\tilde{\rho}}$
is satisfied if and only if~$[\hat{u}^\hol, \hat{v}^\hol](w) \in \Gamma^\hol_{\tilde{\rho}}$.

The latter commutator condition can be verified as follows. We again work
in the chart given by~$\Pert$. According to Proposition~\ref{prpindef}, the vectors
in~$\Gamma^\hol_{\tilde{\rho}}$ are spanned by positive definite eigenvectors of~$U$.
Therefore, by linearity we may assume that
the holomorphic tangent vectors~$u$ and~$v$ are positive definite eigenvectors of~$U$ corresponding to
eigenvalues~$\mu$ and~$\nu$ (which may coincide). In order to obtain
corresponding holomorphic sections,
we apply the projection operator~$\Pi_+$,
\[ u^\hol(\tilde{w}) := -\frac{1}{2 \pi i} \int_{\Gamma_+} (d\Pert|_{\tilde{w}} - \lambda)^{-1}\: u\: d\lambda \:, \]
valid for all~$\tilde{w}$ in a neighborhood of~$w$.
Now we can differentiate in the direction of~$v^\hol$,
\begin{align*}
v^\hol \,u^\hol(w) &= \frac{1}{2 \pi i} \int_{\Gamma_+} (U - \lambda)^{-1}\: D^2\Pert|_{w}
\big( v, (U - \lambda)^{-1}\:u \big)\: d\lambda \\
&= \frac{1}{2 \pi i} \int_{\Gamma_+} \frac{(U - \lambda)^{-1}}{\mu-\lambda} \: D^2\Pert|_{w}(v, u)\: d\lambda \:,
\end{align*}
where in the last step we used that~$U u = \lambda u$.
Antisymmetrizing in~$u$ and~$v$ gives the commutator,
\[ \big[v^\hol, u^\hol \big](w) 
= \frac{1}{2 \pi i} \int_{\Gamma_+} (U - \lambda)^{-1}\:\Big(\frac{1}{\mu-\lambda}
- \frac{1}{\nu-\lambda} \Big)\: D^2\Pert|_{w}(u, v) \;d\lambda\:. \]
This commutator lies in~$\Gamma^\hol$ if and only if
\[ 0 = \frac{1}{2 \pi i} \int_{\Gamma_+} \Pi_- \,(U - \lambda)^{-1}\:\Big(\frac{1}{\mu-\lambda}
- \frac{1}{\nu-\lambda} \Big)\: D^2\Pert|_{w}(u, v) \;d\lambda\:. \]
All the eigenvalues of the operator~$\Pi_- U$ lie outside the contour~$\Gamma_+$.
Therefore, we can compute the contour integral, taking into account only the poles at~$\lambda=\mu$
and~$\lambda=\nu$. A short computation gives the equation
\[ 0 = \Pi_- \,\Big( (U - \mu)^{-1} - (U - \nu)^{-1} \Big) \:D^2\Pert|_{w}(u, v)\:. \]
Using the resolvent identity, we obtain the equivalent condition
\[ \Pi_- \,\frac{\mu - \nu}{(U - \mu)- (U - \nu)} \:D^2\Pert|_{w}(u, v) = 0 \:. \]
This equation is obviously equivalent to the implication~\eqref{commcond}.
\QED

We close with a few remarks. First, the construction could be
generalized to the {\em{infinite-dimensional}} setting by assuming that~$\K$
again has an orthogonal decomposition~\eqref{Kdecomp} into definite invariant
eigenspaces of~$U$, and that that
the spectrum of~$U$ on these invariant subspaces is separated by a spectral gap
in the complex plane, i.e.\
\beq \label{infcond}
\K_\pm = \Pi_\pm \,\K \qquad \text{and} \qquad \text{dist}(\Gamma_+, \Gamma_-) > 0 \:.
\eeq
Under these assumptions, the spectral decomposition~\eqref{Unotation}
can be generalized using the spectral theorem for bounded operators in Hilbert spaces.
Moreover, the above contour integrals
are again well-defined, and the computation in the proof of Proposition~\ref{prpcomplex}
again applies.

We next explain whether the condition~\eqref{commcond} is satisfied in physically
interesting examples. In the Minkowski vacuum, this condition is indeed satisfied.
Namely, in this case the holomorphic jets are composed of positive frequencies.
Using that the product of two
functions of positive frequencies has again positive frequency and that the Green's operator
preserves four-momentum, one finds that
the operator~$D^2\Pert|_0(\phi_\ell, \phi_{\ell'})$ is again formed of
positive frequencies, so that its projection to the negative frequencies vanishes.
However, the condition~\eqref{commcond} does {\em{not}} seem to hold
as soon as~$w$ is non-zero. The reason is that~$w$ in general will involve positive
and negative frequencies, implying that~$D^2\Pert|_w(\phi_\ell, \phi_{\ell'})$
will be composed of mixtures of positive and negative frequencies.
As a consequence, the implication~\eqref{commcond} will be violated.
More generally, this consideration shows that the condition~\eqref{commcond}
is very strong and seems to be violated for most interacting systems of physical interest.

We finally discuss the condition in Proposition~\ref{prpcomplex} that all eigenspaces
of~$U$ must be definite (and similarly in the infinite-dimensional setting that~\eqref{infcond} holds).
If this condition is violated, then the perturbation expansion performed in the proof
of Proposition~\ref{prpcomplex} is more subtle because even arbitrarily small perturbations
can destroy the definiteness of the eigenspaces.
Besides this technical complication, the argument in the proof of Proposition~\ref{prpcomplex}
still goes through, showing that in most physical applications,
there will be {\em{no}} canonical complex structure.

\section{Linear Dynamics on Bosonic Fock Spaces} \label{secfock}
From the condition~\eqref{commcond} in Proposition~\ref{prpcomplex}
we concluded that in most physically interesting
examples, there will be no canonical holomorphic connection which would make
it possible to perform a holomorphic perturbation expansion (see Theorem~\ref{thmholpert}).
But the condition~\eqref{commcond} is satisfied in the Minkowski vacuum,
indicating that in the applications,
the holomorphic perturbation expansion should be valid up to
small error terms which ``mix'' the holomorphic and anti-holomorphic jets.
In this section, we shall make this intuitive picture mathematically precise.
It turns out that this analysis can be carried out most conveniently
in the bosonic Fock space formalism. This has two advantages: First,
the nonlinear dynamics can be reformulated with linear operators on the
Fock space. Second, the bra and ket states entering the complex scalar product on the Fock space
will correspond directly to the holomorphic and anti-holomorphic
components.

\subsection{Preliminaries on Bosonic Fock Spaces} \label{secfockprelim}
We let~$(\h, \la.|.\ra)$ be a separable complex Hilbert space (the {\em{one-particle space}}).
We let~$\h^n = \h \otimes \cdots \otimes \h$ be the $n$-fold tensor product, endowed with
the natural scalar product
\beq \label{sdef}
\la \psi_1 \otimes \cdots \otimes \psi_n \,|\, \phi_1 \otimes \cdots \otimes \phi_n \ra 
:= \la \psi_1 | \phi_1\ra \: \cdots\: \la \psi_n | \phi_n \ra\:.
\eeq
We denote total symmetrization by an index~$\symm$, i.e.\
\[ 
\big( \psi_1 \otimes \cdots \otimes \psi_n \big)_\symm := \frac{1}{n!} \sum_{\sigma \in S_n}
\psi_{\sigma(1)} \otimes \cdots \otimes \psi_{\sigma(n)} \:, \]
where~$S_n$ denotes the group of all permutations.
The totally symmetric tensors form a closed subspace denoted by~$\Fock^n := (\h^n)_\symm \subset \h^n$.
The {\em{bosonic Fock space}}~$(\Fock, \la .|.\ra_\Fock)$ is the direct sum of the $n$-particle spaces,
\[ \Fock = \bigoplus_{n=0}^\infty \Fock^n \:. \]

In order to describe the Fock states more explicitly, we choose an orthonormal basis~$(\phi_\ell)_{\ell=1,\ldots, N}$
with~$N \in \N \cup \{\infty\}$. For ease in notation, we set
\[ \phi_{\ell}^{p} := \underbrace{\phi_\ell \otimes \cdots \otimes \phi_\ell}_{\text{$p$ factors}} \:. \]
Given a finite number of pairs~$(\ell_i, p_i)$ with~$i=1,\ldots, m$
and~$\ell_1 < \ell_2 < \cdots < \ell_m$, we form the Fock space vectors
\beq \label{HF}
\Phi := \big( \phi_{\ell_1}^{p_1} \otimes \cdots \otimes \phi_{\ell_m}^{p_m} \big)_\symm \;\in\; \Fock^n \:,
\eeq
where
\[ n:= p_1 + \cdots + p_m \]
always denotes the number of particles.
According to~\eqref{sdef}, the resulting vectors are orthogonal unless
all the~$\ell_i$ and~$p_i$ coincide. Moreover, by construction of the tensor product,
the vectors of the form~\eqref{HF} are dense in~$\Fock$. In order to determine their normalization,
we compute
\begin{align*}
\la \Phi | \Phi \ra_\Fock &= \big\la \big( \phi_{\ell_1}^{p_1} \otimes \cdots \otimes \phi_{\ell_m}^{p_m} \big)_\symm \,\big|\,
\big( \phi_{\ell_1}^{p_1} \otimes \cdots \otimes \phi_{\ell_m}^{p_m} \big)_\symm \big\ra_\Fock \\
&= \big\la \phi_{\ell_1}^{p_1} \otimes \cdots \otimes \phi_{\ell_m}^{p_m} \,\big|\,
\big( \phi_{\ell_1}^{p_1} \otimes \cdots \otimes \phi_{\ell_m}^{p_m} \big)_\symm \big\ra_\Fock \\
&= \frac{1}{n!} \sum_{\sigma \in S_{p_1+\cdots + p_m}}
\big\la \phi_{\ell_1}^{p_1} \otimes \cdots \otimes \phi_{\ell_m}^{p_m} \,\big|\,
\phi_{j_{\sigma(1)}} \otimes \cdots \otimes \phi_{j_{\sigma(p_1+\cdots+ p_m)}} \big\ra_\Fock \:,
\end{align*}
where the indices~$j_1,\ldots, j_{p_1+\cdots+ p_m}$ count all the vectors in~$\Phi$ with multiplicities.
We get zero unless the vectors in the tensor product coincide pairwise, in which case
we get one. We thus obtain
\[ 
\la \Phi | \Phi \ra_\Fock = \frac{p_1! \cdots p_m!}{n!} \:. \]

We next introduce the creation and annihilation operators and derive their commutation
relations. For a vector~$\phi \in \h$ of the one-particle space, we introduce the
{\em{creation operator}} $a^\dagger(\phi)$ by
\beq \label{createdef}
a^\dagger(\phi) \::\: \h^n_\symm \rightarrow \h^{n+1}_\symm \:,\qquad
\Phi \mapsto c_n\:(\phi \otimes \Phi)_\symm
\eeq
with complex constants~$c_n$ which will be specified below.
Clearly, $a^\dagger(\phi)$ extends uniquely to a mapping from~$\Fock$ to~$\Fock$.
The {\em{annihilation operator}}~$a(\phi)$ is defined as the adjoint of the creation operator,
\[ 
a(\bar{\phi}) := \big(a^\dagger(\phi) \big)^* \]
(here the star denotes the adjoint with respect to the Fock space scalar product~$\la .|. \ra_\Fock$;
the bar~$\overline{\phi}$ indicates that the complex conjugate of~$\phi$ enters).
We now apply these operators to vectors of the form~\eqref{HF}. By definition~\eqref{createdef},
\[ a^\dagger \big (\phi_{\ell_1} \big) 
\big( \phi_{\ell_1}^{p_1} \otimes \cdots \otimes \phi_{\ell_m}^{p_m} \big)_\symm
= c_n\: \big( \phi_{\ell_1}^{p_1+1} \otimes \cdots \otimes \phi_{\ell_m}^{p_m} \big)_\symm \:. \]
Likewise, the annihilation operator reduces the power of~$\phi_{\ell_1}$, i.e.
\beq \label{cfact}
a(\bar{\phi}_{\ell_1} \big) 
\big( \phi_{\ell_1}^{p_1} \otimes \cdots \otimes \phi_{\ell_m}^{p_m} \big)_\symm
= d\: \big( \phi_{\ell_1}^{p_1-1} \otimes \cdots \otimes \phi_{\ell_m}^{p_m} \big)_\symm
\eeq
with a complex prefactor~$d$ (which may depend on~$p_1,\ldots, p_m$).
In order to determine this prefactor, we
compute the following scalar product,
\begin{align}
&\big\la \big( \phi_{\ell_1}^{p_1-1} \otimes \cdots \otimes \phi_{\ell_m}^{p_m} \big)_\symm
\,\big|\, a(\bar{\phi}_{\ell_1} \big) 
\big( \phi_{\ell_1}^{p_1} \otimes \cdots \otimes \phi_{\ell_m}^{p_m} \big)_\symm \big\ra_\Fock \notag \\
&= \big\la a^\dagger(\phi_{\ell_1} \big)  \big( \phi_{\ell_1}^{p_1-1} \otimes \cdots \otimes \phi_{\ell_m}^{p_m} \big)_\symm
\,\big|\, \big( \phi_{\ell_1}^{p_1} \otimes \cdots \otimes \phi_{\ell_m}^{p_m} \big)_\symm \big\ra_\Fock \notag \\
&= \overline{c_{n-1}} \;\big\la \big( \phi_{\ell_1}^{p_1} \otimes \cdots \otimes \phi_{\ell_m}^{p_m} \big)_\symm
\,\big|\, \big( \phi_{\ell_1}^{p_1} \otimes \cdots \otimes \phi_{\ell_m}^{p_m} \big)_\symm \big\ra_\Fock
= \overline{c_{n-1}}\; \frac{p_1! \cdots p_m!}{n!} \:. \label{c1}
\end{align}
On the other hand, computing the same scalar product using the
right side of~\eqref{cfact}, we obtain
\begin{align}
&\big\la \big( \phi_{\ell_1}^{p_1-1} \otimes \cdots \otimes \phi_{\ell_m}^{p_m} \big)_\symm
\,\big|\, a(\bar{\phi}_{\ell_1} \big) 
\big( \phi_{\ell_1}^{p_1} \otimes \cdots \otimes \phi_{\ell_m}^{p_m} \big)_\symm \big\ra_\Fock \notag \\
&= d\: \big\la \big( \phi_{\ell_1}^{p_1-1} \otimes \cdots \otimes \phi_{\ell_m}^{p_m} \big)_\symm
\,\big|\, \big( \phi_{\ell_1}^{p_1-1} \otimes \cdots \otimes \phi_{\ell_m}^{p_m} \big)_\symm \big\ra_\Fock
= d\; \frac{(p_1-1)! \cdots p_m!}{(n-1)!} \:. \label{c2}
\end{align}
The prefactor~$d$ can be read of by comparing~\eqref{c1} and~\eqref{c2}. 
Substituting the result into~\eqref{cfact}, we obtain
\beq \label{nocfact}
a(\bar{\phi}_{\ell_1} \big) 
\big( \phi_{\ell_1}^{p_1} \otimes \cdots \otimes \phi_{\ell_m}^{p_m} \big)_\symm
= \overline{c_{n-1}}\; \frac{p_1}{n}\: \big( \phi_{\ell_1}^{p_1-1} \otimes \cdots \otimes \phi_{\ell_m}^{p_m} \big)_\symm \:.
\eeq
Using~\eqref{createdef} and~\eqref{nocfact}, we can compute products of the
annihilation and creation operators, like for example
\begin{align}
a^\dagger(\phi_{\ell_1} \big) \:a(\bar{\phi}_{\ell_1} \big) 
\big( \phi_{\ell_1}^{p_1} \otimes \cdots \otimes \phi_{\ell_m}^{p_m} \big)_\symm
&= |c_{n-1}|^2\; \frac{p_1}{n}\: \big( \phi_{\ell_1}^{p_1} \otimes \cdots \otimes \phi_{\ell_m}^{p_m} \big)_\symm
\label{ada} \\
a(\bar{\phi}_{\ell_1} \big) \: a^\dagger(\phi_{\ell_1} \big)
\big( \phi_{\ell_1}^{p_1} \otimes \cdots \otimes \phi_{\ell_m}^{p_m} \big)_\symm
&= |c_n|^2\; \frac{p_1+1}{n+1}\: \big( \phi_{\ell_1}^{p_1} \otimes \cdots \otimes \phi_{\ell_m}^{p_m} \big)_\symm \:.
\label{aad}
\end{align}

The complex coefficients~$c_n$ introduced in~\eqref{createdef} can be chosen arbitrarily.
The following choice is most convenient and agrees with common conventions in physics:
First, since~\eqref{ada} and~\eqref{aad} only involve the absolute values of the~$c_n$,
there is no point in choosing these coefficients to be complex
(indeed, a phase in~$c_n$ merely corresponds to introducing irrelevant relative phases
between the subspaces of different particle numbers). Second, the
denominators~$n$ and~$n+1$ in~\eqref{ada} and~\eqref{aad} are unpractical
in longer computations. This leads us to choose
\beq \label{cnrel}
c_n = \sqrt{n+1} \:.
\eeq
Our findings are summarized as follows:

\begin{Lemma} Introducing the annihilation and creation operators by
\begin{align*}
a^\dagger(\phi) &\::\: \h^n_\symm \rightarrow \h^{n+1}_\symm \:,
&\hspace*{-1.5cm} \Phi &\mapsto \sqrt{n+1}\:(\phi \otimes \Phi)_\symm \\
a(\bar{\phi}) &\::\: \h^{n+1}_\symm \rightarrow \h^n_\symm \:,& \hspace*{-1.5cm}
a(\bar{\phi}) &= \big(a^\dagger(\phi) \big)^* \:,
\end{align*}
the following relations hold for any~$k =1,\ldots,m$:
\begin{align}
a^\dagger\big(\phi \big) \:\Phi &= \sqrt{n+1}\; (\phi \otimes \Phi)_\symm \label{crt} \\
a\big(\bar{\phi}_{\ell_k} \big) \:
\big( \phi_{\ell_1}^{p_1} \otimes \cdots \otimes \phi_{\ell_k}^{p_k} \otimes \cdots \otimes \phi_{\ell_m}^{p_m} \big)_\symm
&= \frac{p_k}{\sqrt{n}}\; \big( \phi_{\ell_1}^{p_1} \otimes \cdots 
\otimes \phi_{\ell_k}^{p_k-1} \otimes \cdots
\otimes \phi_{\ell_m}^{p_m} \big)_\symm \label{ann} \\
a^\dagger\big(\phi_{\ell_k} \big) \:a \big(\bar{\phi}_{\ell_k} \big) \:
\big( \phi_{\ell_1}^{p_1} \otimes \cdots \otimes \phi_{\ell_m}^{p_m} \big)_\symm
&= p_k\: \big( \phi_{\ell_1}^{p_1} \otimes \cdots \otimes \phi_{\ell_m}^{p_m} \big)_\symm \label{nop} \\
\big[ a(\bar{\phi}), \:a^\dagger(\psi) \big]
&= \la \phi | \psi \ra\; \1_\Fock \:. \label{ccr}
\end{align}
\end{Lemma}
\Proof The relations~\eqref{crt}--\eqref{nop} follow immediately from~\eqref{ada}
and~\eqref{cnrel}. Likewise, \eqref{ada} and~\eqref{aad} give rise to the
commutation relation
\[ \big[ a(\bar{\phi}_{\ell_k}), \:a^\dagger(\phi_{\ell_k}) \big] = \1_\Fock \:. \]
Moreover, it is obvious that the operators~$a(\bar{\phi}_{\ell_k})$
and~$a(\bar{\phi}_{\ell_l})$ commute if~$k \neq l$.
Writing these relations in a basis-independent form gives~\eqref{ccr}.
\QED
In view of~\eqref{nop}, the operator product~$a^\dagger(\phi) \:a(\bar{\phi})$
is referred to as the {\em{number operator}}.
The relations~\eqref{ccr} are the usual {\em{canonical commutation
relations}} for the creation and annihilation operators.

\subsection{Bosonic Field Operators and the Canonical Commutation Relations} \label{secccr}
In order to apply the bosonic Fock space formalism to causal variational principles,
we again consider the Hilbert space of holomorphic jets~$(\h, ( .|. ))$
as defined in~\eqref{Hprod}.
Applying the Fock space construction of Section~\ref{secfockprelim},
we obtain the corresponding Fock space~$(\Fock, \la.|.\ra_\Fock)$.
In this section we introduce the usual field operators and show that they
satisfy canonical commutation relations, where on the right side the
causal fundamental solution~$G$ in~\eqref{Kdef} appears.
This construction clarifies how the causal structure is built into the Fock space formulation.
Given a compactly supported jet~$\u \in \J^*_0$,
we let~$G\u \in \Glin_\sc$ be the corresponding linearized solution.
Being a real solution, we can decompose it into holomorphic and anti-holomorphic components by
\[ G \u = \phi_\u + \bar{\phi}_u \qquad \text{with} \qquad \phi_\u := \chi^\hol \,G u \in \h \]
(with~$\chi^\hol$ according to~\eqref{chiholdef}).
We introduce the corresponding {\em{field operator}} $\hat{\Phi}(\u)$ by
\beq \label{Phidef}
\hat{\Phi}(\u) := a\big( \bar{\phi}_\u \big) + a^\dagger\big( \phi_\u \big) \:.
\eeq
It is obviously a symmetric linear operator on~$\Fock$.

For two jets~$\u, \v \in \J^*_0$, we can apply~\eqref{Phidef} and~\eqref{ccr} to obtain
\begin{align*}
\big[ &\hat{\Phi}(\u), \hat{\Phi}(\v) \big] = \big[ a\big( \bar{\phi}_\u \big), a^\dagger\big( \phi_\v \big) \big]
- \big[ a\big( \bar{\phi}_\v \big), a^\dagger\big( \phi_\u \big) \big]
= \big( \la \phi_\u | \phi_\v \ra - \la \phi_\v | \phi_\u \ra \big) \:\1_\Fock \\
&= 2i\, \im \big( \la \phi_\u | \phi_\v \ra \big)\: \1_\Fock
= 2i\, \im \Big( \big( \chi^\hol \,G u \:\big|\: \chi^\hol \,G v \big) \Big)\: \1_\Fock \\
&= 2i\, \im \Big( \big( G u \:\big|\: \chi^\hol \,G v \big) \Big)\: \1_\Fock
\overset{\eqref{chiholdef}}{=} 
i\, \im \Big( \big( G u \:\big|\: (\1 - i J) \,G v \big) \Big)\: \1_\Fock \\
&= -i \re \Big( \big( G u \:\big|\:J\,G v \big) \Big)\: \1_\Fock
= i \re \big( \sigma(G u, \,G v) \big)\: \1_\Fock
= i \sigma( G u,\:G v )\: \1_\Fock \:,
\end{align*}
where in the last line we used that, since~$Gu, Gv \in \Glin_\sc$ are real solutions,
their surface layer inner product and symplectic form are both real-valued.
In~\cite[Proposition~5.9]{linhyp} it is shown that
\beq \label{sigmaform}
\sigma(G\u, G\v) = \la \u, G \,\v \ra_{L^2(M)} \:.
\eeq
We thus obtain the commutation relations
\[ \big[ \hat{\Phi}(\u), \hat{\Phi}(\v) \big] 
= i \: \la \u, G \,\v \ra_{L^2(M)}\: \1_\Fock  \:. \]
These are the usual canonical commutation relations in the Heisenberg picture
(see for example~\cite[eq.~(12.37)]{bjorken2} or~\cite[eq.~(2.53)]{peskin+schroeder}).
We remark that, following the path of axiomatic quantum field theory,
one can also take the field operators with their commutation relation as the starting
point. The algebra generated by the field operators is referred to as the algebra
of observables, and representing this algebra on Hilbert spaces gives
Fock spaces (for details see the textbooks~\cite{moretti-book, brunettibook}).
This path can also be taken for causal fermion systems, as is worked out for
linear systems in~\cite{weyl}.

\subsection{The Holomorphic Perturbation Map as a Linear Operator on~$\Fock$} \label{secfockholo}
In preparation for rewriting the perturbation map in the Fock space formalism,
we begin with the situation of Theorem~\ref{thmholpert} in which there is a holomorphic
perturbation expansion.

We point out that the Fock space was constructed starting from the space~$\Glin_\sc$
of linearized solutions. The perturbation map~$\Pert$ in~\eqref{PNpert}, however,
does not map to linearized solutions, but to the space~$\Gamma_\sc$.
This is already clear to second order in perturbation theory, where~$\Pert^{(2)}$
will be an {\em{inhomogeneous}} solutions, where the inhomogeneity takes into account
the interaction term. Therefore, before we can describe~$\Pert$ in the Fock space formalism,
we must consider it as an operator which takes values in the linearized solutions.
To this end, we make use of the restriction operator defined in~\eqref{restrictionmap}.
We introduce the notation
\beq \label{Prestrict}
\Pert^t := \Pert|^t\::\: \Gamma_\sc \rightarrow \Glin_\sc \:.
\eeq
Taking the holomorphic component gives rise to a nonlinear operator
from~$\h$ to~$\h$ which has a perturbation expansion,
\beq \label{Ndef}
\Phol := \chi^\hol \,\Pert^t \::\: \Gamma^\hol \subset \h \rightarrow \h \:,\qquad
\Phol(\lambda z) = \sum_{p=1}^\infty \lambda^p\: \Phol^{(p)}
\big( \!\!\!\underbrace{z, \ldots, z}_{\text{$p$ arguments}} \!\!\!\big) \:,
\eeq
where the operators~$\Phol^{(p)}$ are multilinear and symmetric.
Here the vector~$z \in \h$ is to be considered as the holomorphic component of~$w$, i.e.
\beq \label{zdef}
z := \chi^\hol \,w \:.
\eeq

To any~$z \in \h$ we want to associate a corresponding 
unperturbed Fock state~$\Upsilon(z)$. In order for being able to rewrite the non-linear perturbation
map as a linear operator on the Fock space, it is important that~$\Upsilon(z)$ involves all
tensor powers of~$z$. We make the ansatz
\beq \label{Upsdef}
\Upsilon(z) = \sum_{n=0}^\infty C_n\, z^n \in \Fock
\eeq
with complex coefficients~$C_n$ to be determined below.
Our goal is to construct a linear operator~${\mathfrak{L}} : \Fock \rightarrow \Fock$ with the property that
\[ {\mathfrak{L}} \,\Upsilon(z) = \Upsilon \big( \Phol(z) \big) = \sum_{n=0}^\infty C_n\: \Phol(z)^n \:. \]

Applying~\eqref{ann} to~\eqref{Upsdef}, we obtain
\[ a \big( \bar{\phi} \big)\: \Upsilon(z) = \la \phi | z \ra \;\sum_{n=1}^\infty C_n\: \sqrt{n}\; z^{n-1}
= \la \phi | z \ra \;\sum_{n=0}^\infty C_{n+1}\: \sqrt{n+1}\; z^n \:. \]
Therefore, it seems most convenient to choose
\beq \label{cn}
C_n = \frac{1}{\sqrt{n!}} \:,
\eeq
because we then obtain the simple relation
\beq \label{annphi}
a \big( \bar{\phi} \big)\: \Upsilon(z) = \la \phi | z \ra \: \Upsilon(z) \:.
\eeq
Next, using~\eqref{crt}, we get
\begin{align*}
a^\dagger \big( \phi \big)\: \Upsilon(z) &=
a^\dagger \big( \phi \big)\: \sum_{n=0}^\infty \frac{z^n}{\sqrt{n!}}
= \sum_{n=0}^\infty \frac{\sqrt{n+1}}{\sqrt{n!}}\: \big(\phi \otimes z^n \big)_\symm
= \sum_{n=0}^\infty \frac{n+1}{\sqrt{(n+1)!}}\: \big(\phi \otimes z^n \big)_\symm \\
&= \sum_{n=1}^\infty \frac{n}{\sqrt{n!}}\: \big(\phi \otimes z^{n-1} \big)_\symm 
= \sum_{n=1}^\infty \frac{1}{\sqrt{n!}}\: D_\phi z^n \:,
\end{align*}
where~$D$ denotes the directional derivative. We thus obtain the compact formula
\beq \label{crtphi}
a^\dagger \big( \phi \big)\: \Upsilon(z) = D \Upsilon|_z \,\phi \:.
\eeq

Before going on, we point out that the state~\eqref{Upsdef} with coefficients~$C_n$
given by~\eqref{cn} is well-known in physics and is referred to as a {\em{coherent state}}
(see for example~\cite[eq.~(2.14)]{gazeau} or~\cite[page~10]{combescure}).
Coherent states are quantum states which saturate the uncertainty relations and
are therefore well-suited for the descriptions of semi-classical systems.
However, in our setting the context and the objective
are completely different:
Fock spaces do not appear as a consequence of a ``quantization procedure,'' but
they arise instead simply when ``linearizing'' the nonlinear dynamics on the tensor algebra.
Coherent states are not ``semi-classical states,'' but they are used merely
as a convenient tool for describing the dynamics as described by the causal action principle
in terms of operators on Fock spaces.

For completeness, we now collect the properties of the coherent states which will
be needed later on.
The relations~\eqref{annphi} and~\eqref{crtphi} are very useful for computations, as we now explain.
To begin with, the operator~$\Upsilon(z)$ can be expressed with an exponential
acting on the Fock vacuum.
\begin{Lemma} \label{lemmaPhiexp}
The state~$\Upsilon(z)$ in~\eqref{Upsdef} and~\eqref{cn} can be obtained from the vacuum by
\beq \label{Phiexp}
\Upsilon(z) = \exp\big( a(z)^\dagger \big) \,|0\ra_\Fock \:.
\eeq
\end{Lemma}
\Proof Clearly, the Fock vacuum can be written as~$|0\ra_\Fock= \Upsilon(0)$. Using the exponential series
and applying~\eqref{crtphi} iteratively, we obtain
\[ \exp\big( a(z)^\dagger \big) \,|0\ra_\Fock =
\sum_{n=0}^\infty \frac{1}{n!} \;(a(z)^\dagger)^n \,|0\ra_\Fock = 
\sum_{n=0}^\infty \frac{1}{n!} \;D^n \Upsilon|_0 \,z^n = \Upsilon(z) \:, \]
where in the last step we used the Taylor formula.
\QED
We next compute the Fock scalar product on the image of~$\Upsilon$.
\begin{Lemma} \label{lemmaexpsprod}
For any~$\phi, z \in \h$,
\[ \big\la \Upsilon(\phi) \,\big|\, \Upsilon(z) \big\ra_\Fock = \exp \big( \la \phi | z \ra \big) \:. \]
\end{Lemma}
\Proof For clarity, we give two alternative proofs. First, using~\eqref{Upsdef} and~\eqref{cn}, we obtain
\[ \big\la \Upsilon(\phi) \,\big|\, \Upsilon(z) \big\ra_\Fock
= \sum_{n=0}^\infty \frac{1}{n!}\: \big\la \phi^n \,\big|\, z^n \big\ra_\Fock
\overset{(*)}{=} \sum_{n=0}^\infty \frac{1}{n!}\: \big( \la \phi | z \ra \big)^n
= \exp \big( \la \phi | z \ra \big) \:, \]
where in~$(*)$ we used~\eqref{sdef}.

In the second proof we apply the formula of Lemma~\ref{lemmaPhiexp},
\begin{align*}
\big\la \Upsilon(\phi) \,\big|\, \Upsilon(z) \big\ra_\Fock
&= \la 0 \,|\, \exp\big( a(\overline{\phi}) \big) \: \exp\big( a^\dagger(z) \big)\, | 0 \ra_\Fock \\
&= \sum_{n=0}^\infty \frac{1}{(n!)^2}\;
\la 0 \,|\, a(\overline{\phi})^n\: \big( a^\dagger(z) \big)^n \, | 0 \ra_\Fock \:,
\end{align*}
where we used that we only get a contribution if as many particles are created as are annihilated.
We now iteratively commute the annihilation operators to the right, where
they give zero when acting on the vacuum state.
There are~$n!$ terms (because the first factor~$a$ is commuted $n$ times,
the second factor~$a$ is commuted $n-1$ times, etc.).
According to~\eqref{ccr}, every commutation gives a scalar product.
We thus obtain
\[ \big\la \Upsilon(\phi) \,\big|\, \Upsilon(z) \big\ra_\Fock = 
\sum_{n=0}^\infty \frac{1}{(n!)^2}\; n!\: \la \phi | z \ra = \exp \big( \la \phi | z \ra \big) \:. \]
This concludes the second proof.
\QED

Our next step is to rewrite the nonlinear operator~$\Phol$ 
as a linear operator on the Fock space. To this end, the concept of Wick ordering
will be useful.
\begin{Def} \label{defwick}
A product of creation and annihilation operators is {\bf{Wick ordered}}
by bringing all creation operators to the left and all annihilation operators to the right.
We denote Wick ordered products by putting colons $\wick \cdots \wick$ around them.
\end{Def} \noindent
In the next theorem, we also use the annihilation operator~$a$ without an argument,
to be understood as follows. The operator~$a(\bar{\phi})$ associates to every~$\phi \in \h$
a linear operator on the Fock space. Thus for any two Fock vectors~$\Phi, \tilde{\Phi} \in \Fock$,
we obtain the linear functional
\[ \alpha_{\Phi, \tilde{\Phi}} := \la \Phi \,|\, a(.) \,\tilde{\Phi} \ra_\Fock \::\: \h \rightarrow \C \;. \]
The Fr{\'e}chet-Riesz theorem allows us to identify this functional with a unique vector~$\psi_{\Phi, \tilde{\Phi}} \in \h$ via
\[ \la \psi_{\Phi, \tilde{\Phi}} | \phi \ra = \alpha_{\Phi, \tilde{\Phi}}(\phi) \qquad \text{for all$~\phi \in \h$}\:. \]
In this way, the operator~$a$ gives rise to an operator
\[ a \::\: \Fock \rightarrow \h \times \Fock \:, \]
which with a slight abuse of notation we again denote by~$a$. It is defined by the relation
\[ \la \Phi \,|\, a \,\tilde{\Phi} \ra_\Fock = \psi_{\Phi, \tilde{\Phi}} \in \h \qquad \text{for all~$\Phi, \tilde{\Phi} \in \Fock$} \:. \]
The above relations can be summarized alternatively by the relation
\[ \la \phi | a \ra = a(\bar{\phi}) \in \Lin(\Fock) \:, \]
where both sides of the equation are operators on~$\Fock$.
\begin{Thm} \label{thmlinearfock}
The linear operator
\[ {\mathfrak{L}} = \wick \exp \bigg( a^\dagger \Big( \sum_{p=2}^\infty \Phol^{(p)}(a, \ldots, a) \Big) \bigg)\,\wick 
\;:\; \Fock \rightarrow \Fock \]
linearizes the perturbation map in the sense that
\[ {\mathfrak{L}} \,\Upsilon(z) = \Upsilon \big( \Phol(z) \big) \]
with~$\Upsilon$ according to~\eqref{Upsdef} or~\eqref{Phiexp}.
\end{Thm}
\Proof Iterating~\eqref{crtphi} similar as the proof of Lemma~\ref{lemmaPhiexp},
we obtain
\beq \label{adexp}
\exp \Big( a^\dagger \big( \phi \big) \Big)\, \Upsilon(z)
= \sum_{p=0}^\infty \frac{a^\dagger \big( \phi \big)^p}{p!}\, \Upsilon(z)
= \sum_{p=0}^\infty \frac{1}{p!} \:D^p \Upsilon|_z (\phi^p)
= \Upsilon(z+\phi) \:.
\eeq
Hence
\[ \Upsilon\big(\Phol(z) \big) = \Upsilon\Big(z + \sum_{p=2}^\infty \Phol^{(p)}(z, \ldots, z) \Big)
\overset{(*)}{=} \exp \bigg( a^\dagger \Big( \sum_{p=2}^\infty \Phol^{(p)}(z, \ldots, z) \Big) \bigg)\, \Upsilon(z) \:, \]
where in~$(*)$ we applied the equation~\eqref{adexp} backwards for
\[ \phi =  \sum_{p=2}^\infty \Phol^{(p)}(z, \ldots, z) \:. \]
It remains to write the arguments~$z$ of the operator~$\Phol^{(p)}$ in terms of
field operators. To this end, we iterate~\eqref{annphi} to obtain
\[ 
a \big( \bar{\phi} \big)^p\: \Upsilon(z) = \la \phi | z \ra^p \: \Upsilon(z) \:. \]
Therefore, we may replace each argument~$z$ by an operator~$a$
acting on~$\Upsilon(z)$.
In order to make sure that these operators really act on~$\Upsilon(z)$, we must
Wick order all operator products. This gives the result.
\QED
This proof can be summarized in a more compact form as follows:
\begin{align*}
\Upsilon \big( \Phol (z) \big) &= e^{a^\dagger(\Phol (z))}\, |0\ra_\Fock
= e^{a^\dagger(\Phol (z)) - a^\dagger(z)}\: e^{a^\dagger(z)}\, |0\ra_\Fock \\
& = e^{a^\dagger(\Phol (z)) - a^\dagger(z)}\: \Upsilon(z) \\
&= \wick e^{a^\dagger(\Phol (a)) - a^\dagger(a)} \wick \:\Upsilon(z)
= \wick e^{a^\dagger \big(\Phol (a) - a \big)} \wick \:\Upsilon(z)
\end{align*}

In the remainder of this section, we consider if and how the conservation
laws for scattering systems collected in Corollary~\ref{corconsscatter}
can be formulated with Fock spaces. In order to describe the scattering
process in Figure~\ref{figscatter}, we choose the linearized solution~$z$
as the holomorphic component of~$w$,
\[ z = \chi^\hol \,w \:. \]
Then~$\Phol(z) = \chi^\hol \,w_\text{out}$ is the holomorphic component of the
outgoing jet. Since~$w_\text{out}$ is a linearized solution, the scalar
product~$(w_\text{out},w_\text{out})^{t_\text{out}}$ is well-defined.
However, due to the nonlinear interaction region,
the jet~$w_\text{out}$ does not extend to a linearized solution
in all of spacetime. Therefore, we cannot work with the
surface integrals in~\eqref{cI1nl}--\eqref{cI2symmnl}.
The only surface layer integral which makes sense is the nonlinear
surface layer integral, whose conservation law~\eqref{cosinl} states that
\[ \gamma^{t_\text{in}} \big(w, \rho_\vac \big)
= \gamma^{t_\text{out}} \big(w_\text{out} + \n_\text{out}, \rho_\vac \big) \:. \]
Rewriting~$\gamma^t$ using the formula for linear systems in~\eqref{cons3}
and expressing~$\gamma_{\rho_\vac}(\n_\text{out})$ with the help of~\eqref{innerflux}
as the inner flux, we obtain
\beq \label{consh}
\frac{1}{2}\: (w,w)^{t_\text{in}} = \s\,\mu \big(\n_\text{out}, N^{t_\text{out}} \big)
+ \frac{1}{2}\: (w_\text{out},w_\text{out})^{t_\text{out}} \:.
\eeq
Here the scalar product can be expressed in terms of the
scalar product on the one-particle Hilbert space~$(\h, \la .|. \ra)$,
making it possible to apply the formula of Lemma~\ref{lemmaexpsprod}.
Writing~$w_\text{out} = \Pert(w)$ and~$\n_\text{out} = \mathfrak{N}(w)$
gives the following result:
\begin{Thm} \label{thmNunit}
For any~$z \in \Gamma^\hol \subset \h$,
\beq \label{Nunit}
\big\la {\mathfrak{L}} \,\Upsilon(z) \,\big|\, {\mathfrak{L}} \,\Upsilon(z) \big\ra_\Fock
= \la \Upsilon(z) | \Upsilon(z) \ra_\Fock \;\exp \big\{ \s\, \mu \big(\n_\text{\rm{out}}, N^{t_\text{\rm{out}}} \big) \big\}\:.
\eeq
\end{Thm}
\Proof First, using that the operator~$\Gamma$ is anti-symmetric,
\[ \big( \Pert(w) | \Pert(w) \big) = \frac{1}{2} \: \la \chi^\hol \,\Pert(w) \,|\, \chi^\hol \Pert(w) \ra
= \frac{1}{2} \: \la \Phol(z) \,|\, \Phol(z) \ra \]
and similarly~$(w|w)= \la z | z \ra/2$
(where we again used the notation~\eqref{zdef} as well as the
assumption that the perturbation expansion is holomorphic). Taking~$z=w$ as the
incoming scattering state, we can apply~\eqref{consh} to obtain
\[ \la \Phol(z) \,|\, \Phol(z) \ra = \la z | z \ra + \s\,\mu \big(\n_\text{out}, N^{t_\text{out}} \big) \:. \]
Lemma~\ref{lemmaexpsprod} yields for the Fock space norms of the corresponding coherent states
\[ \big\la \Upsilon(\Phol(z)) \,\big|\, \Upsilon(\Phol(z)) \big\ra_\Fock
= \la \Upsilon(z) | \Upsilon(z) \ra_\Fock \: \exp \big\{ \s\,\mu \big(\n_\text{out}, N^{t_\text{out}} \big) \big\} \:. \]
Applying Theorem~\ref{thmlinearfock} gives the result.
\QED

The main conclusion is that, as a consequence of the inner solutions, the linear dynamics on
the Fock space in general does {\em{not}} preserve the norm on the Fock space.
This is the reason why, although we assume them to be very small (see Section~\ref{secinnersmall}),
the inner solutions must be taken into account.
We will explain in the next section (Section~\ref{secfockFsF}) how to treat this effect
in a way where we do get a norm-preserving time evolution on Fock spaces.
In order to explain another important idea in the present simple setting,
we now explain how one could proceed if the exponential factor in~\eqref{Nunit}
were absent, i.e.\ if
\beq \label{Nunitsimp}
\big\la {\mathfrak{L}} \,\Upsilon(z) \,\big|\, {\mathfrak{L}} \,\Upsilon(z) \big\ra_\Fock
= \la \Upsilon(z) | \Upsilon(z) \ra_\Fock \qquad \text{for all~$z \in \h$}\:.
\eeq
This relation is clearly satisfied if~${\mathfrak{L}}$ is a
unitary operator on~$\Fock$. However, it is not obvious if, conversely,~\eqref{Nunitsimp}
also implies the unitary of~${\mathfrak{L}}$,
because in~\eqref{Nunitsimp} we are only allowed to take the expectation
value for Fock vectors of the form~$\Upsilon(z)$ with
holomorphic one-particle vectors~$z = \chi^\hol w$.
But unitarity can be obtained with the following method:

\begin{Lemma} {\bf{(polarization lemma)}} \label{lemmapolarize}
Assume that an operator~$A$ on the Fock space~$\Fock$ satisfies the relation
\[ \big\la \Upsilon\big( \chi^\hol w \big) \:\big|\: A\, \Upsilon \big( \chi^\hol w \big) \big\ra_\Fock =0\qquad
\text{for all~$w \in \Glin_\sc$}\:. \]
Then~$A$ vanishes.
\end{Lemma}
\Proof Given~$p,q \geq 0$, in generalization of~\eqref{wchoose} we now choose~$w$ as
\[ w = \sum_{\ell=1}^{p+q} \big( e^{i \alpha_\ell}\: v_\ell^\hol + e^{-i \alpha_\ell}\: v_\ell^\ah \big) \]
with vectors~$v_\ell^\hol \in \h$ and phases~$\alpha_\ell \in \R$.
Since the phases can be chosen independently, the contributions with any combination
of the phases vanish separately. In particular, it follows that
\[ 
e^{-i \alpha_1 - \cdots - i \alpha_q + i \alpha_{q+1} + \cdots + i \alpha_{p+q}}
\big\la v_1^\hol \otimes \cdots \otimes v_q^\hol
\:\big|\: A\, v_{q+1}^\hol \otimes \cdots \otimes v_{p+q}^\hol \big\ra_\Fock = 0 \:. \]
Since~$p$ and~$q$ as well as the vectors~$v_\ell^\hol \in \h$
can be chosen arbitrarily, the result follows.
\QED

Applying this polarization lemma to~\eqref{Nunitsimp} gives the following result:
\begin{Corollary} Assume in the setting of a holomorphic perturbation expansion
that the relation~\eqref{Nunitsimp} holds for the linear operator~${\mathfrak{L}}$ introduced in Theorem~\ref{thmlinearfock}.
Then~${\mathfrak{L}}$ is a unitary operator on~$\Fock$.
\end{Corollary}
\Proof After rewriting~\eqref{Nunitsimp} as
\[ \big\la \Upsilon(z) \,\big|\, \big\{ {\mathfrak{L}}^* \,{\mathfrak{L}} - \1 \big\} \,\Upsilon(z) \big\ra_\Fock = 0 \qquad \text{for all~$z \in \h$}\:, \]
one can apply the polarization lemma to conclude that the operator in the curly brackets vanishes.
\QED

\subsection{The Perturbation Map as a Linear Operator on~$\Fock^* \otimes \Fock$} \label{secfockFsF}
The construction in the previous section has two major shortcomings:
First, we concluded in Section~\ref{secwhencomplex} that in
most physical situations there is no canonical complex structure,
so that the assumption of a holomorphic perturbation expansion~\eqref{Ndef} does not hold.
Second, due to the exponential factor in~\eqref{Nunit}, the Fock space norm
in general is not conserved, implying that the scattering operator~${\mathfrak{L}}$
is not unitary. In this section, we explain how to overcome these difficulties.

Thus we return to the general setting of the scattering process as described in Section~\ref{secscatters}.
Since there is no canonical complex structure to our disposal,
we simply work in the scattering regions
with the complex structures of the linear systems (see Section~\ref{seccomplexlinear}).
This is a canonical choice. But we must keep into account that the linearized
time evolution is not compatible with the complex structures, meaning that
holomorphic ingoing jets are mapped to linear combinations of holomorphic
and anti-holomorphic outgoing jets.
Likewise, the perturbation map~$\Pert$ mixes holomorphic and anti-holomorphic parts.
Finally, we must take into account the operator~$\mathfrak{N}$, which generates
an inner solution which contributes to the nonlinear surface layer integral in~\eqref{cosinl}.
The resulting situation is described most conveniently as follows.
Similar to~\eqref{zdef}, we now denote the holomorphic and anti-holomorphic components
of~$w$ by
\[ 
z = \chi^\hol \,w \qquad \text{and} \qquad \overline{z} = \chi^\ah \,w \:. \]
Similar to~\eqref{Ndef} we apply the restriction operator and denote the holomorphic component
of the perturbation map at time~$t$ by
\beq \label{Ndefcomplex}
\begin{split}
\Phol &:= \chi^\hol \,\Pert^t \::\: \h^* \times \h \rightarrow \h \:, \\
\Phol(\lambda \overline{z}, \lambda z) \!&\,= \sum_{p=1}^\infty \sum_{q=0}^\infty \lambda^{p+q}\: \Phol^{(q,p)}
\big( \!\!\!\underbrace{\overline{z}, \ldots, \overline{z}}_{\text{$q$ arguments}};
\underbrace{z, \ldots, z}_{\text{$p$ arguments}} \!\!\!\big) \:.
\end{split}
\eeq

One method of dealing with the anti-holomorphic component would be
to enlarge the Fock space~$\Fock$ by including the anti-holomorphic component
(i.e.\ $\Fock$ could be chosen as the bosonic Fock space generated by~$\Gamma^\C$
instead of~$\Gamma^\hol$).
However, this method would have the shortcoming that the
polarization lemma (Lemma~\ref{lemmapolarize}) would no longer apply,
because both the bra and ket states would involve
both holomorphic and anti-holomorphic components.
As a consequence, we would no longer get operator equations on the Fock space.
This is the reason why it is preferable to work again with the holomorphic Fock space.
The anti-holomorphic contributions give rise to a mixing of the bra and ket state,
as we now describe.

It is convenient to simplify our notation as follows. We let~$(\phi_i)$ be an orthonormal
basis of~$\h$. We write
\[ a^\dagger_i = a^\dagger(\phi_i) \qquad \text{and} \qquad
a^i = a \big(\overline{\phi_i}\big) \:. \]
Then the anti-commutation relations~\eqref{ccr} become
\[ \big[a^i, a^\dagger_j \big] = \delta^i_j \:. \]
Next, we write~\eqref{Ndefcomplex} in components by setting
\[ \Phol^{(q,p)}
\big( \overline{\phi_{j_1}}, \ldots, \overline{\phi_{j_q}};
\phi_{k_1}, \ldots, \phi_{k_p} \big) =
\sum_i {}^i\! (\Phol)^{j_1,\ldots, j_q}_{k_1,\ldots, k_p} \: \phi_i \:. \]
Using~\eqref{annphi}, we can obtain the components of~$z$ by acting
with the annihilation operators on~$\Upsilon(z)$,
\[ 
a^i\: \Upsilon(z) = z^i \:\Upsilon(z) \:. \]
In the case~$q=0$ of a holomorphic expansion, this makes it possible to
rewrite the linear operator of Theorem~\ref{thmlinearfock} as
\[ {\mathfrak{L}} = \wick \exp \Big( \sum_{p=2}^\infty
a^\dagger_i \: {}^i\! (\Phol)_{i_1,\ldots, i_p}\: a^{i_1} \cdots a^{i_p} \Big) \wick \:, \]
where similar to Einstein's summation convention we sum over all Hilbert space indices
which appear twice. In the case~$q \neq 0$, the indices~$j_1,\ldots, j_q$ also need to be
contracted with annihilation operators. However, we cannot work with an operator
acting on~$\Upsilon(z)$, because this only gives holomorphic vectors.
Our method for obtaining anti-holomorphic vectors is to let annihilation operators act on
a bra vector. For example,
\[ \la a_i \Upsilon(z) | \cdots \Upsilon(z) \ra_\Fock = 
\la z_i \Upsilon(z) | \cdots \Upsilon(z) \ra_\Fock = \overline{z_i}\: \la \Upsilon(z) | \cdots \Upsilon(z) \ra_\Fock \:, \]
where~$\cdots$ stands for any other Fock space operators. For notational clarity,
we regard the bra vector as a vector in the dual of the Fock space and write
\[ \la \Upsilon(z) | \; \otimes \; | \Upsilon(z) \ra_\Fock \;\in\; \Fock^* \otimes \Fock \:. \]
Moreover, we introduce the operators
\[ \overline{a}_i \qquad \text{and} \qquad \overline{a^\dagger}^i \]
as the operators~$a^i$ respectively~$a^\dagger_i$ acting on the dual space, i.e.
\begin{align*}
\overline{a}_i \Big( \la \Upsilon(z) | \otimes | \Upsilon(z) \ra_\Fock \Big)
:= \la a^i \,\Upsilon(z) | \otimes | \Upsilon(z) \ra_\Fock \\
\overline{a^\dagger}^i \Big( \la \Upsilon(z) | \otimes | \Upsilon(z) \ra_\Fock \Big)
:= \la a^\dagger_i \,\Upsilon(z) | \otimes | \Upsilon(z) \ra_\Fock \:.
\end{align*}
Then
\beq \label{nowick}
\begin{split}
\Phol^{(q,p)}&
\big( \overline{z}, \ldots, \overline{z};
z,\ldots, z \big) \; \Big( \la \Upsilon(z) | \otimes | \Upsilon(z) \ra_\Fock \Big) \\
&= \phi_i \:{}^i\! (\Phol)^{j_1,\ldots, j_q}_{k_1,\ldots, k_p} \: \overline{a}_{j_1} \cdots \overline{a}_{j_q}\:
a^{k_1} \cdots a^{k_p}\; \Big( \la \Upsilon(z) | \otimes | \Upsilon(z) \ra_\Fock \Big) \:.
\end{split}
\eeq
Next, we introduce {\em{Wick ordering}} for operators acting on~$\Fock^* \otimes \Fock$
in three steps: In the first step, the field operators act on the bra respectively ket states as explained above.
In the second step, all the field operators acting on ket states are Wick ordered in the usual way
by writing annihilation operators to the right. In the third and last step, all the field operators acting on
bra states are Wick ordered as usual. We again denote Wick ordering by~$\wick \cdots \wick$.
Using Wick ordering, the operators in~\eqref{nowick} can be written anywhere, for example
\begin{align*}
\Phol^{(q,p)}& \big( \overline{z}, \ldots, \overline{z};
z,\ldots, z \big) \; \Big( \la \Upsilon(z) | \otimes | \Upsilon(z) \ra_\Fock \Big) \\
&= \wick \Big\la \Upsilon(z) \Big| \otimes \Big| \phi_i \:{}^i\! (\Phol)^{j_1,\ldots, j_q}_{k_1,\ldots, k_p} \: \overline{a}_{j_1} \cdots \overline{a}_{j_q}\: a^{k_1} \cdots a^{k_p}\, \Upsilon(z) \Big\ra_\Fock \wick \\
&= \wick \Big\la a^{j_1} \cdots a^{j_q} \,\Upsilon(z) \Big| \otimes \Big| \phi_i \:{}^i\! (\Phol)^{j_1,\ldots, j_q}_{k_1,\ldots, k_p} \: a^{k_1} \cdots a^{k_p}\, \Upsilon(z) \Big\ra_\Fock \wick \:.
\end{align*}
Here one must only keep in mind that the operators acting on the bra state are complex conjugated,
and that complex conjugation makes upper indices to lower indices and vice versa.

With this notation, Theorem~\ref{thmlinearfock} can be extended to the non-holomorphic
setting as follows:
\begin{Thm} 
The linear operator~${\mathfrak{L}}_\Pert :\Fock^* \otimes \Fock \rightarrow \Fock^* \otimes \Fock$ given by
\beq
\label{Nnohol0}
{\mathfrak{L}}_\Pert = \wick \exp \bigg(
\sum_{(p,q) \neq (1,0)} a^\dagger_i \, {}^i\! (\Phol)^{j_1,\ldots, j_q}_{k_1,\ldots, k_p} \: \overline{a}_{j_1} \cdots \overline{a}_{j_q}\: a^{k_1} \cdots a^{k_p} \bigg)\,\wick 
\eeq
linearizes the perturbation map in the sense that
\[ {\mathfrak{L}}_\Pert \: \big( \la \Upsilon(z) \big| \otimes \big| \Upsilon(z) \ra_\Fock \big)
= \big\la \Upsilon\big(\Phol(\overline{z}, z)\big) \big| \otimes \big| \Upsilon\big(\Phol(\overline{z}, z)\big) \big\ra_\Fock \:. \]
\end{Thm} \noindent
We point out that the fact that~\eqref{Nnohol0} involves the operators~$\overline{a}$ gives rise to a complicated
``mixing'' of the bra and ket states in the dynamics.

Our next step is to build in the inner solutions. The formula~\eqref{Nunit} obtained in the holomorphic
approximation shows that the inner flux gives rise to a prefactor multiplying the
Fock scalar product. In order to obtain a conservation law for the Fock norm, the obvious idea is to
absorb this prefactor into the Fock vectors. In other words, we want to rescale the Fock vectors
by the exponential of the inner flux. This can be achieved simply by multiplying~\eqref{Nnohol}
by this exponential. However, before we can do so, we must rewrite the inner flux as an operator on the Fock space:
Similar to~\eqref{Ndefcomplex} we expand the inner flux in powers of the holomorphic and
anti-holomorphic incoming jets,
\begin{align*}
\mu \big(\n, N^t \big) &\::\: \h^* \times \h \rightarrow \R \:, \\
\mu \big(\n, N^t \big) \!&\,= \sum_{p,q=1}^\infty \lambda^{p+q}\: \mu^{(q,p)}
\big( \!\!\!\underbrace{\overline{z}, \ldots, \overline{z}}_{\text{$q$ arguments}};
\underbrace{z, \ldots, z}_{\text{$p$ arguments}} \!\!\! \big) \:,
\end{align*}
and expand in the orthonormal basis~$(\phi_i)$,
\begin{align*}
\mu^{(q,p)}&
\big( \overline{z}, \ldots, \overline{z};
z,\ldots, z \big) \; \Big( \la \Upsilon(z) | \otimes | \Upsilon(z) \ra_\Fock \Big) \\
&= \mu^{j_1,\ldots, j_q}_{k_1,\ldots, k_p} \: \overline{a}_{j_1} \cdots \overline{a}_{j_q}\:
a^{k_1} \cdots a^{k_p}\; \Big( \la \Upsilon(z) | \otimes | \Upsilon(z) \ra_\Fock \Big) \:.
\end{align*}

This gives the following result:
\begin{Thm} \label{thmlinearfock2}
The linear operator~${\mathfrak{L}} : \Fock^* \otimes \Fock \rightarrow \Fock^* \otimes \Fock$ given by
\beq
\begin{split}
\label{Nnohol}
{\mathfrak{L}} = \wick \exp \bigg(
&\sum_{p,q=1}^\infty \mu^{j_1,\ldots, j_q}_{k_1,\ldots, k_p} \: \overline{a}_{j_1} \cdots \overline{a}_{j_q}\: a^{k_1} \cdots a^{k_p} \\
+ &\sum_{(p,q) \neq (1,0)}  a^\dagger_i \, {}^i\! (\Phol)^{j_1,\ldots, j_q}_{k_1,\ldots, k_p} \: \overline{a}_{j_1} \cdots \overline{a}_{j_q}\: a^{k_1} \cdots a^{k_p} \bigg)\,\wick
\end{split}
\eeq
linearizes the perturbation map in the sense that
\[ {\mathfrak{L}} \: \big( \la \Upsilon(z) \big| \otimes \big| \Upsilon(z) \ra_\Fock \big)
= \big\la \Upsilon\big(\Phol(\overline{z}, z)\big) \big| \otimes \big| \Upsilon\big(\Phol(\overline{z}, z)\big) \big\ra_\Fock
\; e^{\s\,\mu(\n, N^t)} \:. \]
\end{Thm}

Now the linearization takes into account the inner solutions.
Therefore, the conservation law for the nonlinear surface layer integral~\eqref{cosinl}
implies that the Fock norm is conserved in time, as we now make precise.
In order to get a consistent notation, it is preferable to state this result
referring to observables and expectation values.
Exactly as in quantum field theory, an {\em{observable}}~${\mathcal{O}}$ is a symmetric linear operator on~$\Fock$.
The {\em{expectation value}} of an observable with respect
to a {\em{state}}~$\la \Phi | \otimes | \tilde{\Phi} \ra_\Fock \in \Fock^* \otimes \Fock$ is denoted by
\beq \label{eval}
{\mathcal{O}} \big( \la \Phi | \otimes | \tilde{\Phi} \ra_\Fock \big) :=
\la \Phi \,|\, {\mathcal{O}} \,\tilde{\Phi} \ra_\Fock \:.
\eeq
In particular, the scalar product on~$\Fock$ is recovered as the expectation value of the identity,
$\1 \big( \la \Phi | \otimes | \tilde{\Phi} \ra_\Fock \big) :=
\la \Phi | \tilde{\Phi} \ra_\Fock$.

\begin{Thm} \label{thmNunit2}
For any~$z \in \Gamma^\hol \subset \h$,
\beq \label{Nunit2}
\1 \Big( {\mathfrak{L}} \; \big( \la \Upsilon(z) \,| \otimes |\, \Upsilon(z) \ra_\Fock \big) \Big)
= \la \Upsilon(z) | \Upsilon(z) \ra_\Fock \:.
\eeq
\end{Thm}
\Proof First, using that the operator~$\Gamma$ is anti-symmetric,
\[ \big( \Pert(w) | \Pert(w) \big) = \frac{1}{2} \: \la \chi^\hol \,\Pert(w) \,|\, \chi^\hol \Pert(w) \ra
= \frac{1}{2} \: \la \Phol(\overline{z}, z) \,|\, \Phol(\overline{z}, z) \ra \]
and similarly~$(w|w)= \la z | z \ra/2$
(where we again used the notation~\eqref{zdef}). 
Taking~$z=w$ as the incoming scattering state, we can apply~\eqref{consh} to obtain
\[ \la \Phol(\overline{z}, z) \,|\, \Phol(\overline{z}, z) \ra = \la z | z \ra + \s\,\mu \big(\n_\text{out}, N^{t_\text{out}} \big) \:. \]
Lemma~\ref{lemmaexpsprod} yields for the Fock space norms of the corresponding coherent states
\[ \big\la \Upsilon(\Phol(z)) \,\big|\, \Upsilon(\Phol(z)) \big\ra_\Fock
= \la \Upsilon(z) | \Upsilon(z) \ra_\Fock \: \exp \big\{ \s\,\mu \big(\n_\text{out}, N^{t_\text{out}} \big) \big\} \:. \]
Applying Theorem~\ref{thmlinearfock2} gives the result.
\QED

To summarize our findings, in contrast to the setting of quantum
field theory,  the system is not described by a Fock state, but by a pair of two Fock vectors
(one in~$\Fock$ and one in~$\Fock^*$). Likewise, the time evolution operator is not an operator on~$\Fock$,
but an operator on~$\Fock^* \times \Fock$.
This corresponds to a complicated mixing of the bra and ket states in the
time evolution. As a consequence, the time evolution cannot be described by
a unitary operator on~$\Fock$.
Despite this mixing, the conservation law for the nonlinear surface layer integral
implies that the Fock norm is conserved under the time evolution~\eqref{Nunit2}.

\section{The Holomorphic Approximation} \label{secholo}
In Theorem~\ref{thmlinearfock} we saw that the perturbation map gives rise to a
complicated mixing of the holomorphic and anti-holomorphic components of the jets.
In order to analyze this mixing in more detail, we now rewrite the dynamics
as described by the perturbation map as an approximate holomorphic dynamics
with a specific error. The method is to track the jets
while time evolves from~$t_{\min}$ to~$t_{\max}$ in small consecutive
time intervals. In each time step, we approximate the dynamics
by ``projecting onto'' the holomorphic component.
In this way, the mixing of the holomorphic and anti-holomorphic components
disappears while preserving the Fock space norm.

\subsection{Conservation Laws and a Complex Structure in the Time Evolution} \label{secvolterms}
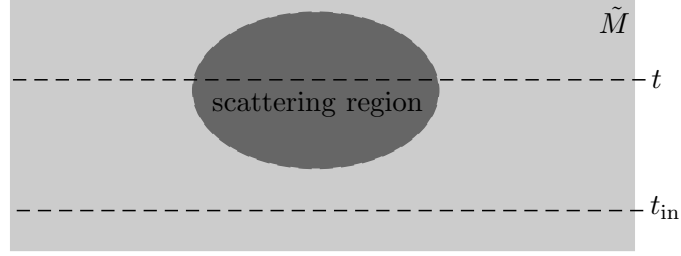
\begin{figure}
%
\psscalebox{1.0 1.0} 
{
\begin{pspicture}(-2.6,-1.675)(13.93,1.675)
\definecolor{colour0}{rgb}{0.8,0.8,0.8}
\definecolor{colour1}{rgb}{0.4,0.4,0.4}
\psframe[linecolor=colour0, linewidth=0.02, fillstyle=solid,fillcolor=colour0, dimen=outer](8.31,1.675)(0.0,-1.675)
\rput[bl](7.83,1.225){\normalsize{$\tilde{M}$}}
\psline[linecolor=black, linewidth=0.02, linestyle=dashed, dash=0.17638889cm 0.10583334cm](0.09,-1.135)(8.42,-1.135)
\rput[bl](8.53,0.495){\normalsize{$t$}}
\rput[bl](8.51,-1.245){\normalsize{$t_\text{\rm{in}}$}}
\psellipse[linecolor=colour1, linewidth=0.02, linestyle=dashed, dash=0.17638889cm 0.10583334cm, fillstyle=solid,fillcolor=colour1, dimen=outer](4.065,0.465)(1.645,1.05)
\rput[bl](2.69,0.115){\normalsize{scattering region}}
\psline[linecolor=black, linewidth=0.02, linestyle=dashed, dash=0.17638889cm 0.10583334cm](0.04,0.605)(8.45,0.605)
\end{pspicture}
}
\caption{Conservation laws at intermediate times.}
\label{figscalet}
\end{figure}

So far, we only analyzed the in- and outgoing scattering
states, but we did not consider the Fock states and the corresponding conservation laws in the
scattering region. To this end, we now consider the dynamics only up to an intermediate
time~$t \in (t_{\min}, t_{\max})$ (see Figure~\ref{figscalet})
and analyze the resulting structures and conservation laws.

As discussed in Section~\ref{secwhencomplex}, in
most physical situations there is no canonical complex structure.
Nevertheless, in order to work with complex Hilbert spaces at time~$t$,
we now introduce a complex structure at time~$t$.
To this end, we simply use the formulas of Definition~\ref{defosi}
for the vacuum measure~$\rho=\rho_\vac$ and evaluate them
for nonlinear jets, i.e.\ in suitable charts
\begin{align}
\big( \Pert(u), \Pert(v) \big)\big|_t &:= \int_{\Omega^t} d\rho_\vac(x) \int_{M \setminus \Omega^t} d\rho_\vac(y) \notag \\
&\qquad\qquad \times
\big( D_{1,\Pert^t(u)} D_{1,\Pert^t(v)} - D_{2,\Pert^t(u)} D_{2,\Pert^t(v)} \big) \L(x,y) \label{nlconserve2} \\
\sigma\big(\Pert(u), \Pert(v) \big) \Big|_t &:=
\int_{\Omega^t} d\rho_\vac(x) \int_{M \setminus \Omega^t} d\rho_\vac(y) \notag \\
&\qquad\qquad \times
\big( \nabla_{1, \Pert^t(u)} \nabla_{2, \Pert^t(v)} - \nabla_{1, \Pert^t(v)} \nabla_{2, \Pert^t(u)} \big) \, \L(x,y) \:,
\label{sympnonlin}
\end{align}
where we again applied the restriction map~\eqref{Prestrict}.
Clearly, these surface layer integrals are {\em{not}} conserved by the nonlinear dynamics.
Having both a scalar product~\eqref{nlconserve2} and a symplectic form~\eqref{sympnonlin}
to our disposal, we can again use the construction of Section~\ref{seccomplexlinear}
to obtain a complex structure on the nonlinear jets at time~$t$
(the holomorphic components of the nonlinear jets give a complex chart on the
nonlinear solution space; this is why we have indeed a complex and not
merely an almost-complex structure).
We again point out that, since this complex structure depends on time,
the time evolution will mix the holomorphic
and anti-holomorphic components.

In the constructions in the previous sections, the conservation of the Fock norm was
based on the conservation law for the nonlinear surface layer integral of Theorem~\ref{thmosinoconserve}.
Moreover, we made use of the fact that, in the scattering regions when the fields are arbitrarily
weak, the nonlinear surface layer integral can be expressed in terms of the inner flux
and the surface layer inner product of the vacuum (see Corollary~\ref{cornl}).
In the interaction region, the nonlinear surface layer integral is still conserved, i.e.\
in analogy to~\eqref{cosinl},
\[ \gamma^t \big(\tilde{\rho}, \rho_\vac \big) \big|_{t_\text{\rm{in}}}^t = 0\:. \]
However, since the jets could be large, it is not obvious that
the nonlinear surface layer integral can
still be written similar to~\eqref{cons3} in terms of surface layers defined in the vacuum.
We now prove that this can indeed be arranged by adding suitable inner solutions.
\begin{Lemma} By adding inner solutions which preserve the total volume
and vanish at initial and final times,
it can be arranged that the formula~\eqref{nl} for the nonlinear surface layer also holds at intermediate times.
\end{Lemma}
\Proof Clearly, \eqref{nl} also holds at intermediate times approximately,
\beq \label{nlapprox}
\gamma^t \big(\tilde{\rho}, \rho_\vac \big) = \s \,\mu\big( \mathfrak{N}(w) ,N^t \big) + \frac{1}{2}\: \big(\Pert(w), \Pert(w) \big)^t + \text{(higher order corrections)} \:.
\eeq
The higher order corrections can be compensated by modifying the inner flux appropriately, i.e.\
by choosing~$\Delta \mathfrak{N}(w)$ with
\[ \s \,\mu\big( \mathfrak{N}(w) ,N^t \big) = - \text{(higher order corrections)} \:. \]
Since the higher order corrections vanish both at initial and final times, the same is true
for the modifications of the inner flux. Therefore, the spacetime integral of the
the scalar component of~$\mathfrak{N}(w)$ vanishes,
\[ \int_M \nabla_{\mathfrak{N}(w)}\:\s = 0 \:. \]
In other words, the inner solution preserves the total volume. This concludes the proof.
\QED
Following the arguments in Section~\ref{secinnersmall} and Appendix~\ref{secapproxinner},
it is sensible to assume that the inner solutions introduced in the previous lemma are small
in the sense of Definition~\ref{defsmallinner}.

In this way, we have arrange that the Fock space construction in Section~\ref{secfockFsF}
also applies at intermediate times. We indicate this simply by adding an superscript~$t$
to the operator~${\mathfrak{L}}$.
\begin{Thm} \label{thmNunitt}
For any~$z \in \Gamma^\hol \subset \h$ and any~$t \in \R$,
linearizing the dynamics up to time~$t$ with a Fock space operator
\beq \label{Ltdef}
{\mathfrak{L}}^t \::\: \Fock^* \otimes \Fock \rightarrow \Fock^* \otimes \Fock \:,
\eeq
the following conservation law holds:
\beq \label{Nunitt}
\1 \Big( {\mathfrak{L}}^t \; \big( \la \Upsilon(z) \,| \otimes |\, \Upsilon(z) \ra_\Fock \big) \Big)
= \la \Upsilon(z) | \Upsilon(z) \ra_\Fock \:.
\eeq
\end{Thm}

\subsection{A Unitary Time Evolution in the Holomorphic Approximation} \label{secholomorphic}
We next consider the evolution of the system in a small time step from~$t$ to~$t+\Delta t$
(by letting~$\Delta t \rightarrow 0$, we will later recover the infinitesimal
time evolution). We again work with the nonlinear jets, which we now denote by
\[ \hat{w}(t) = \Pert^t(w) \qquad \text{and} \qquad \hat{\n}(t) = \mathfrak{N}^t(w) \]
(where the superscript~$t$ again denotes the restriction map~\eqref{Prestrict}).
Similar to~\eqref{Ndefcomplex}, we write the time evolution from~$t$ to~$t+\Delta t$ as
\begin{align*}
\chi^\hol \big( \hat{w}(t+\Delta t) - \hat{w}(t) \big)
&=: \Delta \Phol\big(w(t)) = \Delta \Phol^1\big( w(t) \big) + \Delta \Phol^2\big( w(t), w(t) \big) \\
\chi^\hol \big( \hat{\n}(t+\Delta t) - \hat{\n}(t) \big)
&=: \Delta \hat{\w}_\hol \big(w(t)) = \Delta \hat{\w}_\hol^1\big( w(t) \big) + \Delta \hat{\w}_\hol^2\big( w(t), w(t) \big) \:.
\end{align*}
We here omit the higher orders for two reasons: First for notational convenience, noting
that the higher orders could be treated in a straightforward way.
Second, the higher orders are irrelevant in the physical applications if~$\Delta t$ is chosen
sufficiently small (for example, the coupling term in the Hamiltonian of QED is described by
a quadratic term in the jets formed of a product of a bosonic and a fermionic jet).
Again choosing an orthonormal basis~$(\phi_i)$ of the holomorphic jets,
we decompose the arguments of~$\Delta \Phol$ and~$\Delta \hat{\w}_\hol$
into the holomorphic and anti-holomorphic parts,
\begin{align*}
\Delta \Phol\big(w(t)) &= \phi_l \big( {}^l \!A_j z^j + \: {}^l\! B_{jk} z^j z^k + {}^l\! B^j_k\, \overline{z}_j z^k
+ {}^l\! B^{jk}\, \overline{z}_j \:\overline{z}_k \big)\:\Delta t \\
\Delta \hat{\w}_\hol\big(w(t)) &= \big( E_j z^j + \: F_{jk} z^j z^k + F^j_k\, \overline{z}_j z^k
+ F^{jk}\, \overline{z}_j \:\overline{z}_k \big)\:\Delta t \:. 
\end{align*}

We next consider the corresponding linear time evolution~${\mathfrak{L}}^t$ on~$\Fock^* \otimes \Fock$
in~\eqref{Ltdef}. Taking the linear contributions in~$\Delta t$, we obtain
\begin{align*}
&\Delta \Big( \big\la \Upsilon\big(z(t) \big) \big| \otimes | \Upsilon\big(z(t)\big) \big\ra_\Fock \Big) \\
&= \wick \Big( \big\la (-i {\scrH})\, \Upsilon\big(z(t)\big) \big| \otimes \big| \Upsilon\big(z(t)\big) \big\ra_\Fock
+ \big\la \Upsilon\big(z(t)\big) \big| \otimes \big| (-i {\scrH}) \Upsilon\big(z(t)\big) \big\ra_\Fock \Big) \wick \: \Delta t \\
&\qquad + \O\big( (\Delta t)^2 \big) \:,
\end{align*}
where the operator~${\scrH}$ 
is defined by
\begin{align*}
{\myscr{H}} &= i a^\dagger_l \:{}^l \!A_j a^j + i a^\dagger_l \:\big( {}^l\!
B_{jk} \,a^j a^k + {}^l\! B^j_k\, \overline{a}_j a^k
+ {}^l\! B^{jk}\, \overline{a}_j \:\overline{a}_k \big) \\
&\quad\; + i E_j a^j + i \big( F_{jk} \,a^j a^k + F^j_k\, \overline{a}_j a^k
+ F^{jk}\, \overline{a}_j \:\overline{a}_k \big) \:.
\end{align*}
By decomposing the time evolution from~$\tin$ to a later time~$t$ into time evolutions
on small time intervals~$\Delta t$ and taking the limit~$\Delta t \rightarrow 0$, one finds
that~${\mathfrak{L}}$ is obtained from~${\scrH}$ by exponentiating,
\beq \label{mathfrakLexp}
{\mathfrak{L}}(t) = e^{-i (t-\tin)\: {\scrH}} \:.
\eeq

Due to the complex conjugated field operators, the operator~${\scrH}$ does not act
on the Fock space~$\Fock$, but it acts instead on the tensor product~$\Fock^* \otimes \Fock$ and mixes
the holomorphic and anti-holomorphic components.
In order to obtain a corresponding holomorphic time evolution, it is a canonical procedure to simply replace the
complex conjugations by adjoints. 
We thus introduce the {\em{Hamiltonian}} $H$ as
\beq \label{Hdef}
\begin{split}
H &= i a^\dagger_l \:{}^l \!A_j a^j + i a^\dagger_l \:\big( {}^l\!
B_{jk} \,a^j a^k + {}^l\! B^j_k\, a^\dagger_j \,a^k
+ {}^l\! B^{jk}\, a^\dagger_j \:a^\dagger_k \big) \\
&\quad\:
+i E_j a^j + i \big( F_{jk} \,a^j a^k + F^j_k\, a^\dagger_j \,a^k
+ F^{jk}\, a^\dagger_j \:a^\dagger_k \big)
\::\: \Fock \rightarrow \Fock \:.
\end{split}
\eeq
Let us verify that this Hamiltonian is a symmetric operator on~$\Fock$:
We apply the conservation law for the Fock space dynamics
of Theorem~\ref{thmNunitt}.
To first order in~$\Delta t$, we thus obtain
\beq \label{infcons}
\wick \Big( \big\la {\scrH}\, \Upsilon\big(z(t)\big) \big| \Upsilon\big(z(t)\big) \big\ra_\Fock
+ \big\la \Upsilon\big(z(t)\big) \big| {\scrH} \Upsilon\big(z(t)\big) \big\ra_\Fock \Big) \wick
= 0 \:.
\eeq
By definition of the complex conjugate field operators, the last summand can be
rewritten as
\begin{align}
&\wick \big\la \Upsilon\big(z(t)\big) \big| (-i {\scrH}) \Upsilon\big(z(t)\big) \big\ra_\Fock \wick \notag \\
&= \big\la \Upsilon\big(z(t)\big) \big| \big( a^\dagger_l \:{}^l \!A_j \,a^j + a^\dagger_l \:\big( {}^l\!
B_{jk} \,a^j a^k \big) \Upsilon\big(z(t)\big) \big\ra_\Fock \notag \\
&\quad\:+ \big\la a_j \Upsilon\big(z(t)\big) \,\big|\, a^\dagger_l\: {}^l\! B^j_k\, a^k \Upsilon\big(z(t)\big) \big\ra_\Fock
+ \big\la a_j a_k \Upsilon\big(z(t)\big) \,\big|\, a^\dagger_l\: {}^l\! B^{jk} \,\Upsilon\big(z(t)\big) \big\ra_\Fock 
+ \cdots \notag \\
&= \big\la \Upsilon\big(z(t)\big) \big| \big( a^\dagger_l \:{}^l \!A_j \,a^j + a^\dagger_l \: {}^l\!
B_{jk} \,a^j a^k \big) \Upsilon\big(z(t)\big) \big\ra_\Fock \notag \\
&\quad\:+ \big\la \Upsilon\big(z(t)\big) \,\big|\, a^\dagger_j a^\dagger_l\: {}^l\! B^j_k\, a^k\,
\Upsilon\big(z(t)\big) \big\ra_\Fock
+ \big\la \Upsilon\big(z(t)\big) \,\big|\, a^\dagger_j a^\dagger_k a^\dagger_l\: {}^l\! B^{jk} \,\Upsilon\big(z(t)\big) \big\ra_\Fock + \cdots \notag \\
&= \big\la \Upsilon\big(z(t)\big) \,\big|\, (-i H)\, \Upsilon\big(z(t)\big) \big\ra_\Fock \:, \label{Hcompute}
\end{align}
where for brevity we only considered the operators in the first line of~\eqref{Hdef}.
The computation works the same way for the operators in the second line.
Treating the first summand in~\eqref{infcons} similarly, we obtain
\[ \eqref{infcons} = \big\la (-i H)\, \Upsilon\big(z(t)\big) \big| \Upsilon\big(z(t)\big) \big\ra_\Fock
+ \big\la \Upsilon\big(z(t)\big) \big| (-i H) \Upsilon\big(z(t)\big) \big\ra_\Fock = 0 \:. \]
Applying the polarization lemma (Lemma~\ref{lemmapolarize}), we obtain the following result:
\begin{Thm} \label{thmHsymm}
The Hamiltonian $H$ defined by~\eqref{Hdef} is a symmetric operator on
the Fock space~$\Fock$.
\end{Thm}

\begin{Def} \label{defhol}
The {\bf{holomorphic approximation}} is defined as the unitary
time evolution generated by the Hamiltonian $H$, i.e.
\[ z(t) = S(t,\tin)\, z(\tin)\qquad \text{with} \qquad S(t,t') := e^{-i (t-t') H} \::\: \Fock \rightarrow \Fock\:. \]
\end{Def} \noindent
Denoting the holomorphic time evolution by~$S$ is motivated by the fact that
the operator~$S(\tout, \tin)$ can be identified with the usual {\em{scattering operator}} of quantum field theory.

\subsection{Corrections to the Holomorphic Approximation} \label{secerror}
We now give a systematic procedure for describing the error of the holomorphic approximation.
On the time step from~$t$ to~$t+\Delta t$, the error of the holomorphic approximation is given
by~$E(t)\: \Delta t$, where the error term~$E(t)$ is the operator
\beq \begin{split} \label{EtPhi}
E(t)\: \big\la \Phi(t) \big| \otimes \big| \tilde{\Phi}(t) \big\ra_\Fock
  &:= \wick \;i \Big( \big\la {\scrH}\, \Phi(t) \big| \otimes \big| \tilde{\Phi}(t) \big) \big\ra_\Fock
-\big\la \Phi(t) \big| \otimes \big| {\scrH} \,\tilde{\Phi}(t) \big\ra_\Fock \Big) \wick \\
&\quad -i \Big( \big\la {H} \Phi(t) \big| \otimes \big| \tilde{\Phi}(t) \big\ra_\Fock
- \big\la \Phi(t) \big| \otimes \big| {H} \tilde{\Phi}(t) \big\ra_\Fock \Big) \:.
\end{split}
\eeq
For finite times, the error can be obtained by integrating this expression in a Dyson series:
\begin{Thm} {\bf{(corrections to holomorphic approximation)}} \label{thmerror}
Denoting the holomorphic time evolution on~$\Fock^* \times \Fock$ by~$V(t)$, i.e.
\[ V(t)\: \big\la \Phi \big| \otimes \big| \tilde{\Phi} \ra_\Fock
:= \big\la e^{-i t {H}} \Phi \big| \otimes \big| e^{-i t {H}} \tilde{\Phi} \ra_\Fock \:, \]
the dynamics described by~${\mathfrak{L}}$, \eqref{mathfrakLexp}, can be written as
\beq \label{dyson}
\begin{split}
{\mathfrak{L}}(t) &= V(t) + \int_{\tin}^t V(t-\tau)\: E(\tau)\: V(\tau)\: d\tau \\
&\qquad + \int_{\tin}^t d\tau_1 \int_{\tin}^{\tau_1} d\tau_2\:
V(t- \tau_1)\: E(\tau_1)\:V(\tau_1-\tau_2)\: E(\tau_2)\: V(\tau_2) + \cdots \:.
\end{split}
\eeq
\end{Thm}
\Proof In order to compare the exact dynamics~${\mathfrak{L}}(t)$ with the
approximate dynamics~$V(t)$, we go to the interaction picture, taking~$V(t)$
as the ``free'' dynamics. Thus setting
\beq \label{ipicture}
\la \Phi |\otimes| \tilde{\Phi} \ra^\text{I}(t) := V(t)^{-1} \: \la \Phi(t) |\otimes| \tilde{\Phi}(t) \ra_\Fock
= V(t)^{-1} \: {\mathfrak{L}}(t)\: \la \Phi(\tin) |\otimes| \tilde{\Phi}(\tin) \ra_\Fock \:,
\eeq
the dynamics in the interaction picture is
\begin{align*}
\partial_t  \la \Phi |\otimes| \tilde{\Phi} \ra^\text{I}(t) 
&= i\, V(t)^{-1} \:\Big( 
\wick \big\la {\scrH} \,\Phi(t) \big| \otimes \big| \tilde{\Phi}(t) \big\ra_\Fock \wick
- \wick \big\la \Phi(t) \big| \otimes \big| {\scrH} \,\tilde{\Phi}(t) \big\ra_\Fock \wick \\
&\qquad\qquad\;\;\: - \big\la H\,\Phi(t) \big| \otimes \big| \tilde{\Phi}(t) \big\ra_\Fock
\,-\: \big\la \Phi(t) \big| \otimes \big| H\,\tilde{\Phi}(t) \big\ra_\Fock \Big) \:.
\end{align*}
This equation can be written in a shorter form as
\beq \label{dtint}
\partial_t  \la \Phi |\otimes| \tilde{\Phi} \ra^\text{I}(t) = E^\text{I}(t)\: \la \Phi |\otimes| \tilde{\Phi} \ra^\text{I}(t) 
\qquad \text{with} \qquad E^\text{I}(t):= V(t)^{-1}\, E(t)\, V(t)\:.
\eeq
This ODE can be solved iteratively by
\begin{align*}
&\la \Phi |\otimes| \tilde{\Phi} \ra^\text{I}(t) \\
&= \bigg( \1 + \int_{\tin}^t E^\text{I}(\tau)\: V(\tau)\: d\tau
+ \int_{\tin}^t d\tau_1 \int_{\tin}^{\tau_1} d\tau_2\:
E^\text{I}(\tau_1)\:E^\text{I}(\tau_2) + \cdots \bigg)
\la \Phi |\otimes| \tilde{\Phi} \ra^\text{I}(\tin) \:.
\end{align*}
Transforming back to the Schr\"odinger picture gives the result.
\QED

We finally rewrite this result in terms of the effect on observables.
Our construction is based on the following observation:
\begin{Lemma} The expectation value~\eqref{eval}
of the error term~\eqref{EtPhi} can be written as the expectation value
of an operator involving commutators. More precisely,
\[ {\mathcal{O}} \Big( E(t)\: \big\la \Phi(t) \big| \otimes \big| \tilde{\Phi}(t) \big\ra_\Fock \Big)
= {\mathcal{C}}({\mathcal{O}}) \Big( \big\la \Phi(t) \big| \otimes \big| \tilde{\Phi}(t) \big\ra_\Fock \Big) \:, \]
where~${\mathcal{C}}({\mathcal{O}})$ is the operator
\begin{align*} 
{\mathcal{C}}({\mathcal{O}}) &:= \big[ a^\dagger_j , {\mathcal{O}} \big]\: a^\dagger_l\: {}^l\! B^j_k\, a^k
+ \big[ a^\dagger_j a^\dagger_k, {\mathcal{O}} \big]\: a^\dagger_l\: {}^l\! B^{jk} \\
&\quad\; + a_k^\dagger\, \overline{{}^l\! B^j_k}\, a^l\: \big[ a^j , {\mathcal{O}} \big]
+ a^l\: \overline{{}^l\! B^{jk}}\: \big[ a^k a^j, {\mathcal{O}} \big] \\
&\quad\; + \big[ a^\dagger_j , {\mathcal{O}} \big]\: F^j_k\, a^k
+ \big[ a^\dagger_j a^\dagger_k, {\mathcal{O}} \big]\:  F^{jk} \\
&\quad\; + a^\dagger_k\,\overline{F^j_k}\, \: \big[ a^j , {\mathcal{O}} \big]\: 
+ \overline{F^{jk}} \:\big[ a^k a^j, {\mathcal{O}} \big] \:.
\end{align*}
\end{Lemma}
\Proof Reconsidering the computation~\eqref{Hcompute}, we obtain
\begin{align*} 
&{\mathcal{O}} \Big( \wick \big\la \Phi(t) \,\big|\, (-i {\scrH}) \,\tilde{\Phi}(t) \big\ra_\Fock \wick \Big) \\
&= \big\la \Phi(t) \,\big|\,{\mathcal{O}}\, \big( a^\dagger_l \:{}^l \!A_j a^j + a^\dagger_l \: {}^l\!
B_{jk} a^j a^k \big) \,\tilde{\Phi}(t) \big\ra_\Fock \\
&\quad\:+ \big\la a_j \, \Phi(t) \,\big|\,{\mathcal{O}}\, a^\dagger_l\: {}^l\! B^j_k\, a^k \,\tilde{\Phi}(t) \big\ra_\Fock
+ \big\la a_j a_k \,\Phi(t) \,\big|\,{\mathcal{O}}\, a^\dagger_l\: {}^l\! B^{jk} \,\tilde{\Phi}(t)
\big\ra_\Fock + \cdots \\
&= \big\la \Phi(t) \,\big|\, {\mathcal{O}}\, (-i H)\, \Upsilon\big(z(t)\big) \big\ra_\Fock \\
&\quad\; + \big\la \Phi(t) \,\big|\, \big[ a^\dagger_j , {\mathcal{O}} \big]\: a^\dagger_l\: {}^l\! B^j_k\, a^k\,
\,\tilde{\Phi}(t)\big\ra_\Fock
+ \big\la \Phi(t) \,\big|\, \big[ a^\dagger_j a^\dagger_k, {\mathcal{O}} \big]\: a^\dagger_l\: {}^l\! B^{jk} \,\tilde{\Phi}(t) \big\ra_\Fock + \cdots \:,
\end{align*}
where for brevity we again considered only the operators in the first line of~\eqref{Hdef}.
Treating the other summands in~\eqref{EtPhi} similarly, we obtain
\begin{align*}
& {\mathcal{O}} \Big( E(t)\: \big\la \Phi(t) \big| \otimes \big| \tilde{\Phi}(t) \big\ra_\Fock \Big) \\
&= \big\la \Phi(t) \,\big|\, \big[ a^\dagger_j , {\mathcal{O}} \big]\: a^\dagger_l\: {}^l\! B^j_k\, a^k\,
\,\tilde{\Phi}(t)\big\ra_\Fock
+ \big\la \Phi(t) \,\big|\, \big[ a^\dagger_j a^\dagger_k, {\mathcal{O}} \big]\: a^\dagger_l\: {}^l\! B^{jk} \,\tilde{\Phi}(t) \big\ra_\Fock \\
&\quad\; - \big\la  \big[ a^\dagger_j , {\mathcal{O}} \big]\: a^\dagger_l\: {}^l\! B^j_k\, a^k\,
\Phi(t) \,\big|\,\tilde{\Phi}(t)\big\ra_\Fock
- \big\la \big[ a^\dagger_j a^\dagger_k, {\mathcal{O}} \big]\: a^\dagger_l\: {}^l\! B^{jk} \,\Phi(t) \,\big|\, \tilde{\Phi}(t) \big\ra_\Fock + \cdots \:.
\end{align*}
Bringing the operators in the last line to the right side by taking the adjoints gives the result.
\QED

\begin{Thm} {\bf{(iterated commutators)}} \label{thmcommute}
\[ {\mathcal{O}} \Big( {\mathfrak{L}}(t)\: \la \Phi(\tin) |\otimes| \tilde{\Phi}(\tin) \ra_\Fock \Big)
= {\mathcal{O}}' \Big( {\mathfrak{L}}(t)\: \la \Phi(\tin) |\otimes| \tilde{\Phi}(\tin) \ra_\Fock \Big) \:, \]
where~${\mathcal{O}}'$ is the transformed observable
\begin{align*}
&{\mathcal{O}}' = {\mathcal{O}}
+ \int_{\tin}^t S(t,\tau)\: {\mathcal{C}}\big( S(\tau,t)\,{\mathcal{O}}\, S(t,\tau) \big)\: S(\tau,t)\: d\tau \\
& + \int_{\tin}^t d\tau_1 \int_{\tin}^{\tau_1} d\tau_2\:
S(t,\tau_2)\: {\mathcal{C}} \Big( S(\tau_2,\tau_1)\: {\mathcal{C}}\big( S(\tau_1,t)\,{\mathcal{O}}\, S(t,\tau_1) \big)\: S(\tau_1,\tau_2) \Big)\: S(\tau_2, t) + \cdots \:.
\end{align*}
\end{Thm}
\Proof As in the proof of Theorem~\ref{thmerror}, we again work in
the interaction picture~\eqref{ipicture} and set
\[ {\mathcal{O}}^\text{I}(t) = S(t)^{-1}\, {\mathcal{O}}\, S(t) \:. \]
Then
\[ {\mathcal{O}} \big( {\mathfrak{L}}(t)\: \la \Phi(\tin) |\otimes| \tilde{\Phi}(\tin) \ra_\Fock \big)
= {\mathcal{O}}^\text{I}(t) \Big( \big\la \Phi(\tin) \big|\otimes \big| \tilde{\Phi}(\tin) \big\ra_\Fock^\text{I}(t) \Big) \:. \]
Using again~\eqref{dtint}, we obtain
\begin{align*}
&{\mathcal{O}} \big( {\mathfrak{L}}(t)\: \la \Phi(\tin) |\otimes| \tilde{\Phi}(\tin) \ra_\Fock \big) \\
&= {\mathcal{O}}^\text{I}(t) \bigg(
\big\la \Phi(\tin) \big|\otimes \big| \tilde{\Phi}(\tin) \big\ra_\Fock^\text{I}(\tin) 
+ \int_{\tin}^t \partial_\tau \big\la \Phi(\tin) \big|\otimes \big| \tilde{\Phi}(\tin) \big\ra_\Fock^\text{I}(\tau) \: d\tau \bigg) \\
&= {\mathcal{O}}^\text{I}(t) \bigg(
\big\la \Phi(\tin) \big|\otimes \big| \tilde{\Phi}(\tin) \big\ra_\Fock^\text{I}(\tin) 
+ \int_{\tin}^t E^\text{I}(\tau)\: \big\la \Phi(\tin) \big|\otimes \big| \tilde{\Phi}(\tin) \big\ra_\Fock^\text{I}(\tau) \bigg) \\
&= {\mathcal{O}} \Big( V(t)\:
\big\la \Phi(\tin) \big|\otimes \big| \tilde{\Phi}(\tin) \big\ra_\Fock \Big)
+ \int_{\tin}^t {\mathcal{C}} \big( {\mathcal{O}}^\text{I}(\tau) \big) \Big( \big\la \Phi(\tin) \big|\otimes \big| \tilde{\Phi}(\tin) \big\ra_\Fock^\text{I}(\tau) \Big) \:.
\end{align*}
This relation can again be iterated. Transforming back to the Schr\"odinger picture gives the result.
\QED
The corrections in Theorem~\ref{thmerror} as well as the formula in Theorem~\ref{thmcommute}
will be explained and discussed in Appendix~\ref{seccompare}.

\section{Comparison with Classical $\phi^4$-Theory} \label{secphi4}
We now illustrate our constructions by comparing the obtained structures
with those of classical field theory. In order to work in a concrete example, we consider the
classical $\phi^4$-theory in Minkowski space. As we shall see, the conserved quantities of
classical field theory and the resulting bilinear forms bear a striking similarity to the
structures found for causal variational principles. But there are also major differences,
which indeed make it impossible to apply most of our constructions to classical field theory.

\subsection{Preliminaries} \label{sec2}
We introduce classical $\phi^4$-theory in the Lagrangian formulation.
We consider the Lagrangian~${\mathcal{L}}$
\[ {\mathcal{L}}(\phi, \partial \phi) = \frac{1}{2}\: (\partial_j \phi) (\partial^j \phi) - \frac{\lambda}{4!}\: \phi^4 
\qquad \text{for $\lambda>0$}\:, \]
where~$\phi$ is a real-valued scalar field.
Integrating the Lagrangian over Minkowski space~$(\scrM, g)$ gives the action~${\mathcal{S}}$,
\[ {\mathcal{S}} = \int_\scrM {\mathcal{L}}(\phi, \partial \phi)\: d^4x \:. \]
Considering critical points of the action, one obtains the Euler-Lagrange (EL) equations
\beq \label{cfe}
\Box \phi = -\frac{\lambda}{6}\, \phi^3
\eeq
(where~$\Box = \partial_t^2 - \Delta_{\R^3}$ is the wave operator).
According to Noether's theorem, the symmetries of the Lagrangian correspond to
conserved quantities. In particular, the symmetry under time translations gives rise
to the conserved classical energy~$E$,
\beq \label{E0}
E(\phi) = \int_{t=T} 
\left( \frac{1}{2}\: \dot{\phi}^2 + \frac{1}{2}\: |\nabla \phi|^2 + \frac{\lambda}{4!}\:
\phi^4 \right) d^3x \:.
\eeq

Given smooth and compactly supported initial data~$(\phi, \partial_t \phi)|_{\tin}
\in C^\infty_0(\R^3, \R^2)$ at some initial time~$\tin$,
the Cauchy problem for the nonlinear wave equation~\eqref{cfe} is locally well-posed.
Due to finite propagation speed, the solution has spatially compact support in the
sense that it has compact support at any later time.
Moreover, the solution exists globally and is smooth for sufficiently small initial data.
With our goal of getting a simple explicit example, it suffices to restrict attention to
a finite-dimensional manifold~$\calB \subset \Cisc(\scrM, \R)$ of global solutions of the
nonlinear wave equation, which are all smooth and have spatially compact support.
Then for any base point~$\psi \in \calB$, the tangent space~$T_{\psi}\calB
\subset \Cisc(\scrM, \R)$ is formed of
a finite-dimensional subspace of solutions of the linearized field equations
\beq \label{linfield}
\Box \tilde{\phi} = -\frac{\lambda}{2}\, \psi^2\, \tilde{\phi}\:.
\eeq
On~$T_{\psi}\calB$ one has the following structures: First, the symplectic form defined by
\beq \label{symplectic}
\sigma_\psi(\tilde{\chi}, \tilde{\phi}) := \int_{t=T} \Big( (\partial_t \tilde{\chi})\: \tilde{\phi}
- \tilde{\chi}\: (\partial_t \tilde{\phi}) \Big) d^3x
\eeq
is time independent. This can be verified either explicitly by
differentiating with respect to~$T$, using~\eqref{linfield} and integrating by parts,
or else more abstractly by considering the boundary terms
arising in the variation of the action in a finite time interval
(see for example~\cite[\S2.3]{deligne+freed}).
Next, taking a functional derivative of the energy~\eqref{E0}, one gets the conserved quantity
\beq \label{I1phi4}
\gamma_\psi(\tilde{\phi}) := \frac{1}{2} \int_{t=T} \left( \dot{\psi}\, \dot{\tilde{\phi}} + \nabla \psi \cdot \nabla \tilde{\phi}
+ \frac{\lambda}{3!}\: \psi^3\, \tilde{\phi} \right) d^3x \:.
\eeq
By taking another functional derivative, one
gets an inner product on the linearized solutions. However, the form of this inner product
depends on time, as we now explain in detail:
We consider a two-parameter family~$\phi_{r,s}$ of solutions of the Cauchy problem defined by
the initial conditions
\beq \label{init}
\phi_{r,s}|_{t=\tin} = \psi + r\, \chi^0 + s\, \phi^0 \qquad \text{and} \qquad
\partial_t \phi_{r,s}|_{t=\tin} = \partial_t \psi + r\, \chi^1 + s\, \phi^1
\eeq
(with~$\chi^0, \chi^1, \phi^0, \phi^1 \in C^\infty_0(\R^3)$).
Then the first derivatives give rise to linearized solutions
\[ \tilde{\chi} := \partial_r \psi_{r,s}|_{t=\tin} \qquad \text{and} \qquad \tilde{\phi} := \partial_s \psi_{r,s}|_{t=\tin} \:. \]
We next introduce their energy inner product by
\begin{align}
(\tilde{\chi}, \tilde{\phi})_\psi &\!:= \partial_r \partial_s E \big( \psi_{r,s} \big) \big|_{r=s=0} \notag \\
&= \int_{t=T} 
\left( \dot{\tilde{\chi}} \dot{\tilde{\phi}} + (\nabla \tilde{\chi}) \cdot (\nabla \tilde{\phi})
+ \frac{\lambda}{2}\: \psi^2\: \tilde{\psi}\, \tilde{\phi} \right) d^3x 
+ \gamma_\psi \Big( \partial_r \partial_s \psi_{r,s} \big|_{r=s=0} \Big) \:. \label{en}
\end{align}
The integral in~\eqref{en} has the standard form of an energy, being an integral over
an energy density. It coincides with the energy corresponding to the effective Lagrangian
\beq \label{Leff}
{\mathcal{L}}_\psi(\tilde{\phi}, \partial \tilde{\phi})
= \frac{1}{2}\: (\partial_j \tilde{\phi}) (\partial^j \tilde{\phi})
-\frac{\lambda}{4}\: \psi^2\: \tilde{\phi}^2 \:.
\eeq
However, the corresponding energy is conserved only if the resulting potential~$\psi^2$ is
time-independent. In the general time-dependent setting, however, the energy
corresponding to~\eqref{Leff} is {\em{not}} conserved, explaining
the appearance of the additional term~$\gamma_\psi$ in~\eqref{en}.
In order to compute~$\partial_r \partial_s \psi_{r,s}$, we differentiate~\eqref{cfe}
and~\eqref{init} to obtain the Cauchy problem
\[ \Big( \Box + \frac{\lambda}{3}\: \psi^2 \Big) \partial_r \partial_s \psi_{r,s} \big|_{r=s=0} =
-\frac{\lambda}{2}\, \psi\, \tilde{\chi}\, \tilde{\phi} \:,\qquad
\partial_r \partial_s \phi \big|_{t=\tin} = 0 \:. \]
The solution of this Cauchy problem can be expressed with the help of Green's operators by
\beq \label{exsol}
\big(\partial_r \partial_s \phi\big)(x) =
\frac{\lambda}{2} \int_{\{y^0 >\tin\}} S_\psi(x,y) \:
\big(\psi \,\tilde{\chi}\, \tilde{\phi} \big)(y)\, \: d^4y \:,
\eeq
where~$S_\psi$ is the retarded Green's operator of the linearized wave equation, i.e.
\[ \Big( \Box + \frac{\lambda}{3}\: \psi^2 \Big) S_\psi(x,y) = -\delta^4(x-y) \:. \]
Extending the linearized solutions by zero to times~$t<\tin$,
the $y$-integration can be carried out over all of Minkowski space.
Introducing the operator notation
\[ (S_\psi \phi)(x) := \int_\scrM S_\psi(x,y)\: \phi(y)\: d^4y \]
and using~\eqref{exsol} in~\eqref{en} gives the formula
\begin{align}
(\tilde{\chi}, \tilde{\phi})_\psi &= \int_{t=T} 
\left( \dot{\tilde{\chi}} \dot{\tilde{\phi}} + (\nabla \tilde{\chi}) \cdot (\nabla \tilde{\phi})
+ \frac{\lambda}{2}\: \psi^2\: \tilde{\psi}\, \tilde{\phi} \right) d^3x \label{en1} \\
&\quad + \frac{\lambda}{2} \:\gamma_\psi \Big( 
S_\psi \big(\psi \,\tilde{\chi}\, \tilde{\phi} \big) \Big) \:. \label{en2}
\end{align}
At initial time~$\tin$, the summand~\eqref{en2} vanishes, so that we obtain
the form of the energy as suggested from~\eqref{Leff}.
Since~$\lambda>0$, the bilinear form~$(.,.)_\psi$ is positive definite at time~$\tin$
and thus defines a scalar product.
As a consequence of~\eqref{en2}, the inner product~$(.,.)_\psi$ is
independent of~$T$. We note that, more abstractly, $(.,.)_\psi$ can be understood as the
symmetrized covariant derivative of~$\gamma_\psi$ on~$\calB$
with a connection which is flat at time~$\tin$.

\subsection{Comparison with the Structures of Causal Variational Principles}
The resulting structures are
\begin{align*}
\text{conserved energy~\eqref{E0}} && \qquad E &: \calB \rightarrow \R \\
\text{conserved one-form~\eqref{I1phi4}} &&\qquad \gamma_\psi &: T_\psi\calB \rightarrow \R \\
\text{symplectic form~\eqref{symplectic}} &&\qquad \sigma_\psi
&: T_\psi\calB \times T_\psi\calB \rightarrow \R \\
\text{scalar product~\eqref{en1}, \eqref{en2}}  && \qquad (.,.)_\psi &: T_\psi\calB \times T_\psi\calB \rightarrow \R \:.
\end{align*}
This is very similar to the structures on the jet spaces in the previous sections.
However, there are also differences, mainly related to the fact that
the inner solutions have no correspondence
to classical field theory. More precisely, the analogy and differences
are as follows:
\begin{enumerate}[leftmargin=2em]
\item The conservation of the energy~$E$ bears some similarity with the
nonlinear conservation law of Theorem~\ref{thmosinoconserve} and Corollary~\ref{cornl}.
However, the physical interpretation is different, because~$(\Pert(w), \Pert(w))$
is to be regarded as a probability, not an energy. Nevertheless, from the mathematical
or formal point of view, these conservation laws are analogous in being
positive functionals on the space of nonlinear solutions.
\item The conservation of~$\gamma_\psi$, being the functional derivative of~$E$, is similar to the
conserved one-form~\eqref{gamma} in~\eqref{cI1nl}.
\item The conserved symplectic form~$\sigma_\psi$ corresponds precisely
to the symplectic form; see~\eqref{sigma} and~\eqref{cI2asnl}.
\item The scalar product~$(.,.)_\psi$ on linearized solutions can be regarded as the
analog of the surface layer inner product~\eqref{sprod} in~\eqref{cI2symmnl}.
The volume term~\eqref{en2} plays a similar role as the right side of~\eqref{cI2symmnl}.
\end{enumerate}
The main difference between the structures in classical field theory and those of causal variational
principles is that, 
in contrast to the bilinear form~$(\Pert(w), \Pert(w))$, the energy~$E$
is {\em{not quadratic}} in~$\phi$ and thus does not gives rise to a scalar product on the solution space.
More precisely, $E$ is quadratic only if no interaction is present, in which case we obtain the corresponding
scalar product
\[ (\tilde{\chi}, \tilde{\phi}) = \int_{t=T} 
\left( \dot{\tilde{\chi}} \dot{\tilde{\phi}} + (\nabla \tilde{\chi}) \cdot (\nabla \tilde{\phi}) \right) d^3x \:. \]
In particular, there is a well-defined scalar product on the incoming and outgoing scattering states.
Using the constructions in Section~\ref{seccomplexlinear}, the symplectic form
gives rise to a canonical complex structure on the asymptotic states.
However, there is no scalar product at intermediate times, making it impossible
to apply the constructions in Section~\ref{secholo}.
We regard this shortcoming as a major structural difference between classical field theory
and causal variational principles.
This shortcoming of classical field theory also shows that causal variational principles are
distinguished by providing precisely the structures needed for a
probabilistic interpretation and a formulation in terms of bosonic Fock spaces.

\appendix
\section{General Derivation of the Nonlinear Conservation Law} \label{appconserve}
The goal of this appendix is to show that the nonlinear conservation law
can be arranged in great generality. The construction also sheds light on
the nature of this conservation law. We let~$\rho$ and~$\tilde{\rho}$ be two measures on~$\F$
(not necessarily critical), and denote their supports by~$M:= \supp \rho$ and~$\tilde{M}:= \supp \tilde{\rho}$.
Moreover, we assume that~$\Phi : \tilde{M} \rightarrow M$ is a measurable bijection
whose inverse~$F:=\Phi^{-1} : M \rightarrow \tilde{M}$ is also measurable
(the existence of such a bijection will be shown below).
Given a compact subset~$\Omega \subset M$, in generalization of~\eqref{osinl} we set
\beq \label{osinlgen}
\gamma^\Omega(\tilde{\rho}, \rho) := \int_{F(\Omega)} d\tilde{\rho}(x) \int_{M \setminus \Omega}
d\rho(y) \: \L(x,y) - \int_{\Omega} d\rho(x) \int_{\tilde{M} \setminus F(\Omega)} d\tilde{\rho}(y)\: \L(x,y)
\eeq
In order to characterize when this surface layer integral vanishes, we 
introduce a measure~$\nu$ on~$M$ and a measure~$\tilde{\nu}$ on~$\tilde{M}$ by
\begin{align*}
d\nu(x) &:= \bigg( \int_{\tilde{M}} \L(x,y)\: d\tilde{\rho}(y) \bigg) \: d\rho(x) \\
d\tilde{\nu}(x) &:= \bigg( \int_M \L(x,y)\: d\rho(y) \bigg) \: d\tilde{\rho}(x) \:.
\end{align*}
Intuitively speaking, these measures describe how the measures~$\rho$ and~$\tilde{\rho}$
are connected to each other by the Lagrangian. We refer to them as the {\em{correlation measures}}.
Then we can rewrite~\eqref{osinlgen} as
\begin{align*}
\gamma^\Omega(\tilde{\rho}, \rho) &= \int_\Omega d\big(\Phi_*\tilde{\rho}\big)(x) \int_{M \setminus \Omega} \!\!\!\!
d\rho(y) \: \L\big(F(x),y \big) - \int_{\Omega} d\rho(x) \int_{M \setminus \Omega} \!\!\!\!d\big(\Phi_*\tilde{\rho} \big)(y)\:
\L\big(x,F(y)\big) \\
&= \int_\Omega d\big(\Phi_*\tilde{\rho}\big)(x) \int_{M}
d\rho(y) \: \L\big(F(x),y \big) - \int_{\Omega} d\rho(x) \int_{M} d\big(\Phi_*\tilde{\rho} \big)(y)\:
\L\big(x,F(y)\big) \\
&= \int_\Omega d\big(\Phi_*\tilde{\rho}\big)(x) \int_{M}
d\rho(y) \: \L\big(F(x),y \big) - \int_{\Omega} d\rho(x) \int_{\tilde{M}} d\tilde{\rho}(y)\:
\L(x,y) \\
&= \int_\Omega d\big(\Phi_*\tilde{\nu}\big)(x) - \int_{\Omega} d\nu(x)
= \big( \Phi_* \tilde{\nu} - \nu \big)(\Omega) \:.
\end{align*}
We thus obtain the following result:
\begin{Prp} \label{Greene}
The surface layer integral~\eqref{osinlgen} vanishes for every compact~$\Omega \subset M$
if and only if the correlation measures are mapped to each other,
\beq \label{mutilmu}
\nu = \Phi_* \tilde{\nu} \:.
\eeq
\end{Prp}

The remaining question is whether~\eqref{mutilmu} can be arranged by a suitable choice of~$\Phi$. We consider the measurable and smooth cases separately. 
In the measurable setting, we shall assume that the measures $\nu$ and $\tilde{\nu}$ are both {\emph{non-atomic}}. In that case one can use the methods used in the proof of ~\cite[Lemma~1.4]{continuum} to 
construct a measurable and invertible map~$\Phi$ with measurable inverse such that~\eqref{mutilmu} holds.
Next, in the smooth setting, one can arrange~\eqref{mutilmu} by applying a result of Greene and Shiohama~\cite{greene-shiohama}, which generalizes to the non-compact setting the classical theorem of Moser on volume forms for compact manifolds.
We first quote this result and apply it afterward.
\begin{Prp} \label{prpgreene}
Let $M$ be a non-compact oriented manifold and let $\omega$ and $\tau$ be volume forms on $M$. Assume that
\[ \int_{M}\omega=\int_{M}\tau \leq \infty \:. \]
Furthermore, assume that each end of $M$ has finite $\omega$-volume if it has finite $\tau$-volume, and infinite $\omega$-volume if it has infinite $\tau$-volume. Then there exists a diffeomorphism $\Phi:M\to M$ such that 
\[ \Phi^{*}\omega = \tau \:. \]
\end{Prp}
In the setting of scattering systems in Minkowski space introduced in Section~\ref{secscatter}, 
this proposition has the following consequence:
\begin{Corollary}
For scattering systems in Minkowski space, there is a diffeomorphism~$\Phi : M \rightarrow \tilde{M}$
such that the correlation measures are mapped to each other as in~\eqref{mutilmu}.
\end{Corollary}
\Proof We need to verify that the assumptions of Proposition~\ref{prpgreene} are satisfied.
Since both~$M$ and~$\tilde{M}$ are by assumption diffeomorphic to~$\R^4$,
we have one asymptotic end. Therefore, it suffices to show the equivalence
\[ \nu(M) = \infty \qquad \Longleftrightarrow \qquad \tilde{\nu}(\tilde{M}) = 0 \:. \]
This equivalence follows immediately from the calculation
\begin{align*}
\nu(M) &= \int_M \bigg( \int_{\tilde{M}} \L(x,y)\: d\tilde{\rho}(y) \bigg) \: d\rho(x) \\
&= \int_{\tilde{M}} \bigg( \int_{M} \L(x,y)\: d\rho(x) \bigg) \: d\tilde{\rho}(y) = \tilde{\nu}(\tilde{M}) \;\in\;
\R^+_0 \cup \{\infty\} \:,
\end{align*}
where in the last line we applied Tonelli's theorem (i.e.\ the version of Fubini's theorem for non-negative functions)
and used that the Lagrangian is symmetric.
\QED

We finally explain what this result means infinitesimally.
Assume that~$\Phi$ is given. Then the measure~$\Phi_* \tilde{\rho}$ is supported on~$M$.
In the considered smooth setting, the measures can be related to each other by
\[ d \big(\Phi_* \tilde{\rho}\big) = f\: d\rho \]
with a smooth function~$f \in C^\infty(M, \R^+)$.
Then the measures~$\tilde{\rho}$ and~$\rho$ are again related by~\eqref{rhoFf} with~$F=\Phi^{-1}$.
If~$(f,F)$ can be connected to the identity by a smooth curve~$(f_\tau, F_\tau)$
(with~$(f_0, F_0) = (1, \text{id})$ and~$(f_1, F_1) = (f, F)$),
then the diffeomorphism~$M$ is described infinitesimally in~$\tau$ by a vector field~$u$
which is indeed the same as the vector field of the inner solution constructed in Corollary~\ref{corvmu}, and~$f-1$
goes over infinitesimally to the divergence of~$u$.

\section{Discussion of the Validity of the Approximations} \label{secapprox}
\subsection{Approximation of Small Inner Solutions} \label{secapproxinner}
In Section~\ref{secinnersmall} the approximation of small inner solutions
was introduced. We now explain why in the description of a scattering process,
this limiting case should be an extremely good approximation.
As explained above, inner solutions are not interesting to study by themselves,
because they simply describe infinitesimal symmetries of the universal measure.
In particular, when describing a physical system, it seems most convenient to
choose initial data in~$\Glin_{\rho, \sc}$ with no inner solutions present.
At later times, also for the outgoing dynamics in a scattering process,
there will be inner solutions present, and it is important to keep them into account.
Proceeding in this way, all the inner solutions
have been created either by removing the scalar components of the linearized
solutions (Corollary~\ref{cornoscalar}) or by arranging the nonlinear conservation law
(Theorem~\ref{thmosinoconserve}).
This makes it possible to determine the size of the inner solutions.
We now explain the findings and derive consequences.

Using the methods and results in~\cite{linhyp} one can construct solutions
of the vector component of the linearized field equations using jets with zero scalar components, i.e.
\beq \label{vector}
D_u \int_M (D_{1,v} + D_{2,v}) \L(x,y) \:d\rho(y) = 0 \qquad \text{for all~$\u \in \Jtest$} \:.
\eeq
The scalar component of the linearized field equations is
\beq \label{vec}
\int_M (D_{1,v} + D_{2,v}) \L(x,y) \:d\rho(y) = 0 \:.
\eeq
This equation will in general {\em{not}} be satisfied. But we can satisfy it by introducing
a scalar component~$b$. This leads to the equation
\[ \int_M (D_{1,v} + D_{2,v}) \L(x,y) \:d\rho(y) + \int_M (b(x)+b(y)) \L(x,y) \:d\rho(y) - b(x)\: \s = 0 \:. \]
Using the weak EL equations~\eqref{ELtest}, this equation can be simplified to
\beq \label{it1}
\int_M \L(x,y)\: b(y) \:d\rho(y) = \int_M D_{2,v} \L(x,y) \:d\rho(y) \:.
\eeq
Clearly, the scalar component~$b$ gives rise to an error term in the vector component of the linearized field equations,
which takes the form
\beq \label{it2}
\int_M D_{1,u} \L(x,y) \: b(y) \:d\rho(y) \:.
\eeq

Let us consider the scaling behavior of the terms in~\eqref{it1} and~\eqref{it2}
as worked out in~\cite{action} and~\cite[Appendix~A]{jacobson} for Dirac systems in Minkowski space.
Before beginning, we recall the parameters and their scalings.
We always work in natural units where~$\hbar=c=1$. Then the gravitational coupling
constant~$\kappa$ has dimension length squared. More precisely,
\[ \kappa \simeq \delta^2 \:, \]
where~$\delta \approx 1.6\cdot 10^{-35}\,\text{meters}$ denotes the {\em{Planck length}}.
The rest mass of the Dirac particles determines another length scale,
the {\em{Compton length}}~$m^{-1}$. Next, there is the {\em{regularization length}}~$\varepsilon$.
The simplest and most natural assumption is to identify the regularization length with the Planck length.
However, as is explained in detail in~\cite[Chapter~4]{cfs}, this assumption is too naive,
because the regularization length should be
much smaller than the Planck length. Therefore, we must treat~$\varepsilon$
and~$\delta$ as different parameters. We merely assume that
\[ \varepsilon \ll \delta \ll \frac{1}{m}\:. \]
Finally, there is the length scale~$l_{\text{\tiny{macro}}}$ of macroscopic physics.
Clearly, this length scale depends on the physical system under consideration. Since
energies much larger than the rest masses of the heaviest fermions are not accessible to
experiments, we always assume that
\[ \frac{1}{m} \lesssim l_{\text{\tiny{macro}}}\:. \]

The left side of~\eqref{it1} scales like the integral over the Lagrangian,
\[ \int_M \L(x,y)\: b(y) \:d\rho(y) \sim b\: \int_M \L(x,y) \: d\rho(y) \sim \s\: b \:, \]
where~$\s$ has the scaling behavior (see~\cite[eqs~(A.11) and~(A.17)]{jacobson};
for simplicity we choose the scaling parameters~$\lambda$ and~$\sigma$ to be one)
\beq \label{sscale}
\s \eqsim \frac{1}{\varepsilon^{8}}\:
\bigg( (\varepsilon m)^p + \Big( \frac{\varepsilon}{\delta} \Big)^{8-\hat{s}} \bigg) \qquad \text{with}
\qquad \hat{s} \in \{0,2\} \:.
\eeq
On the right side of~\eqref{it1}, however, the perturbation of the Lagrangian by the
matter fields and the bosonic fields comes into play. It has the scaling behavior
(see~\cite[Proposition~A.1]{jacobson})
\[ \int_M D_{2,v} \L(x,y) \:d\rho(y) \eqsim \frac{1}{\varepsilon^4} \:T(x) \:
\Big( \frac{\varepsilon}{\delta} \Big)^{4-s} \qquad \text{with}
\qquad s \in \{0,2,4\} \:, \]
where~$T$ is the energy-momentum tensor of the respective field.
The energy-momentum tensor typically scales like
\beq \label{Tscale}
T \eqsim \frac{m}{l_{\text{\tiny{macro}}}^3} \lesssim m^4 \:.
\eeq
Comparing these scalings, one concludes that the scalar component~$b$ scales like
\beq \label{bscal}
b \lesssim \frac{m \delta^4}{l_{\text{\tiny{macro}}}^3} \lesssim (m \delta)^4\:.
\eeq
We conclude that the construction before Corollary~\ref{cornoscalar} gives rise to
inner solutions which are by scaling factors of~$m \delta$ smaller than the
original jets in~$\Jlin_0$.

The back reaction of the scalar component on the vector component
as described by~\eqref{it2} and~\eqref{vector} has the following scaling behavior.
Consider the back reaction on the field equation for the electromagnetic field
(for other field equations, the scaling is similar).
The perturbation of the eigenvalues by the electromagnetic potential~$A$ scales
like~$|\delta \lambda| \sim A\, (\varepsilon \,l_{\text{\tiny{macro}}})^{-3}$
(for details see~\cite[Chapter~3]{cfs}).
Since the unperturbed eigenvalues scale like~$|\lambda| \sim \delta^{-4}\, \varepsilon^{-2}$, we
find that the variational derivatives in~\eqref{vector} scale like
\[ D_v \simeq \frac{\delta^4}{\varepsilon\, l_{\text{\tiny{macro}}}^3}\:A \:. \]
Comparing with~\eqref{bscal}, we conclude that~$b$ can be compensated by an electromagnetic
potential~$A$ with the scaling behavior
\[ A \simeq \varepsilon m \:. \]
This argument shows that the back reaction of~$b$
on the vector component of the jet~$\v$ is also extremely small.
This explains why we may disregard this back reaction in the approximation of small inner solutions.

For the inner solutions generated in order to satisfy the conservation law for
the nonlinear surface layer integral, one can argue similarly.
Indeed, in the proof of Theorem~\ref{thmosinoconserve} the scalar component
of the jet~$\v^{(p)}$ is multiplied by~$\s$, which has the scaling behavior~\eqref{sscale}.
The terms which need to be compensated, however, involve third variational
derivatives of the Lagrangian. As a consequence, the scalar component of~$\v^{(p)}$,
and consequently also the resulting inner solution, is again
by scaling factors of~$m \delta$ smaller than the jets in~$\Gamma_{\sc, 0}$.

\subsection{The Holomorphic Approximation} \label{seccompare}
In the holomorphic approximation introduced in Section~\ref{secholomorphic},
the dynamics of critical points of causal variational
principles can be described by a unitary time evolution on a bosonic Fock space
(see Definition~\ref{defhol}), giving a close connection to quantum field theory.
When working out physical applications, it is important to justify the holomorphic approximation.
Moreover, the errors of this approximations are of major interest because
they should give predictions for physical corrections to standard quantum field theory.
With this in mind, we conclude this paper with a discussion of the holomorphic
approximation and its corrections.

We first recall that for non-interacting systems, there is a canonical complex structure which
is preserved by the time evolution (see Section~\ref{seclinear}). As a consequence, the
holomorphic approximation is exact (as is also obvious from Theorem~\ref{thmerror},
keeping in mind that for linear systems the error~$E(t)$ in~\eqref{EtPhi} vanishes).
The question whether the holomorphic approximation is also exact for interacting systems
is equivalent to asking for the existence of a holomorphic connection (see Definition~\ref{defcc}).
The answer to this question depends on the form of the interaction (see Propositions~\ref{prpindef}
and~\ref{prpalmost}), making it necessary to analyze the specific system in detail.
As explained in Section~\ref{secalmostcomplex}, 
we expect that in most physical applications, no holomorphic connections will exist.
In this case, the unitary time evolution merely is an approximation.
In order to justify this approximation, we need to analyze the correction terms as
worked out in Theorems~\ref{thmerror} and~\ref{thmcommute}.

Before discussing these corrections, for clarity we point out
that the corrections to the unitary time evolution do not imply that the probabilistic interpretation
breaks down. Instead, the corrections lead to a mixing of the bra- and ket-state,
as is made precise by the operator~$E(t)$ in~\eqref{EtPhi}.
But this mixing preserves the norm on~$\Fock^* \otimes \Fock$.
Therefore, normalizing by~$\la \Phi | \Phi \ra_\F=1$, the expectation
value~${\mathcal{O}} (\la \Phi | \otimes | \Phi \ra_\Fock)$ defined in~\eqref{eval}
really has a sensible interpretation as the expectation value of a measurement by the
observable~${\mathcal{O}}$. In other words, the corrections to the unitary time evolution
are compatible with the probabilistic interpretation of quantum states.

We now explain the results of Theorems~\ref{thmerror} and~\ref{thmcommute} in some more detail.
The operator~$E(t)$ in~\eqref{EtPhi} describes a mixing of the holomorphic and anti-holomorphic
components of the jets. In other words, $E(t)$ mixes components of the bra- and ket-states of the
Fock space~$\Fock$. According to~\eqref{dyson}, the time evolution of this error is described by
a Dyson series on~$\Fock^* \otimes \Fock$. Since~$E(t)$ preserves the norm, the error
becomes apparent only if the expectation value with an observable~$\O$ is performed.
This is quantified in Theorem~\ref{thmcommute} by iterated commutators involving~$\O$.
These iterated commutators give
a good intuitive understanding of the corrections to the holomorphic approximation,
as we now explain. We consider the situation that we perform a measurement at time~$\tout$.
In this case, the field operators in the commutators in Theorem~\ref{thmcommute}
enter at a time~$\tau$ in the interaction region, whereas the observable~${\mathcal{O}}$
enters at time~$\tout$. As a consequence, the time evolution operators~$S$
in Theorem~\ref{thmcommute} must span at least the time $\tout-\tau$.
This opens the possibility that the error terms become small due to 
decoherence effects.
For simplicity, we first explain this effect for the contribution of first order
\[ \int_{\tin}^t S(t,\tau)\: {\mathcal{C}}\big( S(\tau,t)\,{\mathcal{O}}\, S(t,\tau) \big)\: S(\tau,t)\: d\tau \:. \]
Assume that the commutator at time~$\tau$ involves a phase factor which oscillates rapidly in~$\tau$.
Then the $\tau$-integral becomes small, implying that the error is no longer detectable at time~$t$.
For the higher order corrections, this decoherence effect is even stronger, because the iterated commutators
of order~$p$ involve operators at different times~$\tau_1, \tau_2, \ldots, \tau_p$,
giving more possibilities for destructive interference of phase factors.

In order to justify the holomorphic approximation, one must make this qualitative argument
mathematically precise, and one must quantify it including estimates of the error terms.
Here two effects specific to causal variational principles seem to be essential:
The first effect is that a critical measure~$\rho$ of a causal variational principle need not be diffeomorphic
to Minkowski space or to a spacetime manifold. Instead, it could consist of many
components. This so-called {\em{fragmentation}} of~$\rho$ as introduced in~\cite[Section~5]{perturb}
(see also~\cite[Section~5]{positive}) gives rise to the formula for~${\mathcal{O}}'$
in Theorem~\ref{commute} which involves additional sums over the subsystems.
This gives more freedom for phase factors to appear.
The second effect appears more specifically for the causal action principle for causal fermion systems
(see the textbook~\cite{cfs} and the references therein).
In this setting, the manifold~$\F$ is formed of linear operators on a Hilbert space.
The vectors in this Hilbert space can be represented by wave functions
in spacetime~$M:= \supp \rho$ (the so-called physical wave functions; see~\cite[\S1.1.4]{cfs}).
Likewise, the jets can be expressed by variations of these wave functions
(see~\cite[\S1.4.1]{cfs} and~\cite{action}). Modifying the phases of these wave functions
gives a simple way of obtaining the above-mentioned decoherence effects.
This so-called {\em{microscopic mixing}} of wave functions was introduced in~\cite{qft} 
for causal fermion systems formed of Dirac wave functions.

Clearly, the systematic study of these effects goes beyond the scope
of the present paper. It will be carried out separately in a future paper.

\Thanks{{{\em{Acknowledgments:}} We would like to thank Magdalena Lottner and
Marco Oppio for helpful comments
on the manuscript. We are grateful to the ``Universit\"atsstiftung Hans Vielberth'' for support.
N.K.'s research was also supported by the NSERC grant RGPIN~105490-2018.


\begin{thebibliography}{10}

\bibitem{gazeau}
S.T. Ali, J.-P. Antoine, and J.-P. Gazeau, \emph{Coherent {S}tates, {W}avelets,
  and their {G}eneralizations}, second ed., Theoretical and Mathematical
  Physics, Springer, New York, 2014.

\bibitem{bjorken2}
J.D. Bjorken and S.D. Drell, \emph{Relativistic {Q}uantum {F}ields},
  McGraw-Hill Book Co., New York, 1965.

\bibitem{bognar}
J.~Bogn{\'a}r, \emph{Indefinite {I}nner {P}roduct {S}paces}, Springer-Verlag,
  New York, 1974, Ergebnisse der Ma\-the\-matik und ihrer Grenzgebiete, Band
  78.

\bibitem{brunettibook}
R.~Brunetti, C.~Dappiaggi, K.~Fredenhagen, and J.~Yngvason~(eds),
  \emph{Advances in {A}lgebraic {Q}uantum {F}ield {T}heory}, Math. Phys. Stud.,
  Springer, 2015.

\bibitem{combescure}
M.~Combescure and D.~Robert, \emph{Coherent {S}tates and {A}pplications in
  {M}athematical {P}hysics}, Theoretical and Mathematical Physics, Springer,
  Dordrecht, 2012.

\bibitem{jacobson}
E.~Curiel, F.~Finster, and J.M. Isidro, \emph{Two-dimensional area and matter
  flux in the theory of causal fermion systems}, arXiv:1910.06161 [math-ph],
  Internat. J. Modern Phys. D \textbf{29} (2020), 2050098.

\bibitem{linhyp}
C.~Dappiaggi and F.~Finster, \emph{Linearized fields for causal variational
  principles: {E}xistence theory and causal structure}, arXiv:1811.10587
  [math-ph], Methods Appl. Anal. \textbf{27} (2020), no.~1, 1--56.

\bibitem{weyl}
C.~Dappiaggi, F.~Finster, and M.~Oppio, \emph{The algebra of observables
  generated by linearized fields for causal variational principles}, in
  preparation.

\bibitem{deligne+freed}
P.~Deligne and D.S. Freed, \emph{Classical field theory}, Quantum {F}ields and
  {S}trings: {A} {C}ourse for {M}athematicians, {V}ol. 1 ({P}rinceton, {NJ},
  1996/1997), Amer. Math. Soc., Providence, RI, 1999, pp.~137--225.

\bibitem{fefferman}
C.L. Fefferman, \emph{A sharp form of {W}hitney's extension theorem}, Ann. of
  Math. (2) \textbf{161} (2005), no.~1, 509--577.

\bibitem{continuum}
F.~Finster, \emph{Causal variational principles on measure spaces},
  arXiv:0811.2666 [math-ph], J. Reine Angew. Math. \textbf{646} (2010),
  141--194.

\bibitem{qft}
\bysame, \emph{Perturbative quantum field theory in the framework of the
  fermionic projector}, arXiv:1310.4121 [math-ph], J. Math. Phys. \textbf{55}
  (2014), no.~4, 042301.

\bibitem{cfs}
\bysame, \emph{The {C}ontinuum {L}imit of {C}ausal {F}ermion {S}ystems},
  arXiv:1605.04742 [math-ph], Fundamental Theories of Physics, vol. 186,
  Springer, 2016.

\bibitem{nrstg}
\bysame, \emph{Causal fermion systems: A primer for {L}orentzian geometers},
  arXiv:1709.04781 [math-ph], J. Phys.: Conf. Ser. \textbf{968} (2018), 012004.

\bibitem{positive}
\bysame, \emph{Positive functionals induced by minimizers of causal variational
  principles}, arXiv:1708.07817 [math-ph], Vietnam J. Math. \textbf{47} (2019),
  23--37.

\bibitem{action}
\bysame, \emph{The causal action in {M}inkowski space and surface layer
  integrals}, arXiv:1711.07058 [math-ph], SIGMA Symmetry Integrability Geom.
  Methods Appl. \textbf{16} (2020), no.~091.

\bibitem{perturb}
\bysame, \emph{Perturbation theory for critical points of causal variational
  principles}, arXiv:1703.05059 [math-ph], Adv. Theor. Math. Phys. \textbf{24}
  (2020), no.~3, 563--619.

\bibitem{review}
F.~Finster and M.~Jokel, \emph{Causal fermion systems: An elementary
  introduction to physical ideas and mathematical concepts}, arXiv:1908.08451
  [math-ph], {P}rogress and {V}isions in {Q}uantum {T}heory in {V}iew of
  {G}ravity (F.~Finster, D.~Giulini, J.~Kleiner, and J.~Tolksdorf, eds.),
  Birkh\"auser Verlag, Basel, 2020, pp.~63--92.

\bibitem{fockfermionic}
F.~Finster and N.~Kamran, \emph{Fermionic {F}ock spaces and quantum states for
  causal fermion systems}, arXiv:2101.10793 [math-ph] (2021).

\bibitem{dice2014}
F.~Finster and J.~Kleiner, \emph{Causal fermion systems as a candidate for a
  unified physical theory}, arXiv:1502.03587 [math-ph], J. Phys.: Conf. Ser.
  \textbf{626} (2015), 012020.

\bibitem{noether}
\bysame, \emph{Noether-like theorems for causal variational principles},
  arXiv:1506.09076 [math-ph], Calc. Var. Partial Differential Equations
  \textbf{55:35} (2016), no.~2, 41.

\bibitem{jet}
\bysame, \emph{A {H}amiltonian formulation of causal variational principles},
  arXiv:1612.07192 [math-ph], Calc. Var. Partial Differential Equations
  \textbf{56:73} (2017), no.~3, 33.

\bibitem{osi}
\bysame, \emph{A class of conserved surface layer integrals for causal
  variational principles}, arXiv:1801.08715 [math-ph], Calc. Var. Partial
  Differential Equations \textbf{58:38} (2019), no.~1, 34.

\bibitem{greene-shiohama}
R.E. Greene and K.~Shiohama, \emph{Diffeomorphisms and volume-preserving
  embeddings of noncompact manifolds}, Trans. Amer. Math. Soc. \textbf{255}
  (1979), 403--414.

\bibitem{moretti-book}
V.~Moretti, \emph{Spectral {T}heory and {Q}uantum {M}echanics}, Unitext, vol.
  110, Springer, Cham, 2017, Mathematical foundations of quantum theories,
  symmetries and introduction to the algebraic formulation, Second edition.

\bibitem{peskin+schroeder}
M.E. Peskin and D.V. Schroeder, \emph{An {I}ntroduction to {Q}uantum {F}ield
  {T}heory}, Addison-Wesley Publishing Company Advanced Book Program, Reading,
  MA, 1995.

\end{thebibliography}
\providecommand{\bysame}{\leavevmode\hbox to3em{\hrulefill}\thinspace}
\providecommand{\MR}{\relax\ifhmode\unskip\space\fi MR }
\providecommand{\MRhref}[2]{%
  \href{http://www.ams.org/mathscinet-getitem?mr=#1}{#2}
}
\providecommand{\href}[2]{#2}

\end{document}